\documentclass[iop]{emulateapj}

\usepackage{graphicx}
\usepackage{amsmath,amssymb}
\usepackage{natbib}
\newcommand{\Teff}{T_\mathrm{eff}}
\newcommand{\Msun}{M_\odot}

\newcommand{\dV}{\mathrm{d}V}
\newcommand\av[1]{\overline{\langle{#1}\rangle}}
\newcommand\fav[1]{\widetilde{\langle #1 \rangle}}
\newcommand\br[1]{\langle #1\rangle}

\shorttitle{Turbulent convection in stellar interiors. III.}
\shortauthors{M. Viallet et al.}

\begin{document}


\title{Turbulent convection in stellar interiors. III. Mean-field analysis and stratification effects.}


\author{Maxime Viallet$^{1,2}$, Casey Meakin$^{3,4,5}$, David Arnett$^{4}$, Miroslav Moc\'ak$^{3}$}

%
%


\altaffiltext{1}{Physics and Astronomy, University of Exeter, Stocker Road, Exeter, UK EX4 4QL}
\altaffiltext{2}{Max-Planck-Institut f\"ur Astrophysik, Karl Schwarzschild Strasse 1, Garching, D-85741}
\altaffiltext{3}{Theoretical Division, Los Alamos National Laboratory, Los Alamos, NM 87545}
\altaffiltext{4}{Steward Observatory, University of Arizona, Tucson, AZ 85721}
\altaffiltext{5}{New Mexico Consortium, Los Alamos, NM 87544}


\begin{abstract}
We present 3D implicit large eddy simulations (ILES) of the turbulent convection in the envelope of a 5~$\Msun$ red giant star and in the oxygen-burning shell of a 23~$\Msun$ supernova progenitor. The numerical models are analyzed in the framework of 1D Reynolds-Averaged Navier-Stokes  (RANS) equations. The effects of pressure fluctuations are more important in the red giant model, owing to larger stratification of the convective zone. We show how this impacts different terms in the mean-field equations. We clarify the driving sources of kinetic energy, and show that the rate of turbulent dissipation is comparable to the convective luminosity. Although our flows have low Mach number and are nearly adiabatic, our analysis is general and can be applied to photospheric convection as well. The robustness of our analysis of turbulent convection is supported by the insensitivity of the mean-field balances to linear mesh resolution. We find robust results for the turbulent convection zone and the stable layers in the oxygen-burning shell model, and robust results everywhere in the red giant model, but the mean fields are not well converged in the narrow boundary regions (which contain steep gradients) in the oxygen-burning shell model. This last result illustrates the importance of unresolved physics at the convective boundary, which governs the mixing there.
\end{abstract}

\keywords{}



\section{Introduction}

Since the original publication of the mixing length theory (MLT) by \cite{bohm-vitense_uber_1958}, much effort has been devoted to the improvement of the theory \citep{gough_mixing-length_1977,stellingwerf_convection_1982-1,xiong_evolution_1986,kuhfuss_model_1986,canuto_stellar_1991,canuto_turbulent_1992,gehmeyr_new_1992,wuchterl_simple_1998,deng_anisotropic_2006}. Nevertheless, MLT still remains the standard choice in most state-of-the-art stellar evolution codes.

With the wealth of data coming from asteroseismology missions (CoRot, Kepler), and expected from future observatories (Gaia, JWST), a new generation of stellar models is needed.  The modeling of solar-like oscillations requires reliable models for the Reynolds stresses \citep{belkacem_closure_2006, samadi_amplitudes_2012}. The interaction between convection and pulsations, which sets the location of the red edge of the instability strip, needs a better time-dependent theory of turbulent convection \citep{buchler_turbulent_2000}. In the deep interior, additional mixing is required at convective boundaries across the Hertzprung-Russell diagram, above convective cores \citep{maeder_stellar_1975, matraka_overshooting_1982, schroder_critical_1997}, and below convective envelopes \citep{herwig_evolution_2000, pace_lithium_2012}. Lacking a physically consistent description of this process \citep{renzini_embarrassments_1987}, extra-mixing is currently included in stellar evolution codes with ad-hoc parameterizations, so that predictive powers are hampered by the use of free parameters. Next generation stellar evolution models should rely on a consistent description of convective boundary mixing, together with the effect of internal waves induced by turbulence.

The road to a satisfying theory of turbulent convection is difficult. Stellar convection is highly turbulent, with Reynolds and Rayleigh numbers having ``astronomical" values (Re $> 10^{10}$, Ra $> 10^{20}$). The on-going development of computational physics allows numerical modeling of turbulent systems having an increasing number of degrees of freedom, but 
at present no direct simulation of the problem is possible. Nevertheless, physical insight provided by computer simulations is invaluable in improving our understanding of stellar hydrodynamical processes. The path starting from large 3D data sets and ending with a recipe simple enough to be implemented in stellar evolution codes is not straightforward. Ideally, a common framework should be used both for the analysis of multi-D data and for stellar evolution calculations, strengthening the underlying connection and making the projection from 3D to 1D easier. Reynolds-Averaged Navier-Stokes (RANS) equations are a promising framework for this. Much effort has been already devoted to RANS in the context of stellar hydrodynamics, see e.g. \cite{canuto_compressible_1997,xiong_nonlocal_1997,canuto_stellar_1998,deng_anisotropic_2006,canuto_stellar_2011-3} and references therein. We adopt the same methodology. 

\begin{figure*}[t]
\parbox{0.49\linewidth}{\center \includegraphics[width=0.8\linewidth]{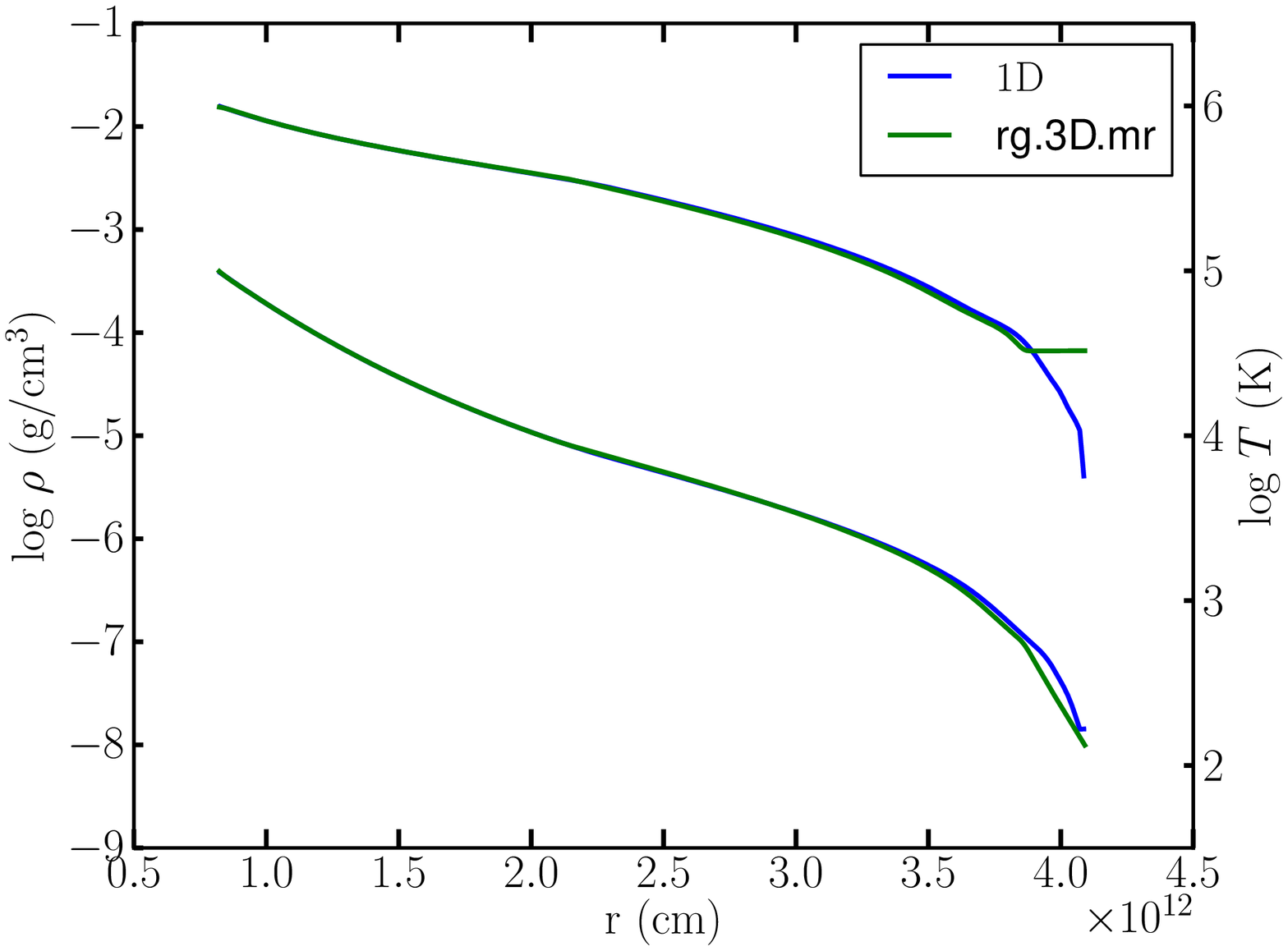}}
\parbox{0.49\linewidth}{\center \includegraphics[width=0.8\linewidth]{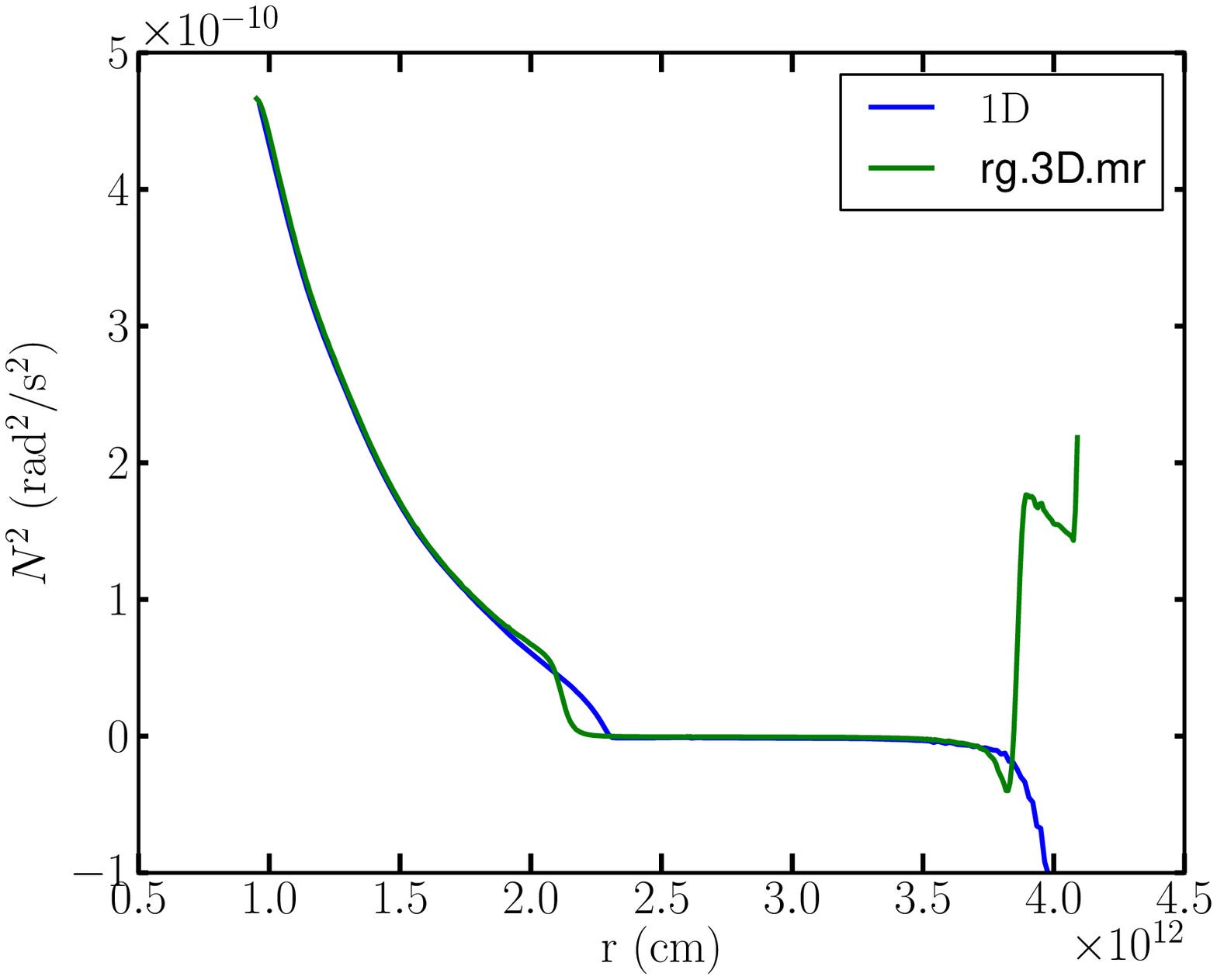}}
\caption{Overview of the simulated region in the red giant model. Left panel: density (bottom curves) and temperature (top curves) stratification in the initial (1D) and 3D model {\sf rg.3D.mr}. Right panel: squared Brunt-V\"ais\"al\"a frequency in the initial (1D) and 3D model {\sf rg.3D.mr}.}
\label{fig:rg_structure}
\end{figure*}

A theory of turbulent convection should have a broad range of applicability: ideally to every type of star at all stages of evolution. Therefore, it is important to identify fundamental properties of the physical process and to understand what changes with different stellar conditions. This is an ambitious project, which we start by considering two very different astrophysical cases: the convection in the envelope of a 5 $\Msun$ red giant star, and the convection in the oxygen-burning shell of a 23 $\Msun$ supernova progenitor. 

The paper is organized as follows: we present in Sect.~\ref{setup} an overview of our stellar models and numerical tools. In Sect.~\ref{1DRANS}, we develop a set of 1D RANS equations as a framework to analyze our data. Results are presented in Sect.~\ref{results}. In Sect.~\ref{conclusion}, we summarize our findings and conclude the paper.

\section{Stellar models and numerical methods}
\label{setup}

In this section we describe the stellar models and the hydrodynamic codes, and summarize the properties of the 3D models.

\subsection{Red giant models}


The red giant model was presented in \cite{viallet_towards_2011}. The initial 1D structure was constructed by integrating the stellar structure equations with the following input parameters: $M=5 \Msun$, $\Teff = 4500$ K, $\log(L/L_\sun)=3.1$. In terms of stellar evolution phase, such a model would correspond to a 5 $\Msun$ star at the end of central He burning, finishing its blue loop and evolving toward lower $\Teff$ and away from the red edge of the Cepheid Instability Strip \citep[see][their Fig. 1 and Tab. 6]{alibert_period-luminosity-color-radius_1999}. 
The stellar structure code is described in \cite{baraffe_evolution_1991}. It uses mixing-length theory (with $\alpha=1.7$) to treat convection, and the extent of the convective region is based on the Schwarzschild criterion. The structure was integrated from the photosphere down to 20 \% of the stellar radius, stopping to avoid the nuclear burning region. This initial stratification is used as an input model for the multi-D hydrodynamic code. The initial stellar structure is shown in Fig. \ref{fig:rg_structure}. The left panel shows the temperature and density stratifications. The model is characterized by a total density stratification $\log(\rho_\mathrm{bottom}/\rho_\mathrm{top}) \sim 4.4$. The total pressure stratification (not shown) is $\log(p_\mathrm{bottom}/p_\mathrm{top}) \sim 6.2$, or equivalently $\sim 14.3$ pressure scale-heights. The right panel shows the radial profile of $N^2$, the Brunt-V\"ais\"al\"a  frequency squared. The convective region extends down to $r \sim 2.3\times 10^{12}$ cm, nearly half of the star in radius. The surface layers are characterized by a strong superadiabatic stratification.

As in \cite{viallet_towards_2011}, we use a proxy for the surface layers. The surface layers are numerically difficult to handle: the decrease in the pressure scale-height yields small scale convective eddies which are difficult to resolve when a single grid is used \citep[see discussion in][]{viallet_towards_2011}. To mimic surface cooling, we apply a Newtonian cooling term in the last 5 \% of the star:

\begin{equation}
\label{eq:surface_cooling}
q = \rho c_v \frac{T - T_0}{\tau} f(r),
\end{equation}

\noindent where $f(r)$ is a spatially varying function that is equal to 1 above a given radius $r_c$ with a smooth transition to zero, $\tau$ is the cooling timescale and $T_0$ is the forcing temperature. We use the same parameters as in \cite{viallet_towards_2011}: $r_c=0.95 R_\star$, $T_0=32\,750$ K, and $\tau = 10^4$ s. About 4.5 pressure scale-heights of the initial 1D model are absorbed into the Newtonian cooling region.

We solve the equations describing the evolution of density, momentum, and total energy for a single fluid, taking into account gravity, radiative diffusion, and surface cooling:

\begin{eqnarray}
\frac{\partial}{\partial t} \rho &=& - \vec \nabla \cdot (\rho \vec u),\\
\frac{\partial}{\partial t} \rho \vec u &=& - \vec \nabla \cdot (\rho \vec u\otimes \vec u)-\vec \nabla p + \rho \vec g,\\ 
\frac{\partial}{\partial t} \rho \epsilon_t &=& -\vec \nabla \cdot (\rho \epsilon_t \vec u + p \vec u) + \rho \vec u \cdot \vec g + \vec \nabla \cdot (\chi \vec \nabla T) - q,
\end{eqnarray}

\noindent where $\rho$ is the density, $\vec u$ the velocity, $\epsilon_t = \epsilon_i + \epsilon_k$ the specific total energy ($\epsilon_i$ is the specific internal energy and $\epsilon_k$ the specific kinetic energy), $p$ the gas pressure, $T$ the temperature, $\vec g$ the gravitational acceleration, and $\chi$ the thermal conductivity. For photons, the thermal conductivity is given by

\begin{equation}
\label{eq:chirad}
\chi = \frac{16 \sigma T^3}{3\kappa \rho},
\end{equation}

\noindent where $\kappa$ is the Rosseland mean opacity, and $\sigma$ the Stefan-Boltzmann constant. 

\begin{figure*}[t]
\parbox{0.49\linewidth}{\center \includegraphics[width=0.8\linewidth]{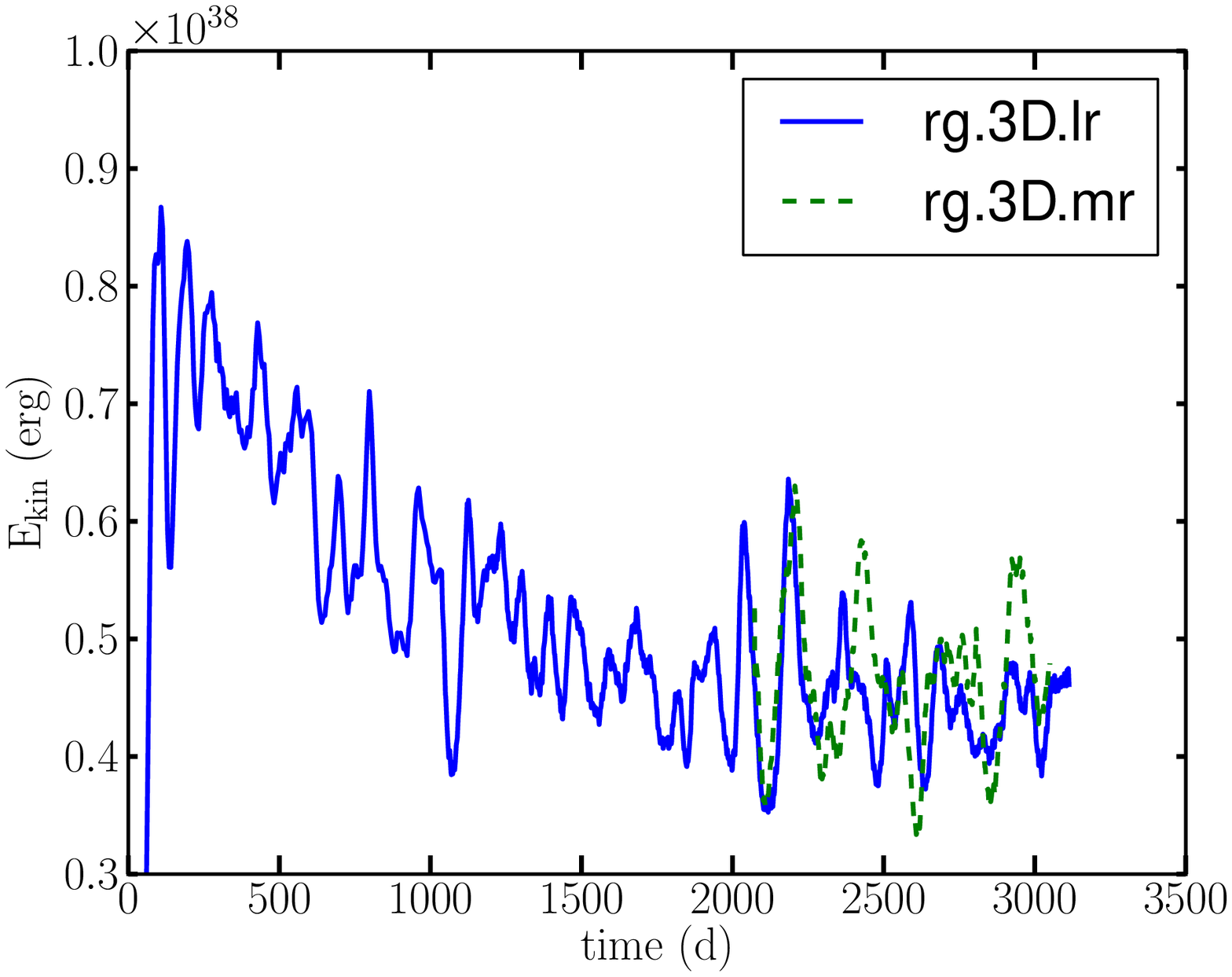}}
\parbox{0.49\linewidth}{\center \includegraphics[width=0.8\linewidth]{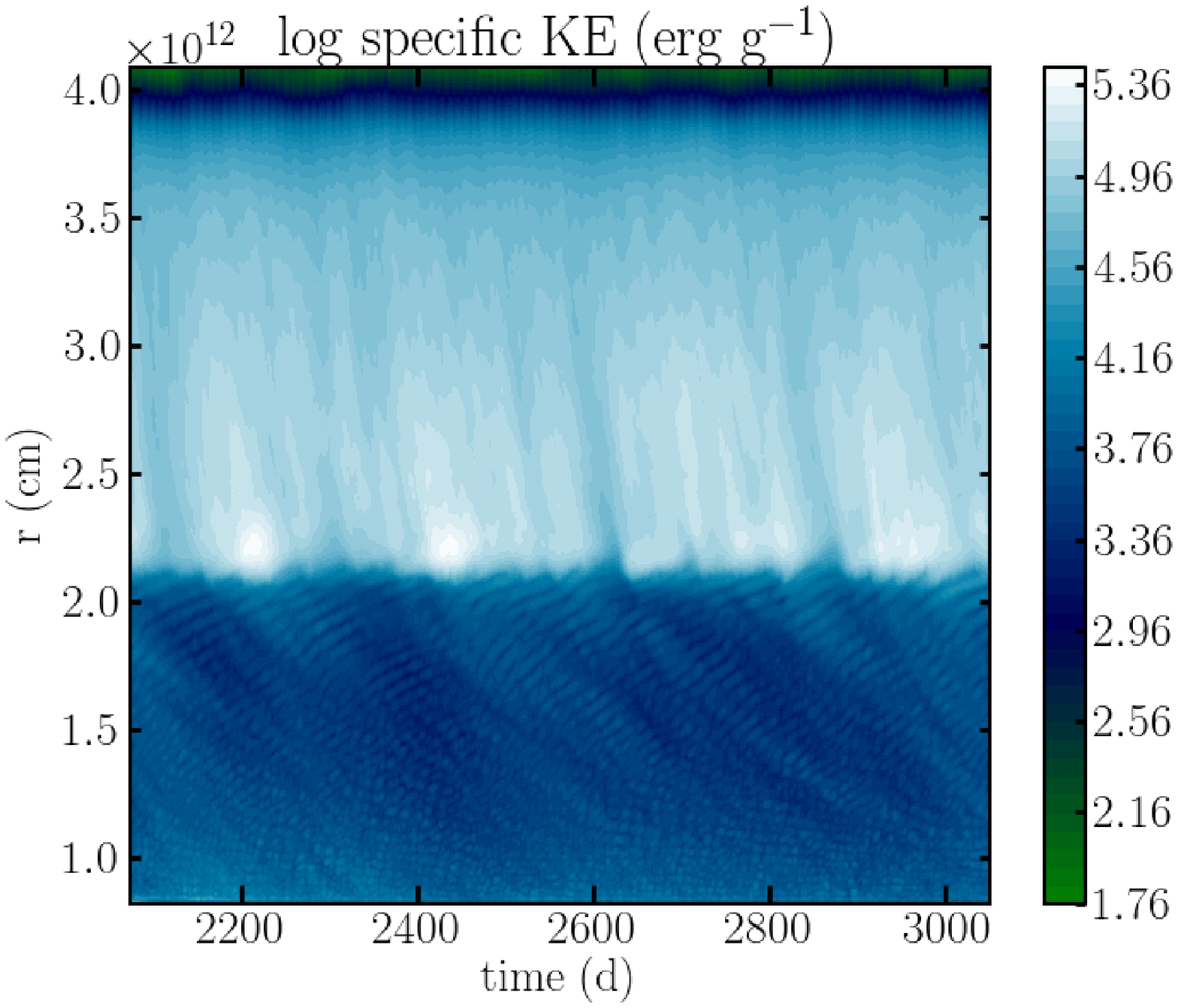}}
\caption{Left panel: evolution of the total kinetic energy in the red giant models. Right panel: space-time diagram of the specific kinetic energy in model {\sf rg.3D.mr}.}
\label{fig:rg_ekin}
\end{figure*}

The 3D simulations were performed with the MUSIC code, as described in \cite{viallet_towards_2011} and \cite{viallet_comparison_2013} (submitted to A\&A). The spatial method is based on a finite volume discretization of a spherical wedge (``box in a star" approach). The method is based on staggered velocity components, and has been extended to 3D for this work. The time-marching scheme used for the models presented in this paper is based on the Minimum Residual Approximate Implicit Scheme from \cite{botchev_stability_1999}. The MUSIC code is optimized to run on parallel computers and it uses domain-decomposition to distribute the computation over the computational nodes. The Message Passing Interface (MPI) is used to handle communications of boundary data. We use periodic boundary conditions in horizontal directions, and non-penetrative stress-free conditions at the bottom and top of the domain. As the nuclear burning region is not included in the computational domain, a radiative flux corresponding to the stellar luminosity is imposed at the inner boundary.

We do not model viscosity explicitly in the equations. The expected value of the molecular viscosity in stellar interiors implies huge Reynolds numbers (larger than $10^{10}$). It is therefore impossible to model all scales of the flow, from the stellar scale down to the dissipation scale, on current generation of  computers. We adopt the Implicit Large Eddy Simulation paradigm \citep[ILES][]{ILES_grinstein} and solve the inviscid equations to model the turbulent flow. The underlying motivation of ILES is that monotonic, finite-volume based methods have physical properties of the Navier-Stokes equations ``built-in'' within the numerics (unlike spectral and finite difference methods). The conservation properties of finite volumes schemes and the monotonicity preserving property enforce correct physical behavior at the grid scale. As a result, the loss of information that takes place at the grid scale mimics turbulent dissipation (see \ref{rans:analysis_ek}).

\begin{figure}[t]
\center \includegraphics[width=0.8\linewidth]{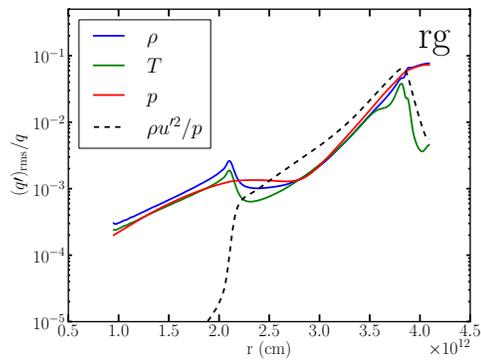}
\caption{Radial profiles of the rms fluctuations of density, temperature, and pressure (continuous lines) in model {\sf rg.3D.mr}. The black dashed line shows $\rho u'^2 / p$ for comparison (see discussion in Sect. \ref{pressure_fluctuations}).}
\label{fig:rg_thermodynamical_perturbations}
\end{figure}

\begin{figure*}
\centering
\parbox{0.49\linewidth}{\center \includegraphics[width=0.8\linewidth,clip=true,trim=20 20 20 20]{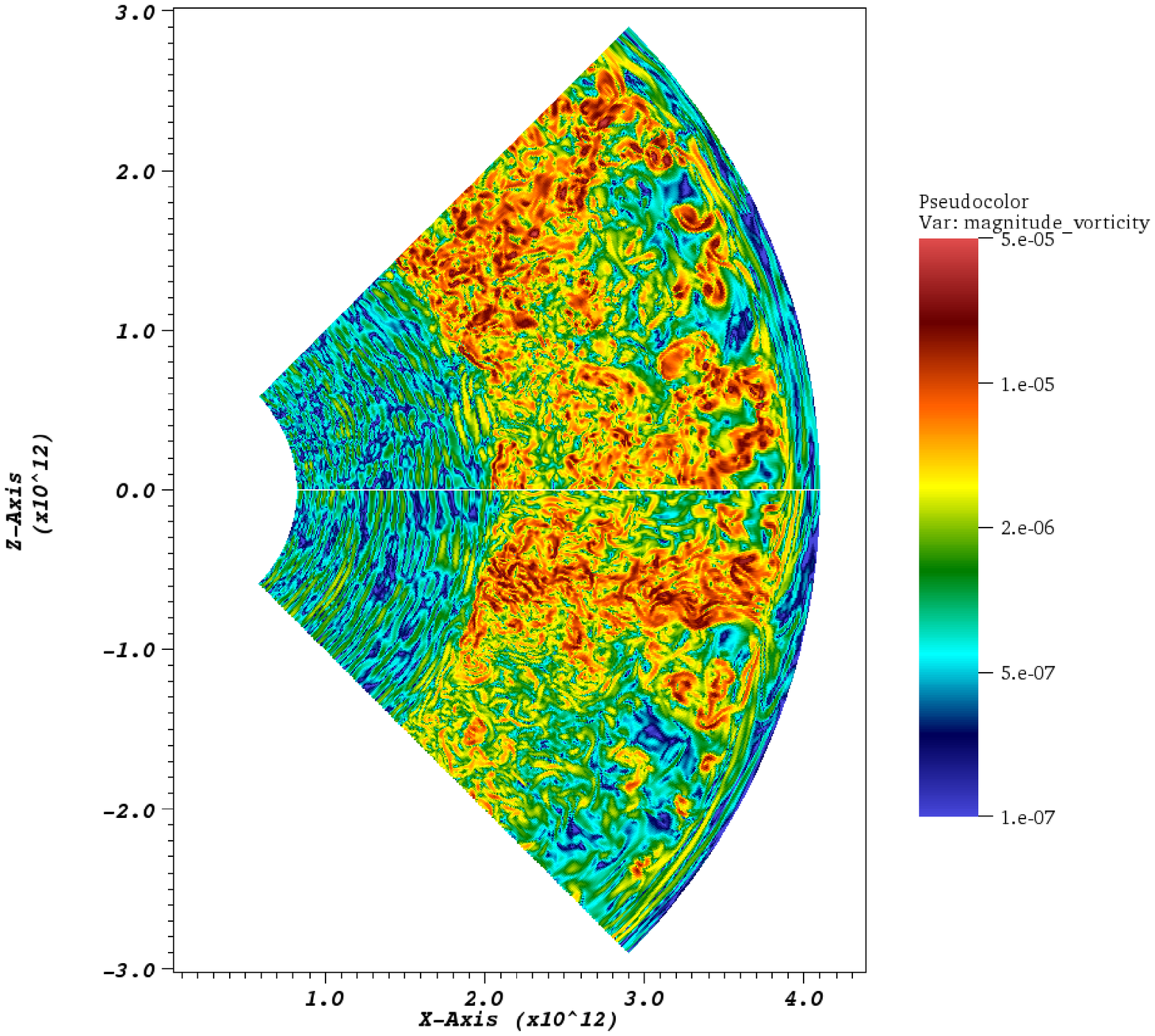}}
\parbox{0.49\linewidth}{\center \includegraphics[width=0.8\linewidth,clip=true]{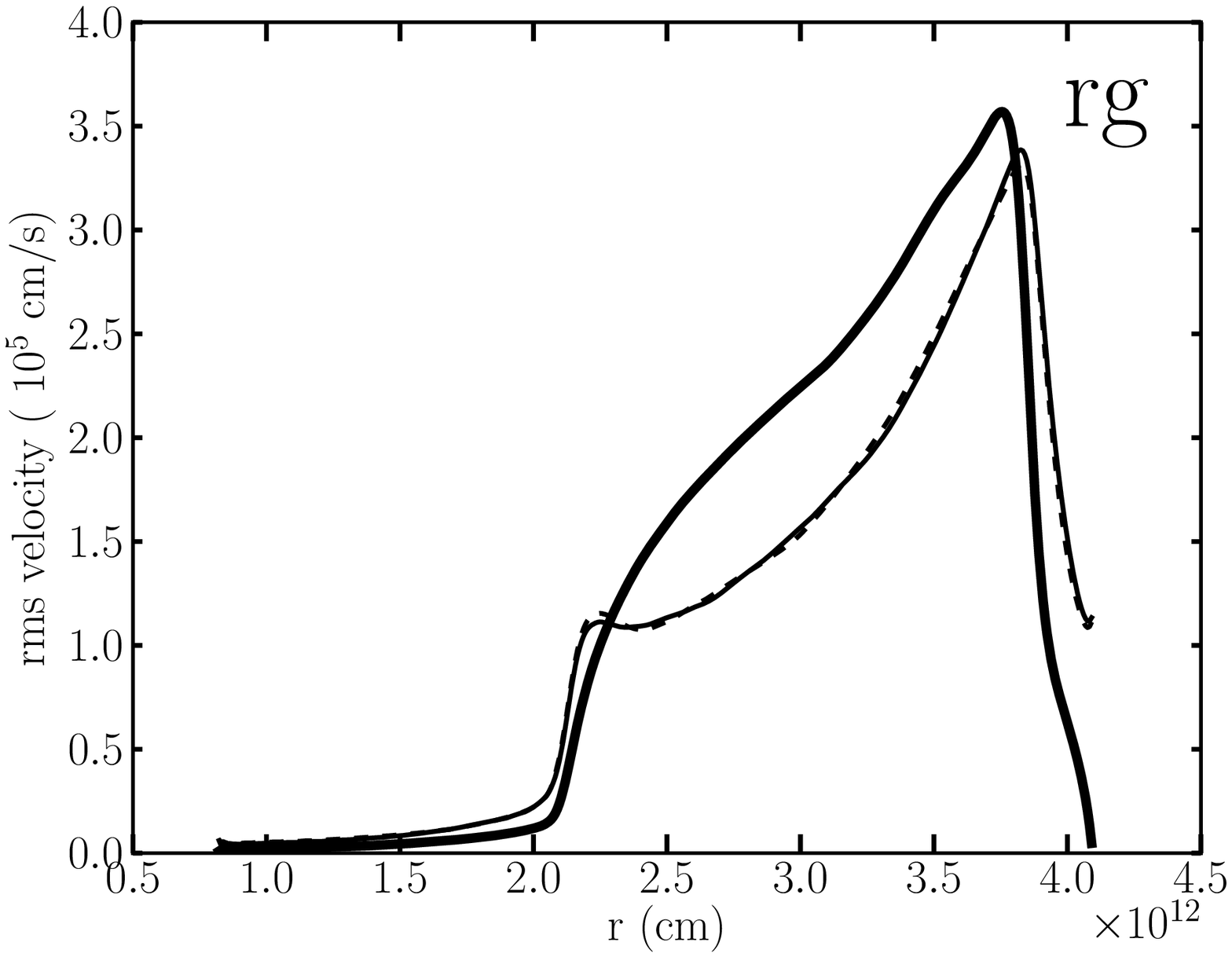}}
\caption{Left panel: snapshots of the computational domain for model {\sf rg.3d.mr}. The magnitude of the vorticity $|| \vec \omega ||$ is shown (two perpendicular cuts in the vertical direction were taken and are plotted together). Right panel: Radial profiles of rms velocity components for model {\sf rg.3d.mr} : $u_r$ (thick solid line), $u_\theta$ (thin solid line), and $u_\phi$ (thin dashed line).}
\label{fig:rg_velocity}
\end{figure*}

Models with two different resolutions were computed: model {\sf rg.3D.lr} with a $216\times128^2$ resolution, and model {\sf rg.3D.mr} with a $432\times256^2$ resolution. The main properties of these models are summarized in Table \ref{table:rg_runs}. We use an opening angle of $45^\circ \times 45^\circ $. Model {\sf rg.3D.lr} was evolved for 3100 days of model time (for $\sim 3.6\times 10^4$ CPU hours on 512 cores) and model {\sf rg.3D.mr} was evolved for 1050 days of model time (for $\sim 2\times 10^5$ CPU hours on 2048 cores), starting from the solution at $t=2000$ d of model {\sf rg.3D.lr}. The computations were performed on the IBM iDataPlex supercomputer ``Hydra" at the Rechen-zentrum Garching, and on the SGI Altix ICE 8200 supercomputer ``Zen" at the University of Exeter.  At the beginning of the computation, the Newtonian cooling term modifies the top layers and this triggers the convective instability. Downdrafts form and sink. They break up rapidly due to hydrodynamical instabilities, and the flow becomes turbulent. Left panel of Fig. \ref{fig:rg_ekin} shows the evolution of the total kinetic energy in the domain.  The initial rise of the total kinetic energy is strong, followed by a slow decay toward a quasi steady state. The left panel of Fig. \ref{fig:rg_structure} shows the temperature and density stratifications obtained in the 3D model when a quasi-steady state is reached. The effect of the Newtonian cooling is to drive an isothermal region at the top. The right panel of Fig. \ref{fig:rg_structure} shows the profile of $N^2$ in the 3D model. The most striking feature is the modification of the radial profile of $N^2$, which shows a much steeper slope at the bottom boundary of the convective zone in the 3D model than in the initial 1D model. This is an evidence for overshooting, and will be discussed in Sect. \ref{penetration}. In the 3D model, the density stratification is $\log(\rho_\mathrm{bottom}/\rho_\mathrm{top}) \sim 2.3$ in the convective zone. The pressure stratification is $\log(p_\mathrm{bottom}/p_\mathrm{top}) \sim 3.4$ in the convective zone, or equivalently $\sim 7.8$ pressure scale-heights. Clarifying the influence of such a large stratification on the properties of the turbulent convection is an important focus of this work. Figure \ref{fig:rg_thermodynamical_perturbations} shows the radial profiles of the rms values of the density, temperature, and pressure fluctuations, normalized by their mean values. In the bulk of the convective region, all fluctuations have the same relative order of magnitude.



Computer animations and snapshots of the models illustrate the strong asymmetry of the flow, with the presence of plumes triggered by cooling at the surface that sink in the convective zone within much slower upflowing material. This is due to the degree of stratification, as emphasized by previous 3D simulations \citep{stein_topology_1989,cattaneo_turbulent_1991,brummell_turbulent_1996,porter_three-dimensional_2000,brummell_penetration_2002-4}. These plumes form a ``network" of downflows in strong interaction, and from time to time plumes coalesce to form a strong downdraft which sinks through the whole stratification. The right panel of Fig. \ref{fig:rg_ekin} shows a space-time diagram of the specific kinetic energy, which shows the imprint of these sporadic events. The left panel of Fig. \ref{fig:rg_velocity} is a snapshot of the flow, as seen in the magnitude of the vorticity vector field $\vec \omega = \vec \nabla \times \vec u$ and emphasizes the prominence of small scale structures, characteristic of 3D turbulence. The right panel of Fig. \ref{fig:rg_velocity} shows the rms values of the velocity components $u_r$, $u_\theta$, $u_\phi$. The tangential and azimuthal components are roughly equal as the angular directions are homogeneous. Including rotation and/or magnetic fields would break this symmetry. The corresponding Mach number goes from $\sim 0.1$ at the top of the convective zone to $\sim 0.01$ at the bottom. As in \cite{arnett_turbulent_2009}, we define a global rms velocity such as 

\begin{equation}
\frac{1}{2}M_\mathrm{CZ} v_\mathrm{rms}^2 = E_\mathrm{k,CZ},
\end{equation}

\noindent where $M_\mathrm{CZ}$ and $E_\mathrm{k,CZ}$ are the total mass and kinetic energy in the convective zone (CZ), respectively. The convective zone is taken as the region between $r_\mathrm{in} = 2\times 10^{12}$ cm and $r_\mathrm{in} = 4 \times 10^{12}$ cm, but we have checked that results are not too sensitive to these values. We find $v_\mathrm{rms} = 2.34\times 10^5$ cm/s for model {\sf rg.3D.mr}. We define the convective turn-over timescale as

\begin{equation}
\tau_\mathrm{conv} = 2\frac{l_\mathrm{CZ}}{v_\mathrm{rms}},
\end{equation}

\noindent which yields $\tau_\mathrm{conv} = 198$ d for model {\sf rg.3D.mr}.

Finally, Fig. \ref{fig:rg_velocity} and the right panel of Fig. \ref{fig:rg_ekin} hint at the presence of g-modes in the underlying radiative zone. The modes are excited at the convective boundary layer. In this paper, we will focus on the dynamics in the convective region, and leave the study of wave excitation and dynamics for future work.
\begin{deluxetable}{l c c}
2250 3050
\tablecaption{\label{table:rg_runs} Summary of the Red Giant Simulations.}
\tablehead{Parameter  & {\sf rg.3D.lr} & {\sf rg.3D.mr} }
\startdata
Grid zoning & 216$\times 128^2$ & 432$\times 256^2$ \\
$r_\mathrm{in}$, $r_\mathrm{out}$ ($10^{12}$ cm) & 0.82, 4.09 & 0.82, 4.09 \\
$\Delta \theta$, $\Delta \phi$ & 45$^\circ$, 45$^\circ$ & 45$^\circ$, 45$^\circ$\\
CZ stratification ($H_p$) & 7.8 & 7.8 \\
$t_\mathrm{av} (\Delta t_\mathrm{av})$ (days) & $2650 (800)$ & $2650(800)$\\
$v_\mathrm{rms}$ ($10^5$ cm/s) & 2.27 &  2.34\\
$\tau_\mathrm{conv}$ (days) &  204 & 198 \\
$L_\mathrm{\star}$ ($10^{36}$ erg/s) & 4.9 & 4.9\\
$L_\mathrm{d}$ ($10^{36}$ erg/s) & 7.33 & 7.39\\
$l_\mathrm{d}$ ($10^{11}$ cm) & 7.0 & 7.7\\
$\tau_\mathrm{d}$ (days) & 18 & 19\\
Pe & 4900 & 5200
\enddata
\tablecomments{$v_\mathrm{rms}$: global rms velocity, $\tau_\mathrm{conv}$: convective turnover timescale, $L_\star$: luminosity of the stellar model, $L_d$: rate of kinetic energy dissipation, $l_d$: dissipation length-scale, $\tau_d$: dissipation time-scale, Pe: P\'eclet number.}
\end{deluxetable}

\subsection{Oxygen-burning shell models}
\label{ob}

\begin{figure*}[t]
\parbox{0.49\linewidth}{\center \includegraphics[width=0.8\linewidth]{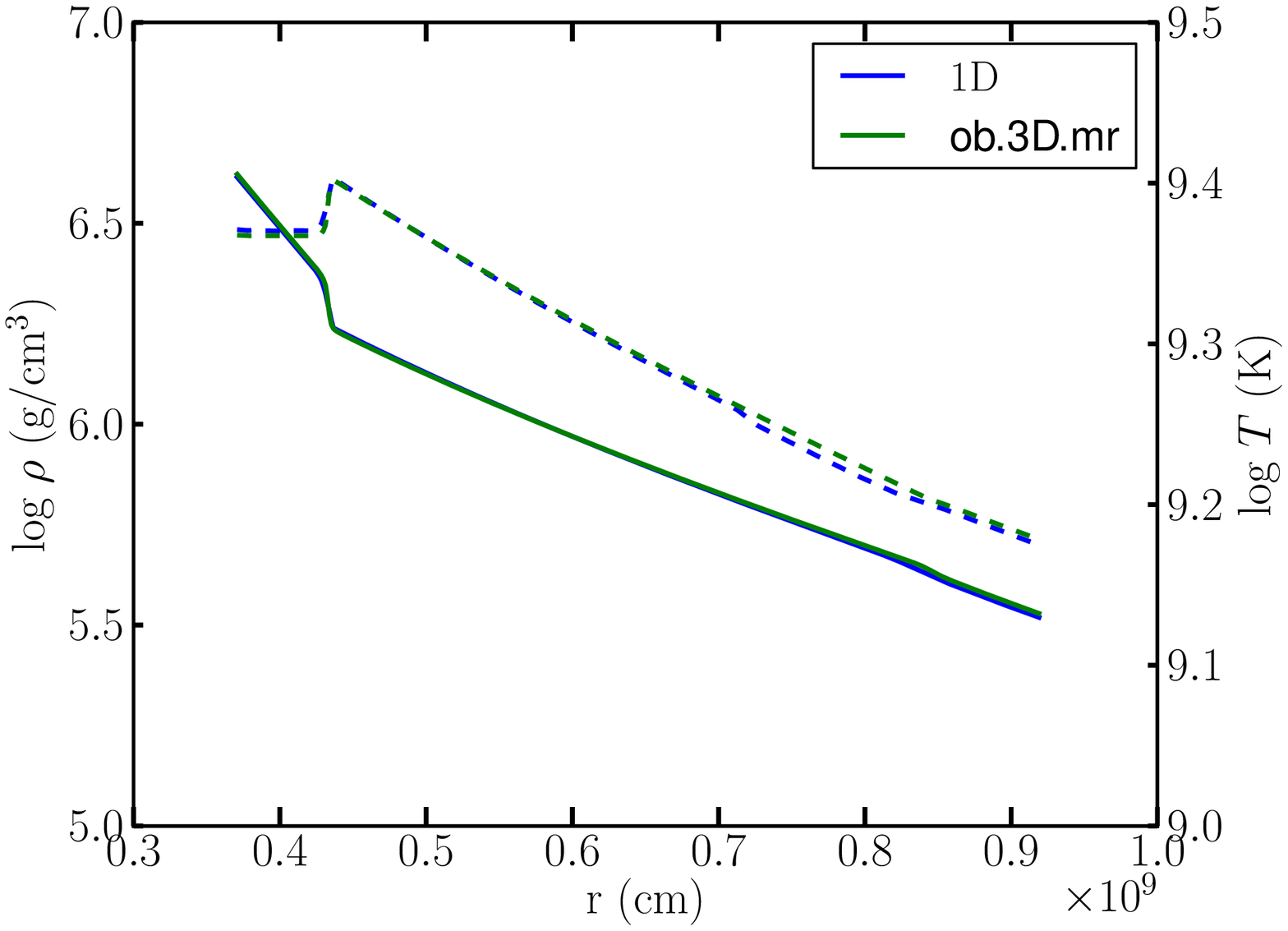}}
\parbox{0.49\linewidth}{\center \includegraphics[width=0.8\linewidth]{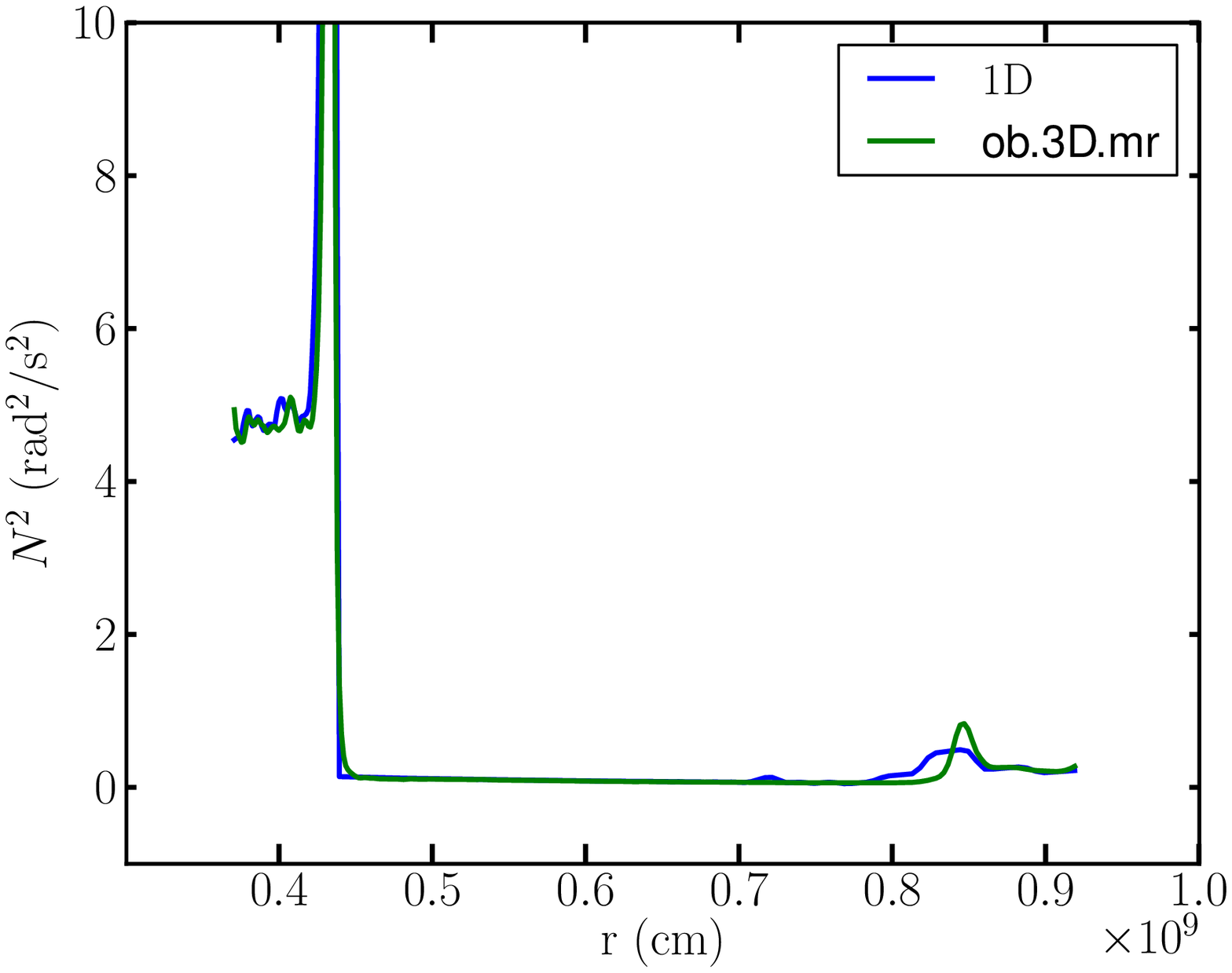}}
\caption{Overview of the simulated region in the oxygen-burning shell model for the initial model (1D) and for the 3D model {\sf ob.3D.mr}. Left panel: density (continuous line) and temperature (dashed line) stratification. Right panel: squared Brunt-V\"ais\"al\"a frequency.}
\label{fig:ob_structure}
\end{figure*}

\par We simulate the turbulent flow in a convective oxygen burning shell using as initial conditions a 23 M$_{\odot}$ stellar model of solar composition that was  previously evolved with the TYCHO stellar evolution code \citep{arnett_turbulent_2009} up to a point where oxygen, neon, carbon, helium, and hydrogen are burning in concentric shells about a silicon-sulfur rich core.  Additional details can be found in \cite{meakin_active_2006,meakin_turbulent_2007}.  As in \cite{meakin_turbulent_2007}, we study only the oxygen burning convection zone and its stably stratified bounding layers. 

\par Reactive-hydrodynamic evolution was simulated with the PROMPI code \citep{meakin_turbulent_2007}, a distributed memory adaptation of the PROMETHEUS code \citep{fryxell_computation_1989}, an Eulerian implementation of the piecewise parabolic method (PPM) hydrodynamics scheme \citep{colella_piecewise_1984} extended to treat realistic equations of state \citep{colella_efficient_1985}, multi-species advection, and a general nuclear reaction network. The setup for these simulations are nearly identical to that described in \cite{meakin_turbulent_2007}, including the 25 species followed to track nuclear evolution, and differ only in terms of resolution and domain size.  As with the MUSIC code, PROMPI solves the inviscid equations, and we rely on the ILES paradigm to model sub-grid scale dissipation. The major differences between the two is that PROMPI is multi-fluid, includes nuclear burning, and that radiative diffusion is negligible in this context \citep{arnett_supernovae_1996}. Figure \ref{fig:ob_structure} shows the initial density and temperature stratifications (left panel), and the initial profile of $N^2$ (right panel). The latter is characterized by a narrow peak at the bottom boundary of the convective zone, due to the sharp composition gradient that characterizes the boundary of the nuclear burning region. We refer the reader to  \cite{meakin_turbulent_2007} for similar figures as those shown earlier for the red giant model.

\par As with the red giant models, the oxygen shell burning models studied in this paper use a $45^\circ \times 45^\circ $ opening angle.  Models with three different spatial resolutions were  computed: model {\sf ob.3D.lr} with $192\times128^2$ zones, model {\sf ob.3D.mr} with $384\times256^2$ zones, and model {\sf ob.3D.hr} with $768\times512^2$ zones. The main properties of these models are summarized in Table \ref{table:ob_runs}. Models {\sf ob.3D.lr} and {\sf ob.3D.mr}  are used for comparison with the red giant models, whereas model {\sf ob.3D.hr} is used to further assess the numerical convergence of our results with resolution (see Sect. \ref{resolution_effects}).  All of the oxygen burning models were computed on the University of Tennessee's Kraken Cray XT5.  Model {\sf ob.3D.lr} was evolved for approximately 600~s of model time (on 768 cores for $\sim 5\times 10^4$ CPU-hours). Model {\sf ob.3D.mr} was evolved for approximately 280~s of model time (on 12,288 cores for $\sim 3.5\times 10^5$ CPU-hours) starting from the solution at $t=300$~s of model {\sf ob.3D.lr}. Model {\sf ob.3D.hr} was evolved for approximately 200~s of model time (on 24,576 cores  for $\sim 4\times 10^6$ CPU-hours) starting from the solution at $t=310$ s of model {\sf ob.3D.mr}. 

\begin{figure}[t]
\center \includegraphics[width=0.8\linewidth]{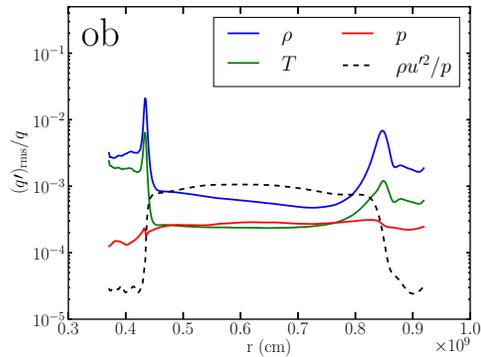}
\caption{Same as Fig. \ref{fig:rg_thermodynamical_perturbations}, but for the oxygen-burning shell model {\sf ob.3D.mr}.}
\label{fig:ob_thermodynamical_perturbations}
\end{figure}

At the beginning of the computation, the convective instability is triggered by a band of random, small amplitude density perturbations ($\rho'/\rho \sim 10^{-4}$) imposed within the convection zone. The flow rapidly becomes turbulent as it fills the convectively unstable region. The early transient evolution of the models is characterized by a strong penetration of the flow in the top stable region. After roughly 300 s, a quasi steady-state obtains with a slow evolution resulting from a net heating due to the imbalance between nuclear burning and neutrino cooling
(a common feature of neutrino-cooled stages of stellar evolution which results in growth of the convective region). As a result, global characteristics of the model (e.g., total nuclear burning, total kinetic energy, etc.) increase slowly with time. The nuclear evolution time scale for this phase is roughly $5\times 10^3$ s. The convective region is characterized by a density stratification $\rho_\mathrm{bottom}/\rho_\mathrm{top} \sim 6$, and a pressure stratification $p_\mathrm{bottom}/p_\mathrm{top} \sim 7.5$ or two pressure scale-heights. The global rms velocity and turnover timescale are computed as for the red giant (see previous section), and shown in Table \ref{table:ob_runs}. Figure \ref{fig:ob_thermodynamical_perturbations} shows the radial profiles of the rms values of the density, temperature, and pressure fluctuations, normalized by their averaged values. The largest values in the relative magnitude of the fluctuations occur at the boundaries, and are of the order of a percent.

\section{Horizontally-averaged mean-field equations}
\label{1DRANS}

Using the RANS methodology, we develop in this section a framework of mean-field equations which are consistent with spherical geometry. A similar approach has been applied to the kinetic energy equation; see \cite{hurlburt_nonlinear_1986,hurlburt_penetration_1994-1,meakin_turbulent_2007}. We show here how this can be extended, as the kinetic energy is only one of the several mean-field equations that can be derived from the hydrodynamical equations. Although we refer to our equations as ``Reynolds-Averaged Navier-Stokes" equations, we actually introduce the effect of viscosity only through the kinetic energy dissipation rate $\epsilon_d$ (units: erg s$^{-1}$ g$^{-1}$).  In the large Reynolds number regime, the viscous dissipation is the dominant contribution of viscosity in the mean-field equations. This is the so-called ``dissipation anomaly" characteristic of 3D turbulence, see discussion in Sect. \ref{resolution_effects}. Other terms, such as the viscous flux of kinetic energy, are neglected.

To obtain our set of 1D equations, we introduce two types of averaging: 1) statistical averaging, denoted by an overbar; 2) horizontal averaging, denoted by brackets. Practically, statistical averages are computed by performing a time average (the ergodic hypothesis). The horizontal average of quantity $q$ is defined as:

\begin{equation}
\br{q} = \frac{1}{\Delta \Omega}\int_{\Delta \Omega} q(r,\theta,\phi) d \Omega,
\end{equation}

\noindent where $d \Omega = \sin \theta d \theta d \phi$ is the solid angle in spherical coordinates.

We decompose the flow quantities into mean and fluctuating components:

\begin{equation}
q = \av{q} + q',
\end{equation}

\noindent so that $\av{q'} = 0$, by construction. We also introduce Favre averaging, which is a density-weighted average:

\begin{equation}
\fav{q} = \frac{\av{\rho q}}{\av{\rho}},
\end{equation}

\noindent and which leads to the decomposition

\begin{equation}
q = \fav{q} + q''.
\end{equation}

We stress the difference between $q'$ and $q''$: the first denotes the fluctuation around the Reynolds average $\av{q}$, the second denotes the fluctuation around the Favre average $\fav{q}$.

\begin{deluxetable}{l c c c c}
 \tablecaption{\label{table:ob_runs} Summary of the Oxygen Burning Simulations.}
\tablehead{Parameter & {\sf ob.3D.lr} & {\sf ob.3D.mr} & {\sf ob.3D.hr} }
\startdata
Grid zoning &   192$\times 128^2$ & 384$\times 256^2$ & 786$\times 512^2$\\
$r_\mathrm{in}$, $r_\mathrm{out}$ ($10^{9}$ cm)  & 0.3, 1.0 & 0.3, 1.0 & 0.3, 1.0 \\
$\Delta \theta$, $\Delta \phi$  &  45$^\circ$, 45$^\circ$ & 45$^\circ$, 45$^\circ$ &45$^\circ$, 45$^\circ$\\
CZ stratification ($H_p$) & 2 & 2 & 2\\
$t_\mathrm{av}  (\Delta t_\mathrm{av})$ (s) & $494(230)$ & $494 (230)$ & $429(165)$\\
$v_\mathrm{rms}$  ($10^6$ cm/s) & 9.2 & 9.57 & 9.2\\
$\tau_\mathrm{conv}$  (s) & 96 & 92 & 96\\
$L_\mathrm{nuc}$ ($10^{46}$ erg/s) & 2.69 & 2.63 & 2.47\\
$L_\mathrm{d}$ ($10^{46}$  erg/s) & 0.30 &0.29 &0.30\\
$l_\mathrm{d}$  ($10^{8}$ cm) & 4.9 & 5.5 & 4.7\\
$\tau_\mathrm{d}$  (s) & 26 & 29 & 26
\enddata
\tablecomments{$v_\mathrm{rms}$: global rms velocity, $\tau_\mathrm{conv}$: convective turnover timescale, $L_\mathrm{nuc}$: total energy released by nuclear burning, $L_d$: rate of kinetic energy dissipation, $l_d$: dissipation length-scale, $\tau_d$: dissipation time-scale.}
\end{deluxetable}

The density and energy equations lead to their averaged counter-part. From the momentum equations, the three mean-field equations which are consistent with spherical geometry concern the mean radial velocity $\fav{u_r}$, the mean specific angular momentum along the $z$-axis $\fav{j_z}$, and the mean specific kinetic energy $\fav{\epsilon_k}$. By ``consistent with spherical geometry" we mean that the resulting equations do not have terms which have an explicit angular dependance, as would for instance have the mean-field equations for $u_\theta$ or $u_\phi$. The energy equation can be formulated either for the mean specific internal energy $\fav{\epsilon_i}$, the mean specific total energy $\fav{\epsilon_t}$, or the mean specific entropy $\fav{s}$. For completeness, we show all forms. The resulting equations are summarized in Table \ref{fig:1DRANS}. We show the equation for $\fav{j_z}$ although it is ``trivial" when rotation is not included; we will not discuss it further here. Note, however, that when rotation is included, the horizontal average as introduced here is not suited because the angular directions are not homogeneous anymore. In this case, similar 1D averaged equations cannot be obtained.

We denote by $\nabla_r$ the radial part of the divergence operator in spherical coordinates, i.e. $\nabla_r f= \frac{1}{r^2} \partial_r \big ( r^2 f \big)$. The mean-field equations are characterized by second-order correlations, which stem from the Reynolds/Favre decompositions, as, e.g., the turbulent fluxes which are of the form $\fav{q''u_r''}$ or $\av{q'u_r'}$. The equations in Table \ref{fig:1DRANS} are written in terms of the averaged Lagrangian derivative

\begin{equation}
\fav{D_t} q = \partial_t q + \fav{u_r} \partial_r q.
\end{equation}

The connection with the Eulerian, conservative form is immediate:

\begin{equation}
\av{\rho} \fav{D_t} q = \partial_t ( \av{\rho} q) + \nabla_r \big ( q \fav{u_r} \big ),
\end{equation}

\noindent where we used the 1D averaged continuity equation.

\begin{table*}[t]
\caption{1D RANS equations in Lagrangian form.}
\label{fig:1DRANS}
\begin{align}
\fav{D_t} \av{\rho} =& - \av{\rho} \nabla_r \fav{u_r} \\
\av{\rho} \fav{D_t} \fav{\epsilon_i} = & - \nabla_r  \av{\rho} \fav{\epsilon_i'' u_r''} + \nabla_r  \av{\chi} \partial_r \av{T}  - \av{p}\nabla_r  \av{u_r}  + \nabla_r \av{\chi' \partial_r T'}  - \av{p' \vec \nabla \cdot \vec u'} + \av{\rho \epsilon_\mathrm{nuc}} + \av{\rho \epsilon_d}\\
\av{\rho} \fav{D_t} \fav{\epsilon_t} =& - \nabla_r  \av{\rho} \fav{h'' u_r''}   - \nabla_r \av{\rho} \fav{\epsilon_k'' u_r''}   - \av{p} \nabla_r \fav{u_r} + \nabla_r  \av{\chi} \partial_r \av{T} + \nabla_r  \av{\chi' \partial_r T'}  + \av{\rho \epsilon_\mathrm{nuc}}\ \label{eq:rans_etot} \\
\av{\rho} \fav{D_t} \fav{s} =& - \nabla_r  \av{\rho} \fav{s'' u_r''}  - \av{\frac{1}{T} \vec \nabla \cdot \vec F_r} + \av{\rho \frac{\epsilon_\mathrm{nuc} + \epsilon_d}{T}}  \label{eq:rans_entropy}\\
\av{\rho} \fav{D_t} \fav{\epsilon_k} =& - \nabla_r  \av{\rho} \fav{\epsilon_k'' u_r''}  
 - \nabla_r  \av{p' u_r'} + \av{p' \vec \nabla \cdot \vec u'} + \av{\rho' \vec u' \cdot \vec g} - \av{\rho \epsilon_d} \label{eq:rans_ekin} \\
\av{\rho} \fav{D_t} \fav{u_r} =& - \nabla_r \av{\rho} \fav{ u_r''^2}  - \partial_r \av{p} - \av{\rho} \fav{g} + \frac{\av{\rho}}{r} \big ( 2\fav{\epsilon_k} - \fav{u_r}^2 - \fav{u_r''^2} \big) \label{eq:rans_ur} \\
\av{\rho} \fav{D_t} \fav{j_z} =& - \nabla_r \av{\rho} \fav{ j_z'' u_r''}
\end{align}
\tablecomments{Definitions: density $\rho$, temperature $T$, pressure $p$, velocity $\vec u$, radial velocity component $u_r$, specific internal energy $\epsilon_i$,  specific kinetic energy $\epsilon_k$, specific total energy $\epsilon_t$,  specific entropy $s$, specific enthalpy $h$, $z$-component of the specific angular momentum $j_z=r\sin \theta u_\phi$, thermal conductivity $\chi$, radiative flux $\vec F_r$, gravitational acceleration $\vec g$, rate of nuclear energy production $\epsilon_\mathrm{nuc}$, rate of viscous dissipation $\epsilon_\mathrm{d}$. Reynolds decomposition: $q = \av{q} + q'$, Favre decomposition: $q = \fav{q} + q''$.}
\end{table*}

\begin{table*}[t]
\caption{1D RANS equations in Lagrangian mass coordinate (same definitions as in Table \ref{fig:1DRANS}).}
\label{fig:1DRANS_lagrangian}
\begin{align}
 \partial_t r |_m =& \fav{u_r}\\
 \partial_t \fav{\epsilon_i} |_m = & - \partial_m \big( 4 \pi r^2 \av{\rho} \fav{\epsilon_i'' u_r''} \big)+ \partial_m \big( 4\pi r^2 \av{\chi} \partial_r \av{T} \big ) -  \av{p} \partial_m ( 4 \pi r^2 \av{u_r}) +  \partial_m ( 4\pi r^2 \av{\chi' \partial_r T'} ) - \frac{\av{p' \vec \nabla \cdot \vec u'}}{\av{\rho}} + \fav{\epsilon_\mathrm{nuc}} + \fav{\epsilon_d}\\
\partial_t \fav{\epsilon_t}|_m = &- \partial_m (4 \pi r^2 \av{\rho} \fav{h'' u_r''} )  - \partial_m (4 \pi r^2 \av{\rho} \fav{\epsilon_k'' u_r''} ) + \partial_m ( 4\pi r^2 \av{\chi} \partial_r \av{T} ) + \partial_m ( 4\pi r^2 \av{\chi' \partial_r T'} ) - \av{p} \partial_m \big (4 \pi r^2 \fav{u_r} \big ) + \fav{\epsilon_\mathrm{nuc}}\\
\partial_t \fav{s} |_m = &- \partial_m (4 \pi r^2 \av{\rho} \fav{s'' u_r''} ) + \frac{1}{\av{\rho}} \av{\frac{1}{T} \vec \nabla \cdot \vec F_r}  +  \frac{1}{\av{\rho}} \av{\rho \frac{ \epsilon_\mathrm{nuc}+\epsilon_d}{T}} \\
\partial_t \fav{\epsilon_k}|_m = & - \partial_m ( 4\pi r^2 \av{\rho} \fav{\epsilon_k'' u_r''} ) - \partial_m ( 4\pi r^2 \av{p' u_r'})  + \frac{\av{p' \vec \nabla \cdot \vec u'}}{\av{\rho}} + \frac{\av{\rho' \vec u' \cdot \vec g}}{\av{\rho}}  - \fav{\epsilon_d} \\
\partial_t \fav{u_r}|_m = &- \partial_m \big ( 4 \pi r^2 \av{\rho} \fav{ u_r''^2} \big ) - 4\pi r^2 \partial_m \av{p} - \fav{g}  + \frac{1}{r} \big ( 2\fav{\epsilon_k} - \fav{u_r}^2 - \fav{u_r''^2} \big) \\
\partial_t  \fav{j_z}|_m = & - \partial_m \big ( 4\pi r^2 \av{\rho} \fav{u_r'' j_z''} \big )
\end{align}
\end{table*}

Connection with the common form of the stellar evolution equations can be made using:

\begin{align}
\partial_r  &= 4 \pi r^2 \av{\rho} \partial_m,\\
\fav{D_t} &= \partial_t |_m
\end{align}

\noindent The first equation is the standard relation between the Eulerian derivative at constant radius and the Lagrangian derivative at constant mass shell. The second equation states that $\fav{D_t}$ is exactly the Lagrangian derivative at constant mass. The resulting equations are summarized in Table \ref{fig:1DRANS_lagrangian}.


\section{Results}
\label{results}

Unless stated otherwise, the results in this section are based on models {\sf rg.3D.mr} and {\sf ob.3D.mr}. In Sect.~\ref{rans:analysis}, we use the mean-field equations developed in Sect.~\ref{1DRANS} to analyze our 3D data. These 1D equations provide a powerful framework to analyze the physical processes characterizing the numerical models, and to assess the consistency of the models with the physical equations. Section ~\ref{turbulent_velocity_field} is devoted to the analysis of the turbulent velocity field. We motivate the distinction between ``deep convection", relevant for the red giant models, and ``shallow convection", relevant for the oxygen-burning shell models. We analyze  the anisotropy of the Reynolds stresses, which characterizes the asymmetry of the flow, and which motivates a decomposition of the velocity field into a large scale component, characterizing the plumes, and an isotropic component at small scales. We show that the isotropic component provides a natural interpretation of the kinetic energy damping in terms of a dissipation in a turbulent cascade. In Sect.~\ref{pressure_fluctuations}, we show how the magnitude of pressure fluctuations is related to the stratification of the convective zone, which explains the main differences in the mean field analysis between the red giant model and the oxygen-burning shell model. In Sect.~\ref{turbulent_fluxes}, we analyze the turbulent fluxes. These are the quantities that should be modeled in a theory of turbulent convection. In Sect.~\ref{KEdriving}, we discuss the kinetic energy driving and the turbulent dissipation. We show that in steady state the rate of turbulent dissipation is of the order of the convective luminosity. Section \ref{penetration} is devoted to the red giant models, for which we discuss thermal effects and characterize the overshooting at the bottom of the convective region. Finally, we discuss in Sect. \ref{resolution_effects} the convergence of our results with resolution, and consider the implications regarding the global dynamics in the numerical models.


\subsection{Mean-field analysis}
\label{rans:analysis}

\begin{figure*}[t]
\centering
\parbox{0.49\linewidth}{\center \includegraphics[width=0.8\linewidth]{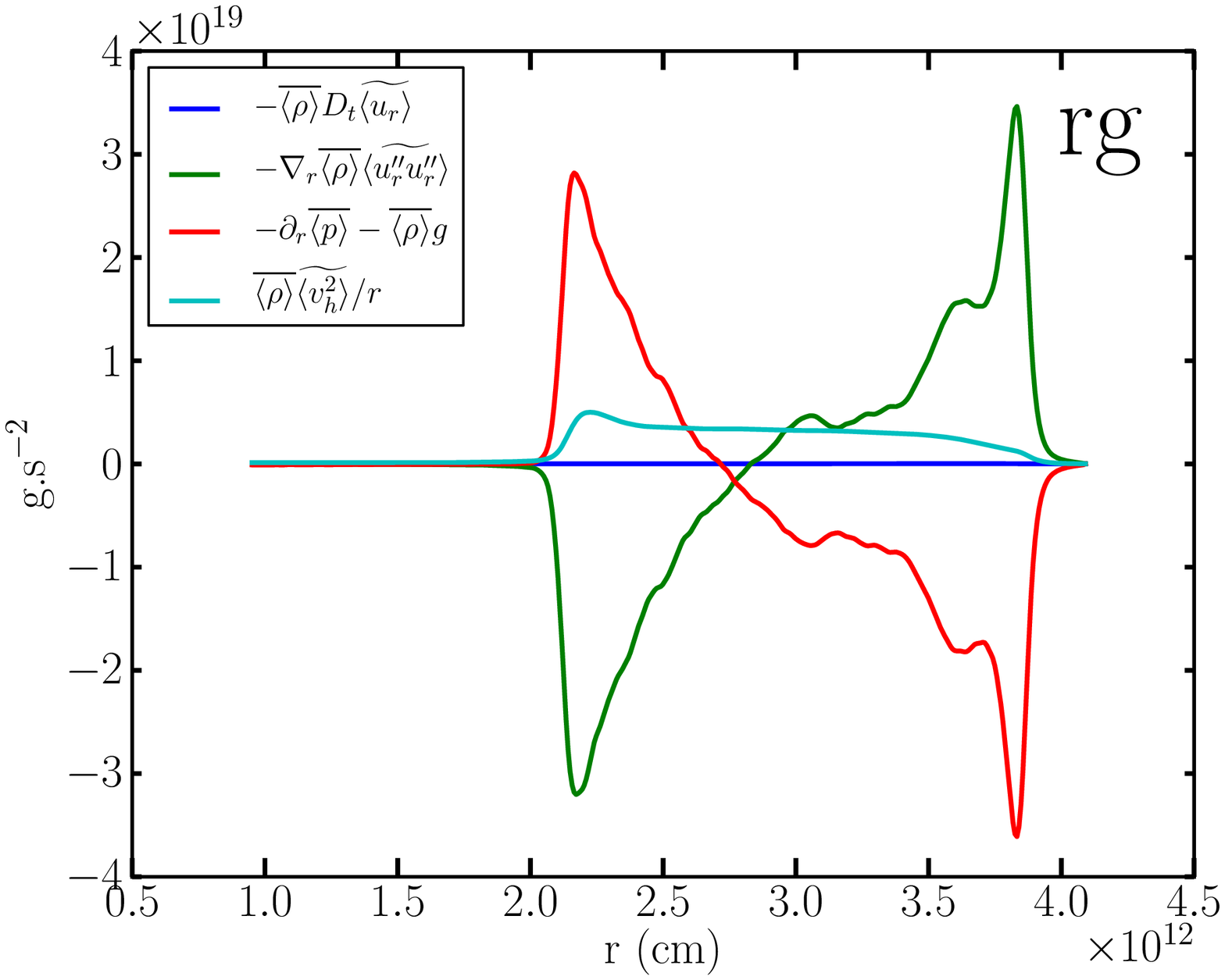}}
\parbox{0.49\linewidth}{\center \includegraphics[width=0.8\linewidth]{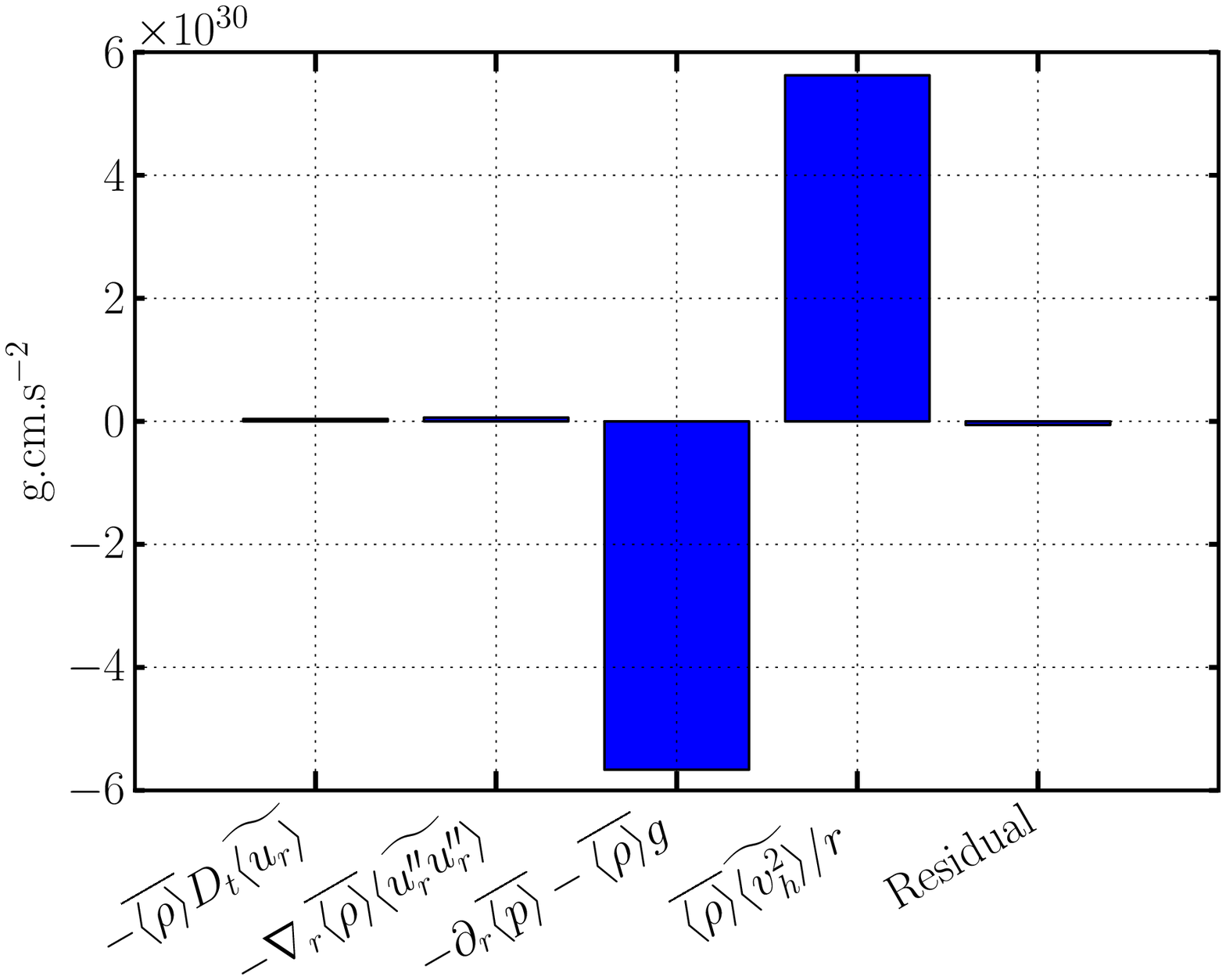}}
\parbox{0.49\linewidth}{\center \includegraphics[width=0.8\linewidth]{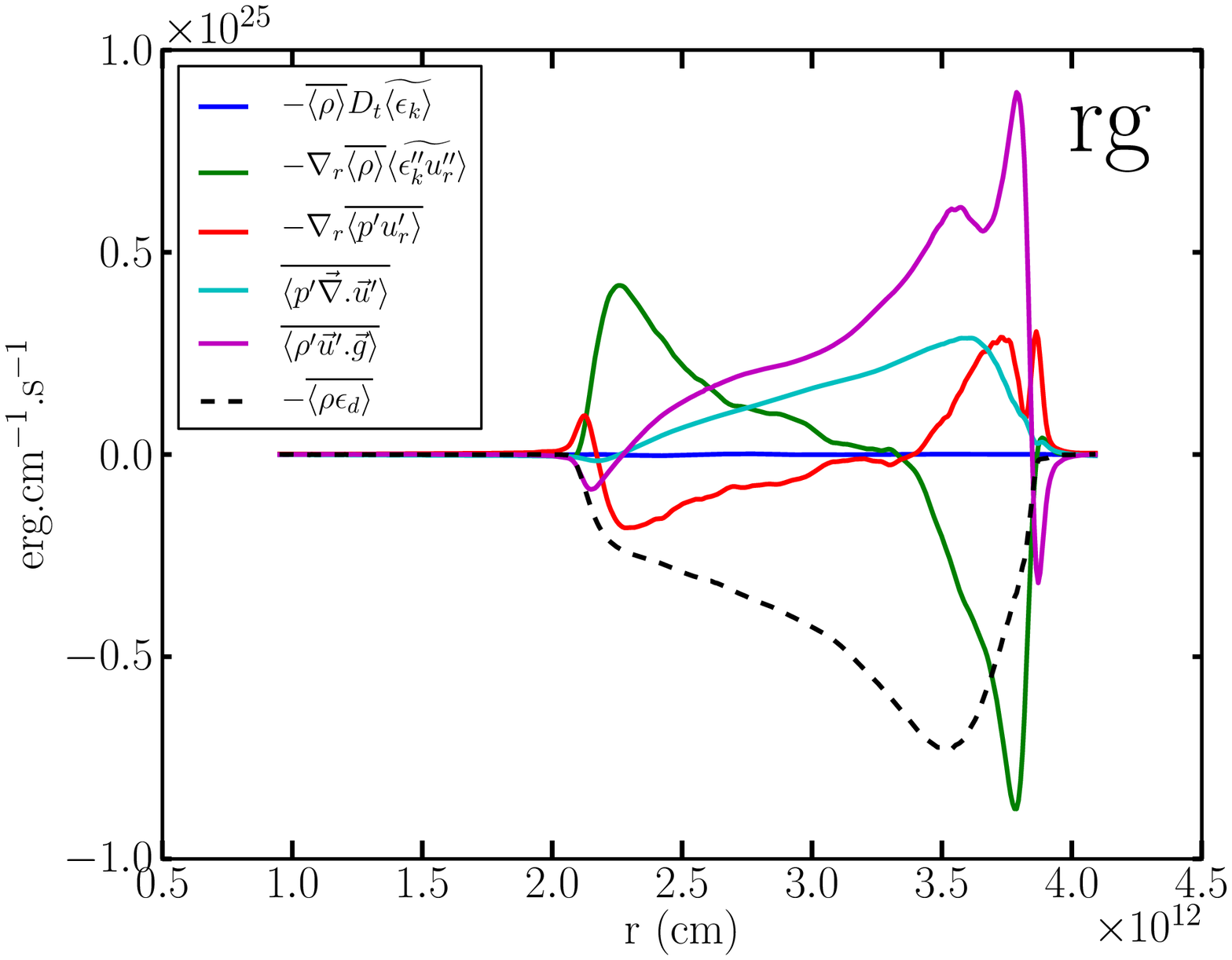}}
\parbox{0.49\linewidth}{\center \includegraphics[width=0.8\linewidth]{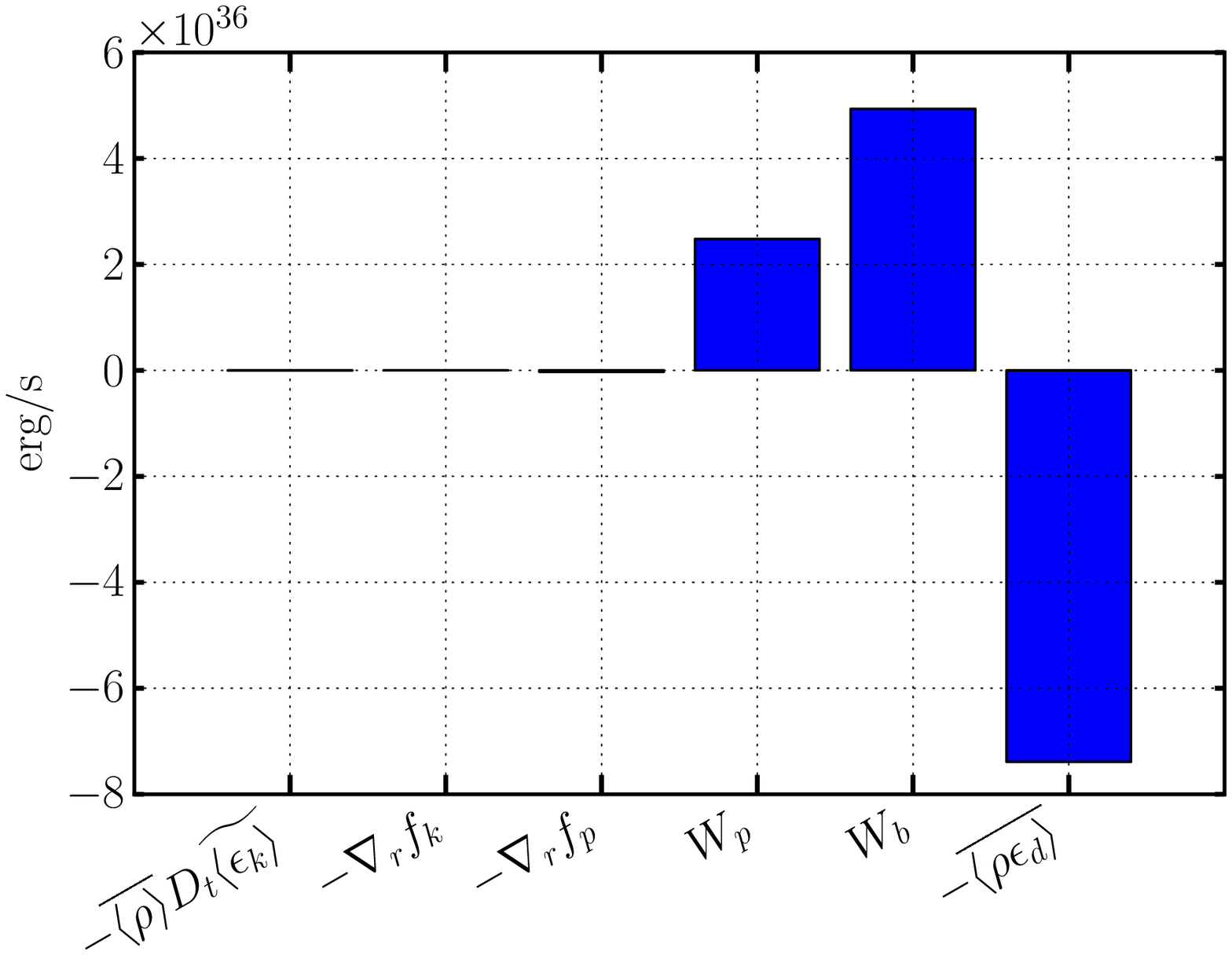}}
\parbox{0.49\linewidth}{\center \includegraphics[width=0.8\linewidth]{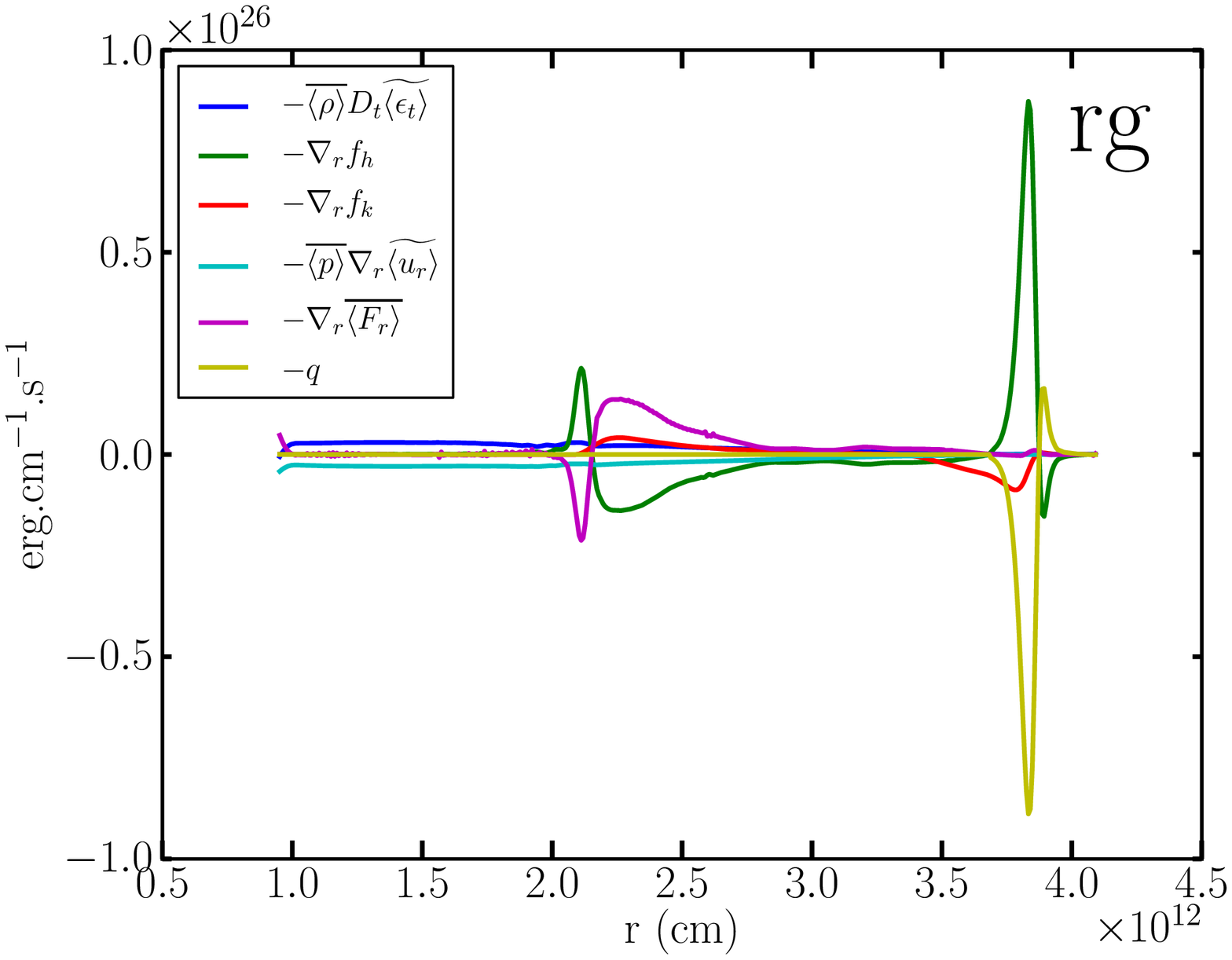}}
\parbox{0.49\linewidth}{\center \includegraphics[width=0.8\linewidth]{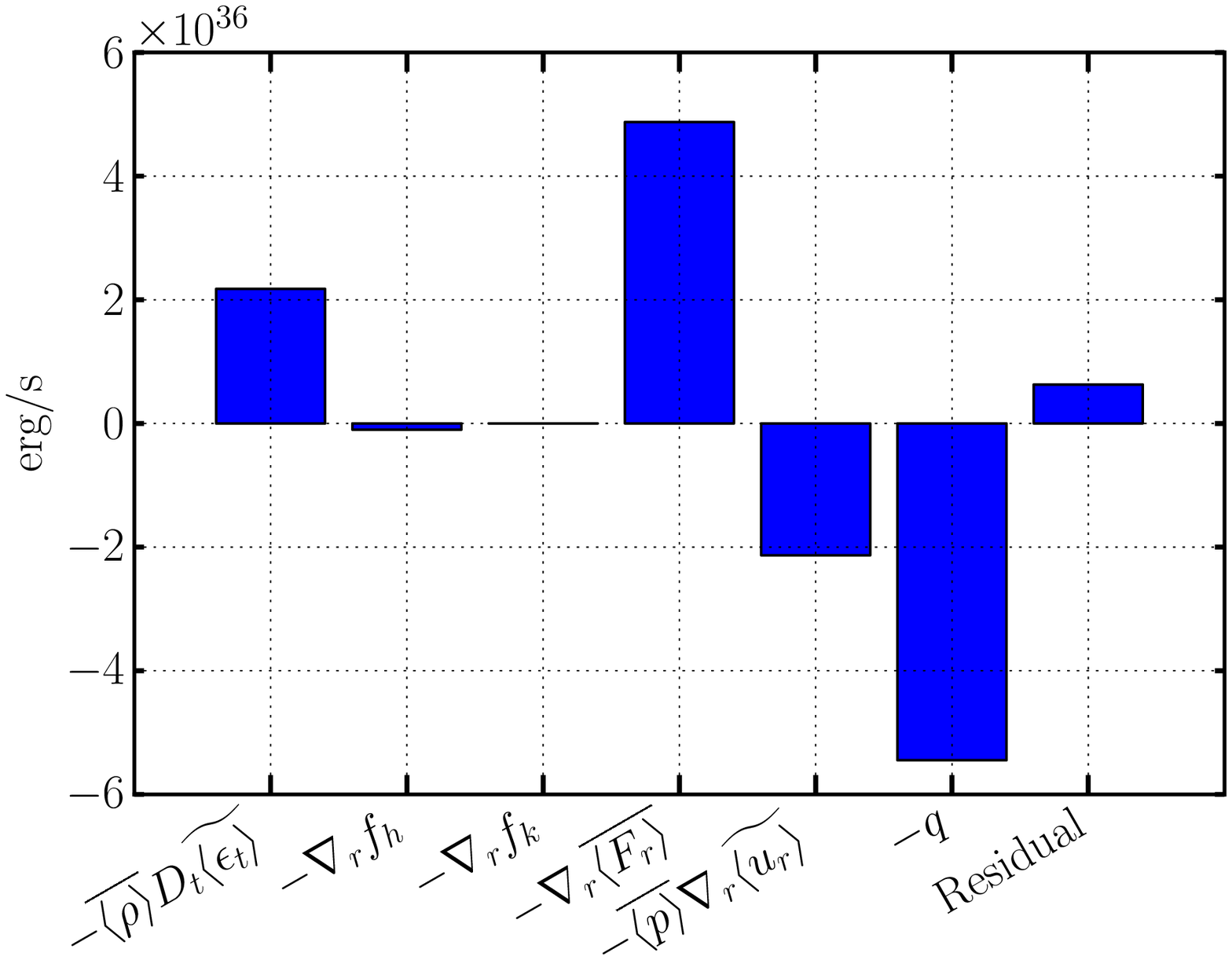}}
\caption{Mean-field analysis for model {\sf rg.3D.mr}, time averaging is performed over 800 days. From top to bottom: radial expansion velocity, kinetic energy, and total energy balance. Left panel: Radial profiles of the terms in the balance. For the sake of clarity, each term is multiplied by $4 \pi r^2$. As a result, the volume integral of a term is equal to the area below the corresponding curve, and it can be appreciated visually. Right panel: Integral budget over the convective zone.}
\label{fig:rans_rg}
\end{figure*}

\begin{figure*}[t]
\centering
\parbox{0.49\linewidth}{\center \includegraphics[width=0.8\linewidth]{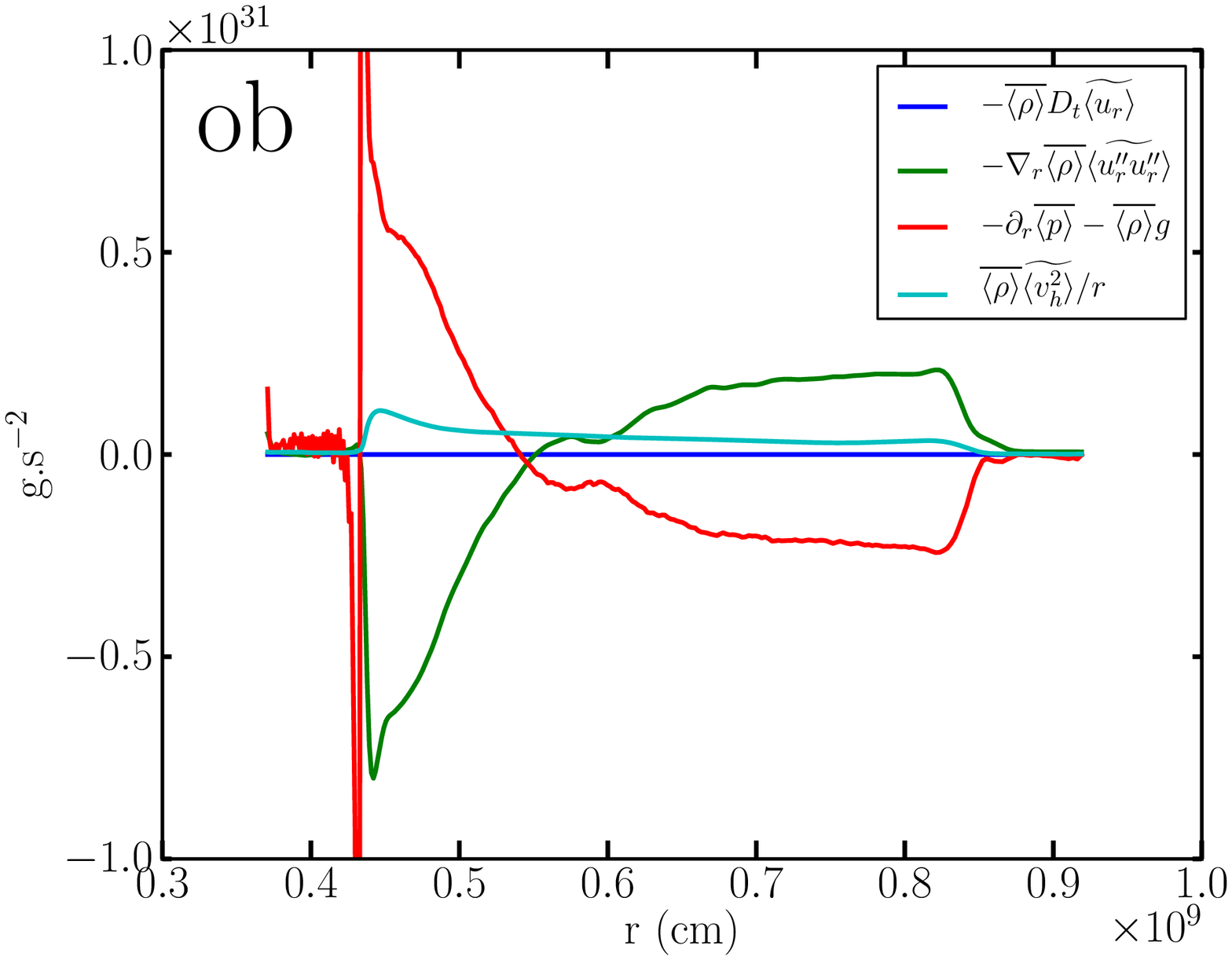}}
\parbox{0.49\linewidth}{\center \includegraphics[width=0.8\linewidth]{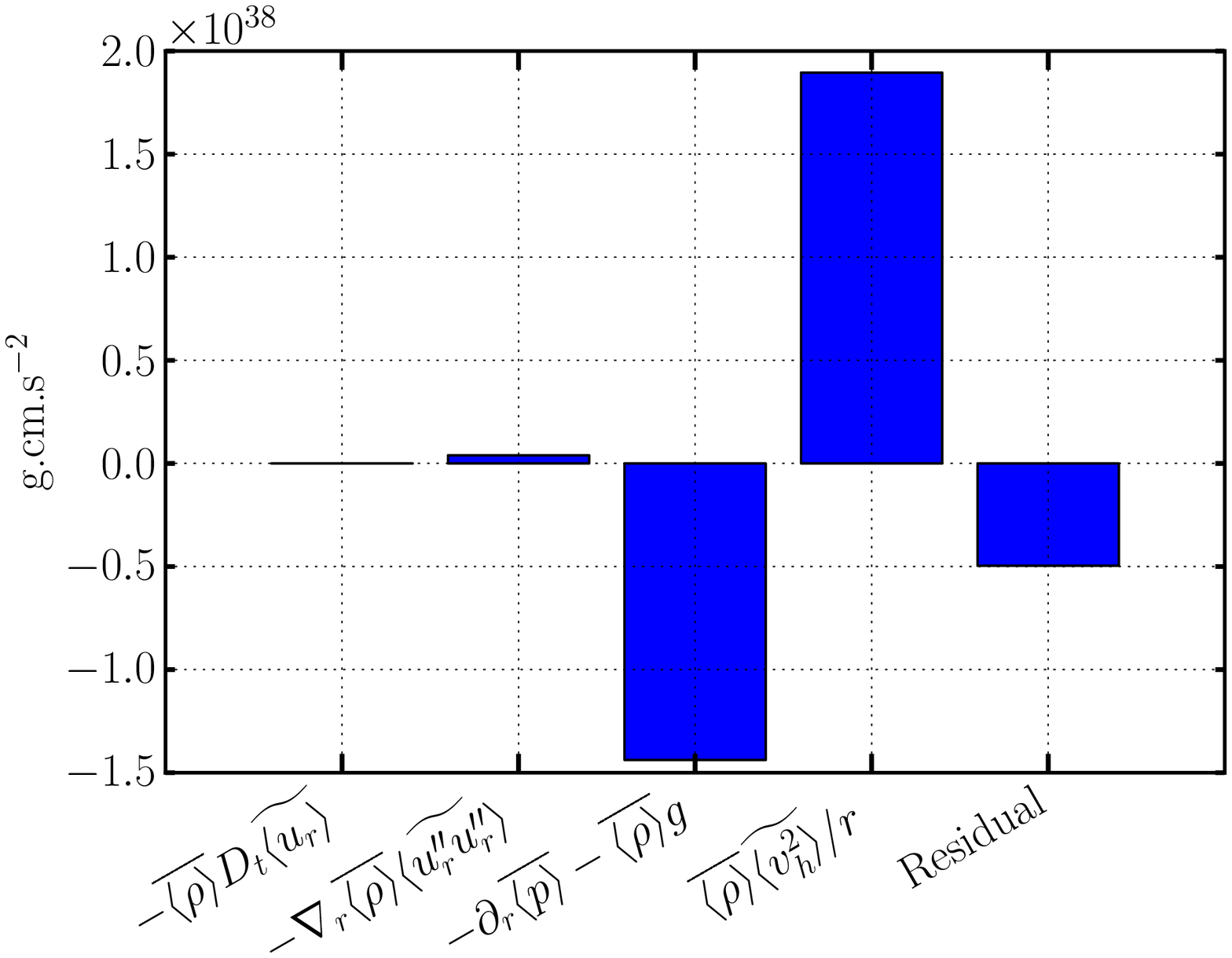}}
\parbox{0.49\linewidth}{\center \includegraphics[width=0.8\linewidth]{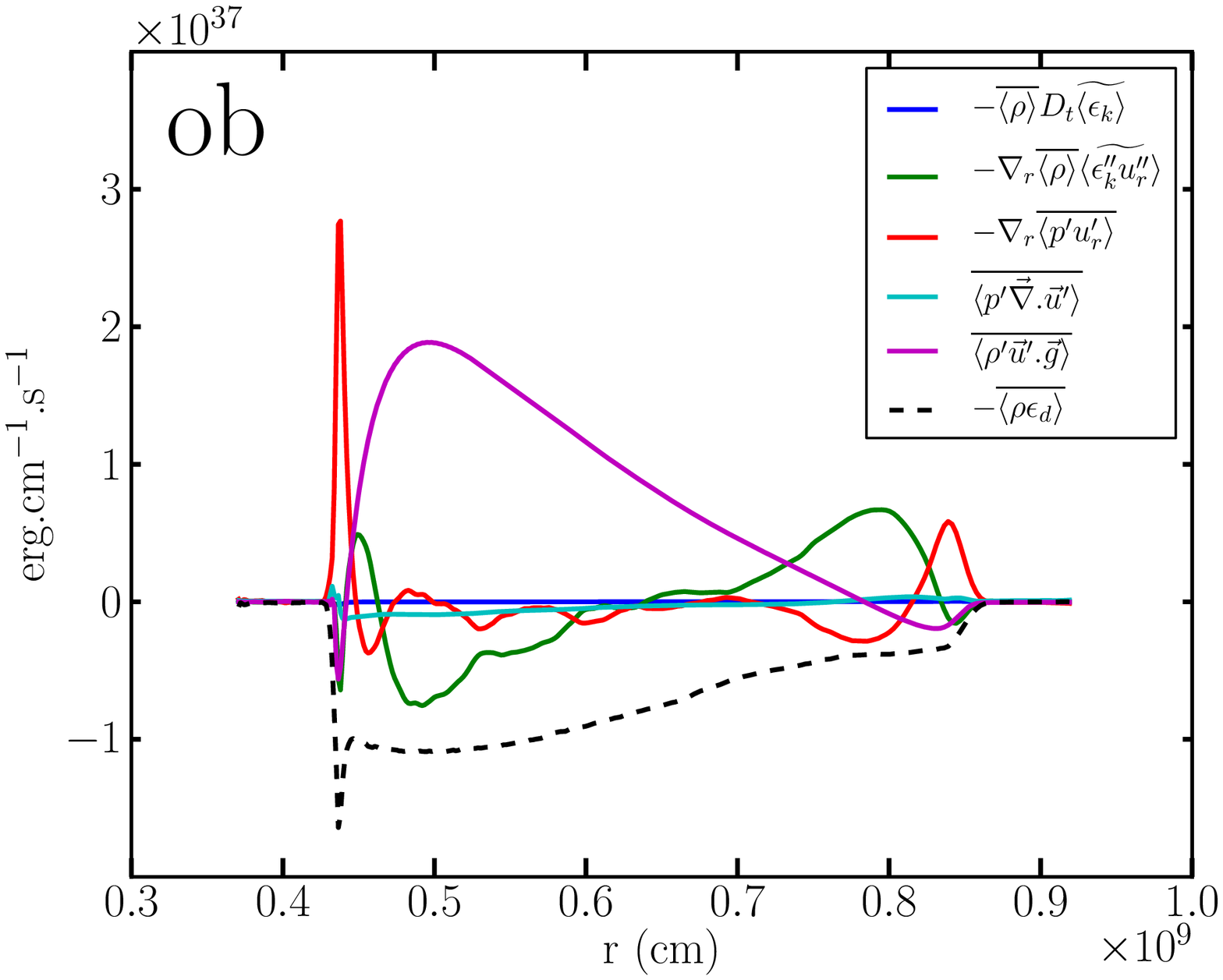}}
\parbox{0.49\linewidth}{\center \includegraphics[width=0.8\linewidth]{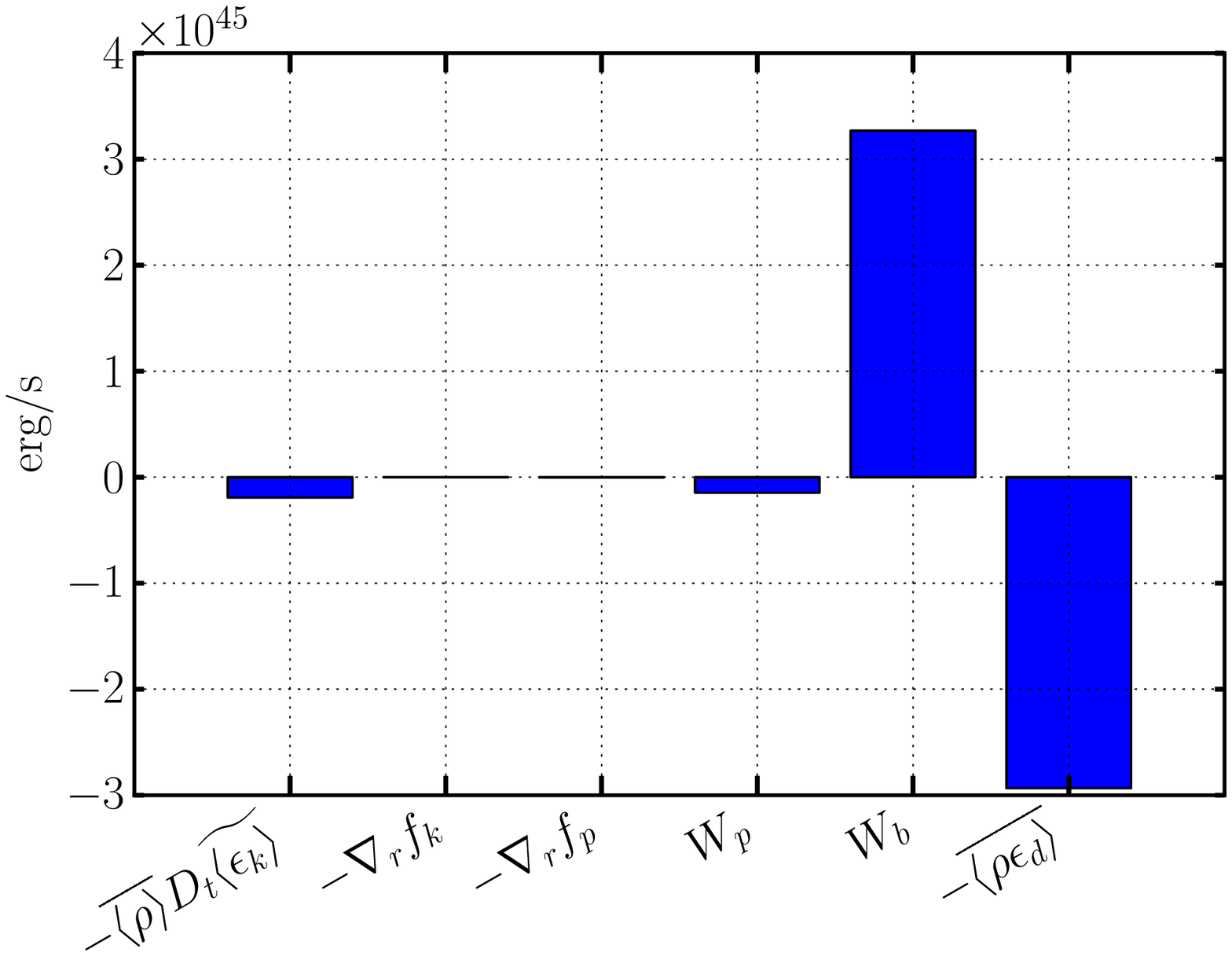}}
\parbox{0.49\linewidth}{\center \includegraphics[width=0.8\linewidth]{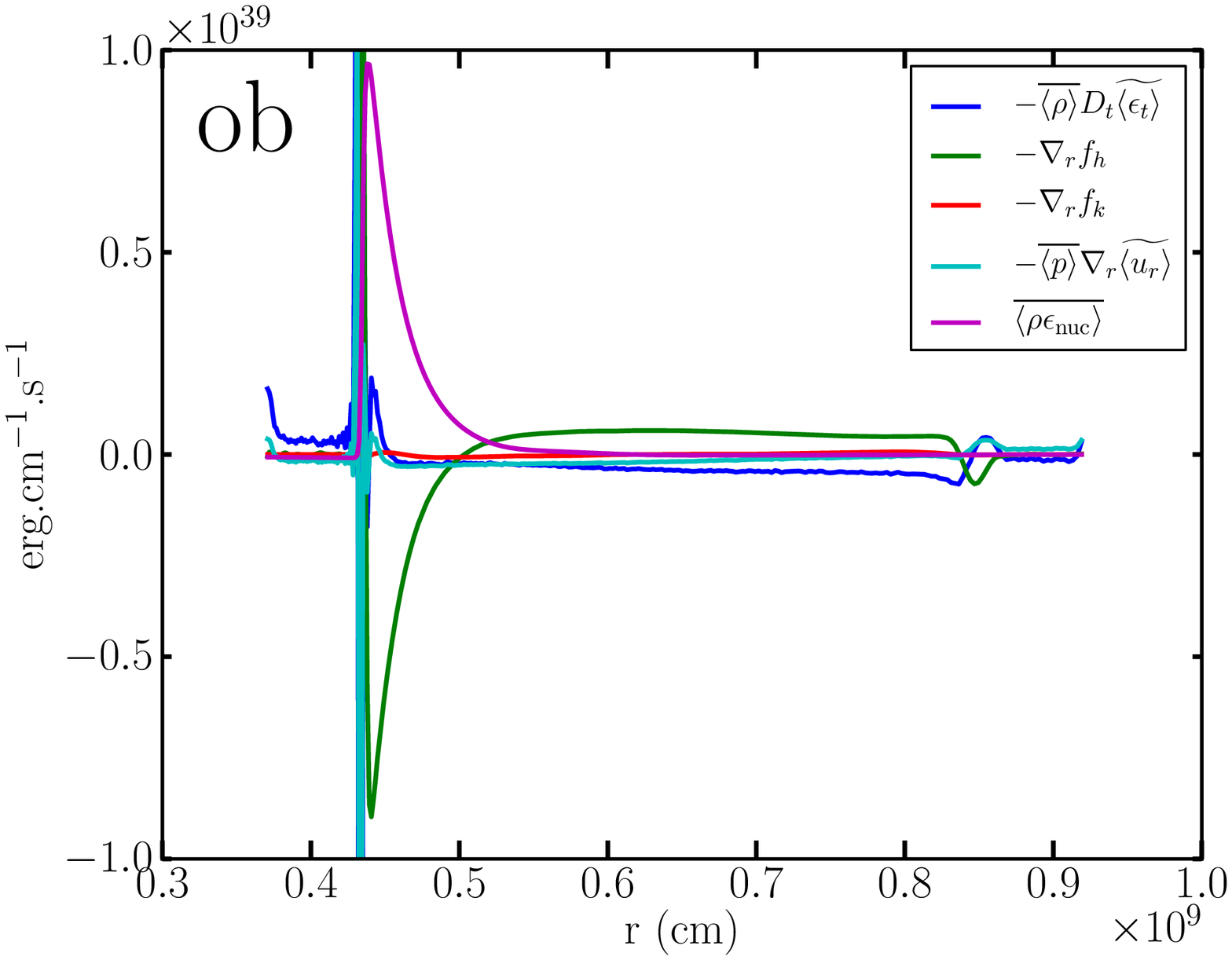}}
\parbox{0.49\linewidth}{\center \includegraphics[width=0.8\linewidth]{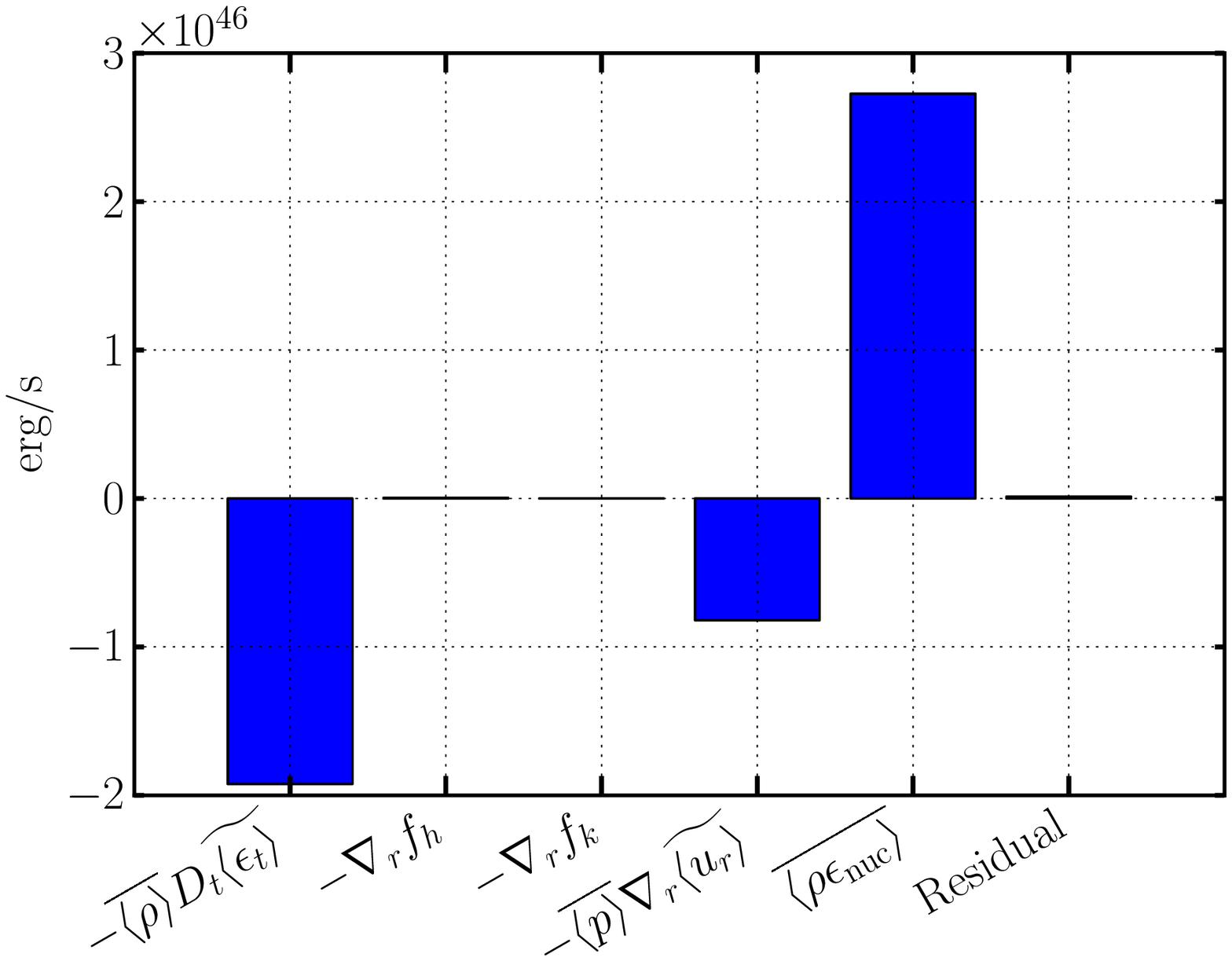}}
\caption{Same as Fig. \ref{fig:rans_rg} for model {\sf ob.3D.mr}, with time averaging performed over 230 s.}
\label{fig:rans_ob}
\end{figure*}

We use the framework developed in Sect. \ref{1DRANS} to analyze the data from our multi-D simulations, in terms of the radial momentum, kinetic energy, and total energy balances. The temporal averaging is performed on the interval $[t_\mathrm{av}-\Delta t_\mathrm{av}/2, t_\mathrm{av}+\Delta t_\mathrm{av}/2]$, with $t_\mathrm{av}$ and $\Delta t_\mathrm{av}$ given in Tables \ref{table:rg_runs} and \ref{table:ob_runs} for each model. In practice, we find that averaging the data over more than two turn-over timescales yields very similar results. In this section, the models were averaged over the common time period available at different resolutions, namely four turnover timescales (800 d) for the red giant model, and roughly 2.5 turnover timescales (230 s) for the oxygen-burning shell model.

Throughout the paper, we use the lower case letter $f$ to denote turbulent fluxes, $f = \av{\rho} \fav{u_r'' q''}$ or $f = \av{u_r' q'},$ depending on the quantity $q$.

\subsubsection{Radial momentum balance}

\begin{deluxetable}{c c c c}[t]
\tablecaption{\label{table:rg_ur_budget} Radial momentum balance for model {\sf rg.3D.mr} - integral budget over the convective zone (g.cm.s$^{-2}$).}
\tablehead{Term & Value & Term & Value}
\startdata
$- \int  \av{\rho} D_t \fav{u_r} \dV$ &  3.85(28) &  $\int \av{\rho}\frac{\fav{v_h^2}}{r} \dV$ & 5.62(30)\\
$- \int \nabla_r  \av{\rho} \fav{ u_r''^2}  \dV$ & 6.03(28)  &  Residual & -5.91(28) \\
$- \int \big( \partial_r \av{p} - \av{\rho} \fav{g} \big) \dV $ &  -5.66(30)
\enddata
\end{deluxetable}

\begin{deluxetable}{c c c c}[t]
\tablecaption{\label{table:ob_ur_budget} Radial momentum balance for model {\sf ob.3D.mr} - integral budget over the convective zone (g.cm.s$^{-2}$).}
\tablehead{Term & Value & Term & Value}
\startdata
$- \int  \av{\rho} D_t \fav{u_r} \dV$ &  -4.55(31) &  $\int \av{\rho}\frac{\fav{v_h^2}}{r} \dV$ & 1.88(38)\\
$- \int \nabla_r  \av{\rho} \fav{ u_r''^2}  \dV$ & -2.13(36)  &  Residual & -4.39(37) \\
$- \int \big( \partial_r \av{p} - \av{\rho} \fav{g} \big) \dV $ &  -1.42(38)
\enddata
\end{deluxetable}

Equation (\ref{eq:rans_ur}) can be written as:

\begin{align}
\label{eq:rans_ur2}
 \av{\rho}  \fav{\partial_t} \fav{u_r} =&  - \nabla_r  \av{\rho} \fav{ u_r''^2}  - \partial_r \av{p} - \av{\rho} \fav{g} \notag \\
 & + \av{\rho}\frac{\fav{v_h^2}}{r}.
\end{align}

\noindent This equation is an equation for the mean radial component of the mass flux since by definition $\av{\rho}\fav{u_r} = \av{\rho u_r}$. The right-hand side involves the divergence of the radial component of the Reynolds stress $R_{rr}$, the balance between the gradient of the mean pressure and the mean gravity force, and a geometric term characteristic of spherical geometry in which a horizontal velocity $v_h$ induces a radial acceleration $v_h^2/r$ due to inertia.

The top-left panels in Fig. \ref{fig:rans_rg} and Fig. \ref{fig:rans_ob} show the radial profiles of these terms in models {\sf rg.3D.mr} and {\sf ob.3D.mr}. In both cases, we find the Lagrangian time derivative to be negligible, showing that the balance is in a statistically steady state. The figures show that a slight hydrostatic imbalance is due to the turbulent ram-pressure, with a smaller, but not negligible, contribution from the inertial acceleration due to horizontal motions. Defining

\begin{equation}
p_\mathrm{turb} = \av{\rho} \fav{u_r''^2},
\end{equation}

\noindent we find that the ratio $p_\mathrm{turb}/p$ decreases from $\sim 2\times10^{-2}$ at the top of the CZ to $\sim 2 \times 10^{-4}$ at the bottom in the red giant model. In the oxygen burning-shell model, $p_\mathrm{turb}/p \sim 10^{-3}$ in the bulk of the convective zone, and it decreases rapidly at the boundaries.

In the red giant model, the hydrostatic imbalance corresponds to an inward acceleration of $\sim 3.5$ \% $g$ at the top of the convective zone, and an upward acceleration of $\sim 0.05$ \% at the bottom. For the oxygen-burning shell, the results are similar, with values within $\pm 0.2$ \% of $g$ throughout the convective zone. Thus, as expected for the deep interior, hydrostatic equilibrium holds, typically to the order of a percent or less. In the photospheric regions the turbulent pressure can significantly affect the hydrostatic balance
\citep{stein_topology_1989}.

The top-right panels in Fig. \ref{fig:rans_rg} and Fig. \ref{fig:rans_ob} (see also Tables \ref{table:rg_ur_budget} and \ref{table:ob_ur_budget}) show the balance upon integration over the convective zone. The residual is very small in the red giant models, showing good consistency with the physical equations. The oxygen-burning shell models show spurious oscillations at the inner boundary, which are responsible for the non-zero residual seen in the figure. This is due to the steep gradients at this boundary (see Sect. \ref{resolution_effects}), and we have checked that the consistency is very good everywhere else. Progress has been made in dealing with multi-fluids in PPM \citep{plewa_mueller,woodward_ppb}, which may help.


\subsubsection{Kinetic energy balance} 
\label{rans:analysis_ek}

\begin{deluxetable}{c c c c}[t]
\tablecaption{\label{table:rg_ek_budget} Kinetic energy balance for model {\sf rg.3D.mr} - integral budget over the convective zone (erg s$^{-1}$).}
\tablehead{Term & Value & Term & Value}
\startdata
$- \int \av{\rho} D_t \fav{\epsilon_k} \dV$ & -3.47(33)  &  $\int W_p \dV$ & 2.48(36) \\
$- \int \nabla_r f_k \dV$ & -1.61(32)  &  $\int W_b \dV$  & 4.94(36) \\
$- \int \nabla_r f_p \dV$ &  -2.68(34) & $- \int \av{\rho \epsilon_d} \dV$ & -7.39(36)
\enddata
\end{deluxetable}

\begin{deluxetable}{c c c c}[t]
\tablecaption{\label{table:ob_ek_budget} Kinetic energy balance for model {\sf ob.3D.mr} - integral budget over the convective zone (erg s$^{-1}$).}
\tablehead{Term & Value & Term & Value}
\startdata
$- \int \av{\rho} D_t \fav{\epsilon_k} \dV$ & -1.91(44)  &  $\int W_p \dV$ & -1.45(44) \\
$- \int \nabla_r f_k \dV$ & -1.96(41)  &  $\int W_b \dV$  & 3.27(45) \\
$- \int \nabla_r f_p \dV$ &  -1.04(42) & $- \int \av{\rho \epsilon_d} \dV$ & -2.93(45)
\enddata
\end{deluxetable}

At the first sight, the temporal behavior of the kinetic energy (KE) shows a high level of complexity (see Figure \ref{fig:rg_ekin} for the red giant model). The mean kinetic energy balance, Eq. (\ref{eq:rans_ekin}), allows insight in the dynamics of the turbulent convection \citep{hurlburt_nonlinear_1986,hurlburt_penetration_1994-1,meakin_turbulent_2007}. The KE equation involves two turbulent fluxes: the KE turbulent flux $f_k = \av{\rho} \fav{u_r'' \epsilon_k''}$ and the acoustic flux $f_p = \av{p' u_r'}$. These transport terms characterize the non-locality of turbulent convection. Source terms for KE are the gravitational work due to density fluctuations $W_b = \av{\rho'  \vec u' \cdot \vec g}$, and the work done by pressure fluctuations $W_p = \av{p' \vec \nabla \cdot \vec u'}$, also called ``pressure-dilatation". There is some freedom in the formulation of the kinetic energy equation, especially in the expression/splitting of the driving terms, see e.g. discussions in \cite{lrsp-2009-2} and \cite{arnett_turbulent_2011}. Here we choose to split the driving into $W_p$ and $W_b$ as these are thermodynamically the most fundamental quantities: $W_b$ measures the reversible exchange between kinetic energy and potential energy, $W_p$ measures the reversible exchange between kinetic energy and internal energy. Furthermore, this splitting will highlight the differences between the red giant and oxygen-burning shell models. $W_b$ is often called the buoyancy work, an appellation inherited from the Boussinesq literature. In Sect. \ref{KEdriving} we present another expression which emphasizes the underlying physical origin of the driving.

As discussed in Sect. \ref{setup}, in the ILES paradigm we rely on the numerical dissipation at the grid scale to mimic the effect of viscosity. As a result, kinetic energy is dissipated at the grid scale, and a sink term $- \rho \epsilon_d$ is introduced in the analysis. We compute each term in the balance equation, and identify the residual with the dissipation. This is valid for conservative methods of advection.

\begin{figure*}[t]
\centering
\parbox{0.49\linewidth}{\center \includegraphics[width=0.8\linewidth]{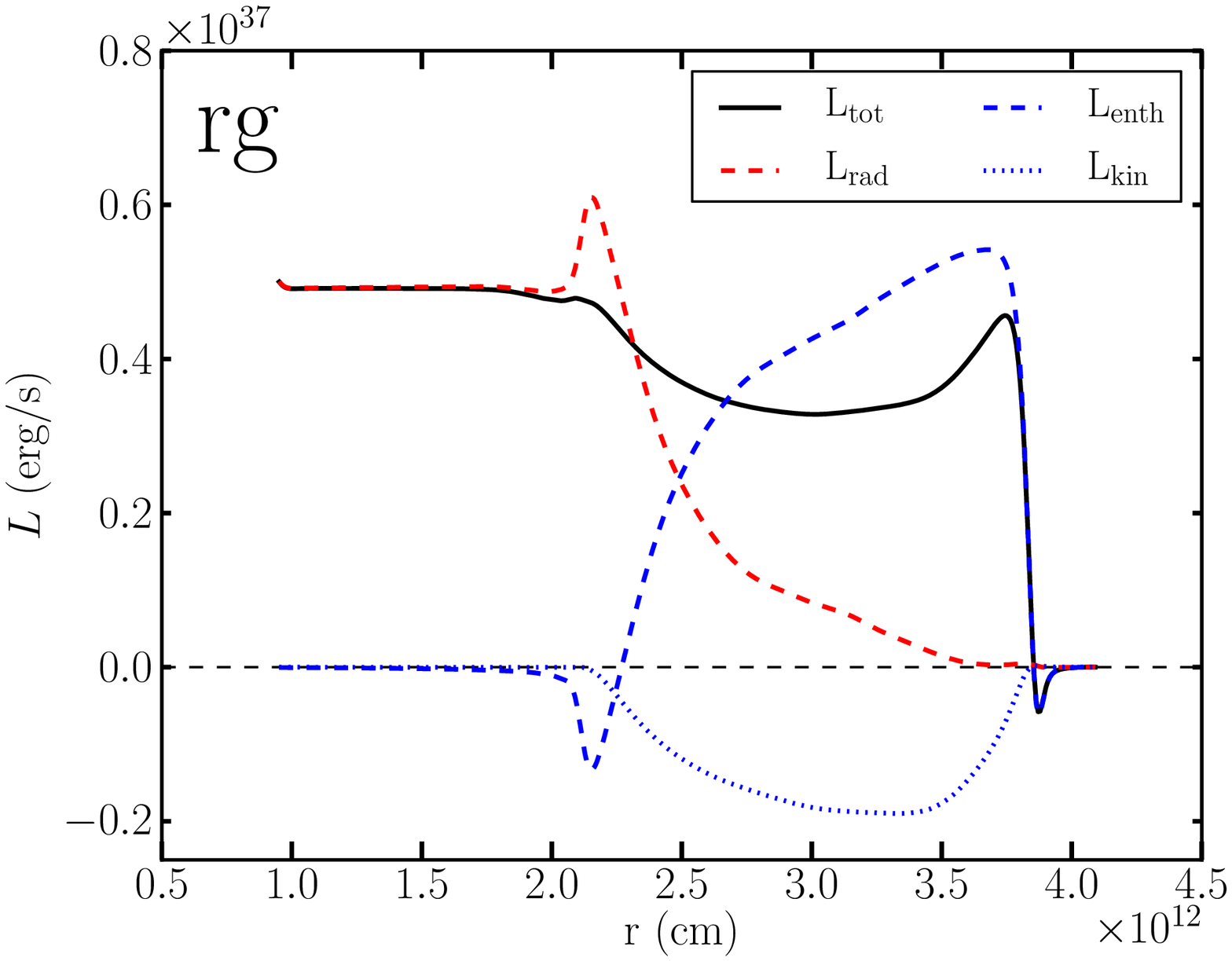}}
\parbox{0.49\linewidth}{\center \includegraphics[width=0.8\linewidth]{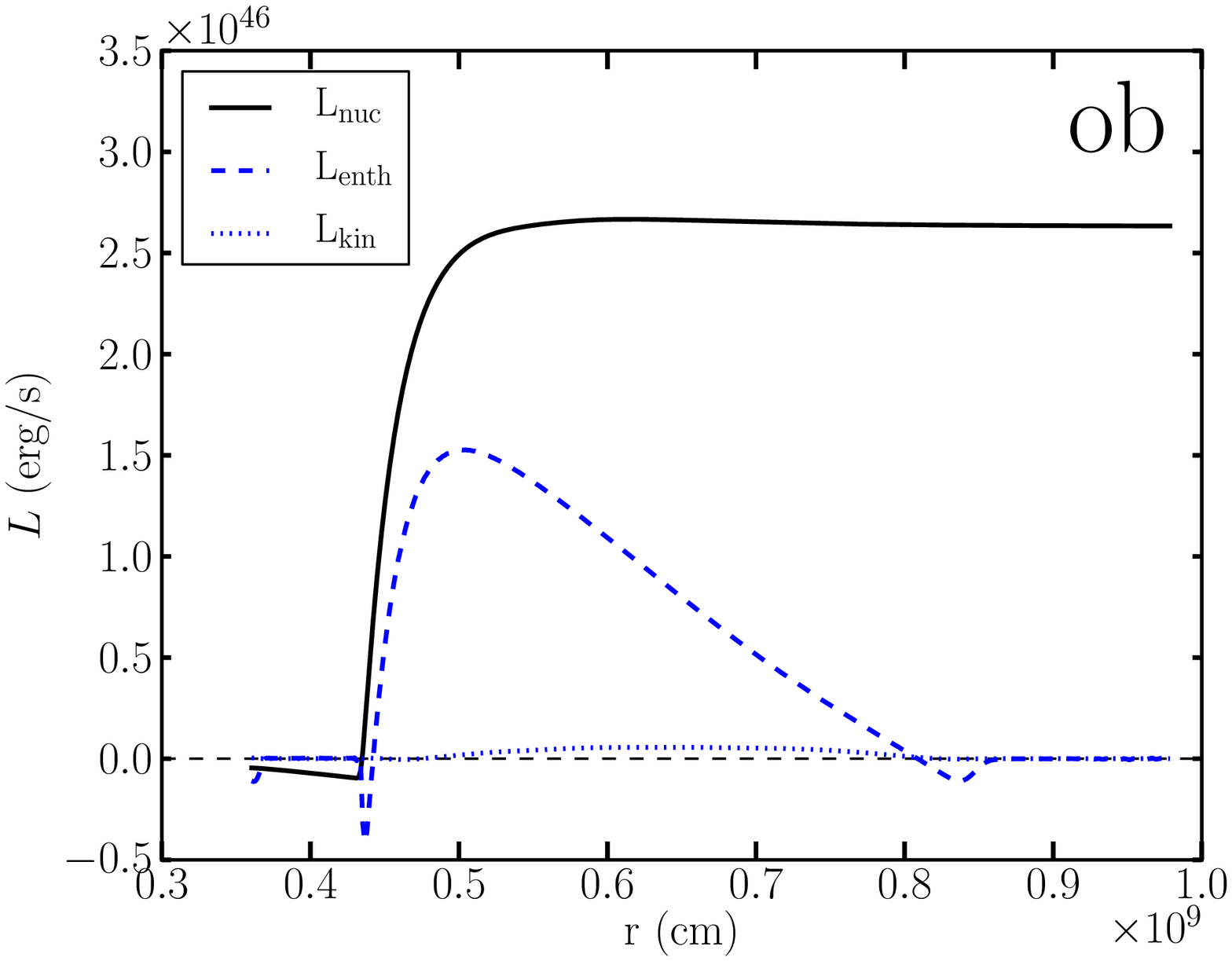}}
\caption{Radial profiles of different luminosities in {\sf rg.3D.mr} (left panel) and in {\sf ob.3D.mr}  (right panel).}
\label{fig:energyflux_balance}
\end{figure*}

The middle-left panels of Fig. \ref{fig:rans_rg} and Fig. \ref{fig:rans_ob} show the radial profiles of the different terms in the KE balance for models {\sf rg.3D.mr} and {\sf ob.3D.mr}. The inferred dissipation is shown as a black dashed line. The kinetic energy balance is in a statistically steady state, as the contribution from the time derivative is negligible. In the red giant model, we find that both $W_b$ and $W_p$ drive (i.e. positive contribution) turbulent motion within the convective zone. $W_b$ dominates the driving but the pressure-dilatation term is significant and cannot be neglected. This contrasts with the oxygen-burning shell models where driving is largely dominated by $W_b$, with no significant contribution from pressure-dilatation. We attribute the importance of $W_p$ in the red giant model to the degree of stratification, which is larger than in oxygen-burning model (see Sects. \ref{pressure_fluctuations} and \ref{KEdriving}). The convective region is bounded by regions where $W_b <0$. The KE dissipation profile has a large amplitude and is rather evenly distributed over the convective region. There is no local balance between the driving and the dissipation, due to transport by the turbulent fluxes. In the red giant model, the net effect of the turbulent KE flux is to transport energy downward in the convective zone, whereas in the oxygen-burning shell model KE is transported upward. In the red giant model, the acoustic flux has an opposite effect to the KE flux as it transports KE upwards. It has a more complex behavior close to the boundaries, due to an important contribution from waves. In the oxygen-burning shell model the acoustic flux is important close to the boundaries, but it has a negligible effect in the bulk of the convective zone.

The middle-right panel of Fig. \ref{fig:rans_rg} and Fig. \ref{fig:rans_ob} (see also Tables \ref{table:rg_ek_budget} and \ref{table:ob_ek_budget}) show the KE integral budget over the convective zone. Upon integration on the convective zone, we are left with a balance between driving and dissipation:

\begin{equation}
\label{eq:ek_balance}
\int W_b \dV + \int W_p \dV \approx \int \av{\rho \epsilon_d} \dV.
\end{equation}

\noindent  To obtain the balance described by Eq. (\ref{eq:ek_balance}), we have integrated over the unstable region and the convective boundary layers, where dissipation is taking place. The rough balance between driving and dissipation obtained over this region implies that only a small amount of kinetic energy in transmitted to the stable zones in form of internal waves, so that the dissipation of waves energy in these regions is small.

\subsubsection{Total energy balance}
\label{rans:analysis_et}

\begin{deluxetable}{c c c c}
\tablecaption{\label{table:rg_et_budget} Total energy balance for model {\sf rg.3D.mr} - integral budget over the convective zone (erg s$^{-1}$).}
\tablehead{Term & Value & Term & Value}
\startdata
$- \int \av{\rho} D_t \fav{\epsilon_t} \dV $ &  2.18(36) & $ - \int \av{p} \nabla_r \fav{u_r} \dV$ & -2.13(36) \\
$- \int \nabla_r f_h \dV$ & -1.01(35)     &   $- \int q \dV$ & -5.45(36)\\
$- \int \nabla_r f_k \dV$ &  -1.61(32)  & Residual & 6.30(35)\\
$- \int \nabla_r \av{F_r} \dV$  & 4.88(36)   \\
\enddata
\end{deluxetable}

\begin{deluxetable}{c c c c}
\tablecaption{\label{table:ob_et_budget} Total energy balance for model {\sf ob.3d.mr} - integral budget over the convective zone (erg s$^{-1}$).}
\tablehead{Term & Value & Term & Value}
\startdata
$- \int \av{\rho} D_t \fav{\epsilon_t} \dV $ &  -1.92(46) & $ - \int \av{p} \nabla_r \fav{u_r} \dV$ & -8.20(45) \\
$- \int \nabla_r f_h \dV$ & 4.26(43)    &   $- \int \av{\rho \epsilon_\mathrm{nuc}} \dV$ & 2.73(46)\\
$- \int \nabla_r f_k \dV$ &  -1.96(41)  & Residual & 1.20(44)
\enddata
\end{deluxetable}

We now discuss the total energy balance, Eq. (\ref{eq:rans_etot}). The fluxes that appear in this equation are the radiative flux, the enthalpy flux, and the kinetic energy flux:

\begin{align}
F_r &= - \av{\chi} \partial_r \av{T}  \label{eq:fr},\\
f_h &=  \av{\rho} \fav{h'' u_r''} \label{eq:fh},\\
f_k &= \av{\rho} \fav{\epsilon_k'' u_r''}.\label{eq:fk}
\end{align}

We discuss this form of the energy equation because of its relevance for stellar evolution calculations. For instance, let us consider the simplest case by assuming a steady state and $\fav{u_r}=0$. Multiplying Eq. (\ref{eq:rans_etot}) by $4 \pi r^2$ and integrating over radius, one obtains:

\begin{equation}
\label{eq:et_balance}
4 \pi r^2 \big ( f_h + f_k + F_r ) = L(r),
\end{equation}

\noindent where $L(r) = \int_0^r  \av{\rho \epsilon_\mathrm{nuc}} \dV$. Stellar evolution codes use the mixing-length theory to compute $f_h$ and ignore $f_k$. Figure \ref{fig:energyflux_balance} illustrates the different terms in Eq. (\ref{eq:et_balance}) for models {\sf rg.3D.mr} and {\sf ob.3D.mr}. The radiative flux is negligible in the oxygen-burning shell models and is therefore ignored. In the red giant model, the radiative luminosity is equal to the stellar luminosity in the radiative zone, and it decreases in the convective zone where a large and outward directed enthalpy flux takes over. Furthermore, the red giant model is characterized by a downward directed kinetic energy flux reaching an amplitude of roughly $35$ \% of the maximum enthalpy flux. The convective boundary is characterized by a negative, i.e. downward directed, enthalpy flux. This is somewhat counterbalanced by a bump in the radiative luminosity at the same location. The total luminosity is not constant, which means that the model is not in thermal equilibrium. The oxygen-burning shell model is characterized by an upward kinetic energy flux, with a maximum amplitude roughly equal to 5 \% of the maximum enthalpy flux. Neutrino cooling is responsible for the shallow region of negative luminosity. Nuclear burning is more important and dominates $L_\mathrm{nuc}$. A net heating results and the system is in thermal imbalance (Eq. \ref{eq:et_balance} is not valid in this case).

The bottom-left panels in Fig. \ref{fig:rans_rg} and Fig. \ref{fig:rans_ob} show the different terms of Eq. (\ref{eq:rans_etot}) for models {\sf rg.3D.mr} and {\sf ob.3D.mr}. In the red giant case, we introduced in the r.h.s. the Newtonian cooling term $q$, see Eq. (\ref{eq:surface_cooling}). This last term is only important at the top of the CZ, where it mimics the radiative cooling that would take place if we had a realistic photosphere. This convection is driven by cooling at the top. At the bottom of the convective zone, there is a rough balance between the radiative and the enthalpy flux. Radiation cools the gas in a shallow layer below the convective zone, and heats it at the bottom of the convective zone. The radiative cooling in a shallow layer below the convective zone is a different manifestation of the bump seen in the radiative luminosity, this is discussed further in Sect. \ref{penetration}. The discussion of kinetic energy flux remains the same as in the previous section: kinetic energy is transported downward in the convective zone. The oxygen-burning shell model shows a rough balance between nuclear heating and the divergence of the enthalpy flux: convection is driven by a heating from below. It should be noticed that the time derivative of the total energy contributes significantly to the balance, as a result from the thermal imbalance. Spurious oscillations are also present at the location of the steep gradients (at the convective boundaries), but we have checked that the consistency was good everywhere else.

\begin{figure*}[t]
\centering
\parbox{0.49\linewidth}{\center \includegraphics[width=0.8\linewidth]{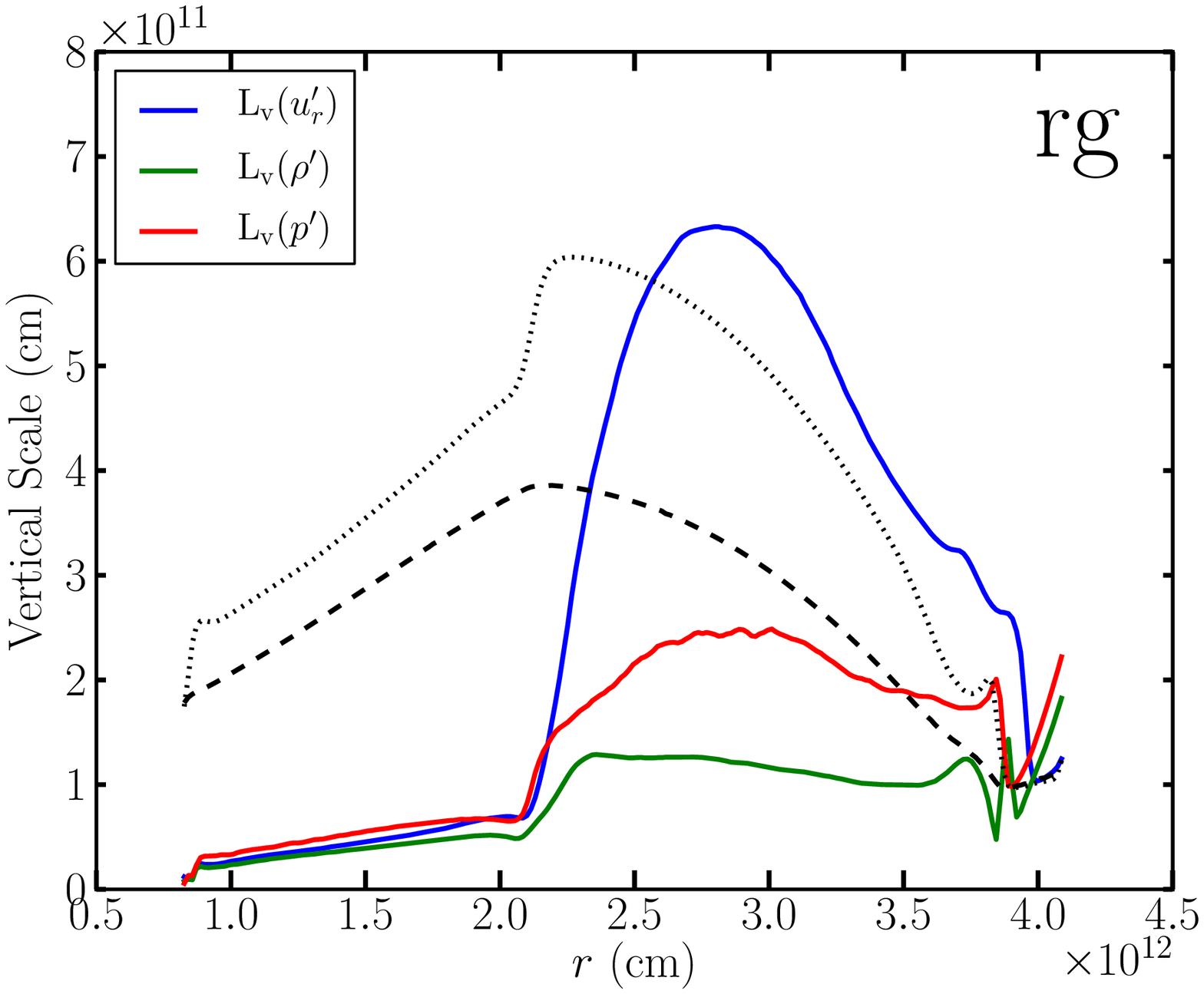}}
\parbox{0.49\linewidth}{\center \includegraphics[width=0.8\linewidth]{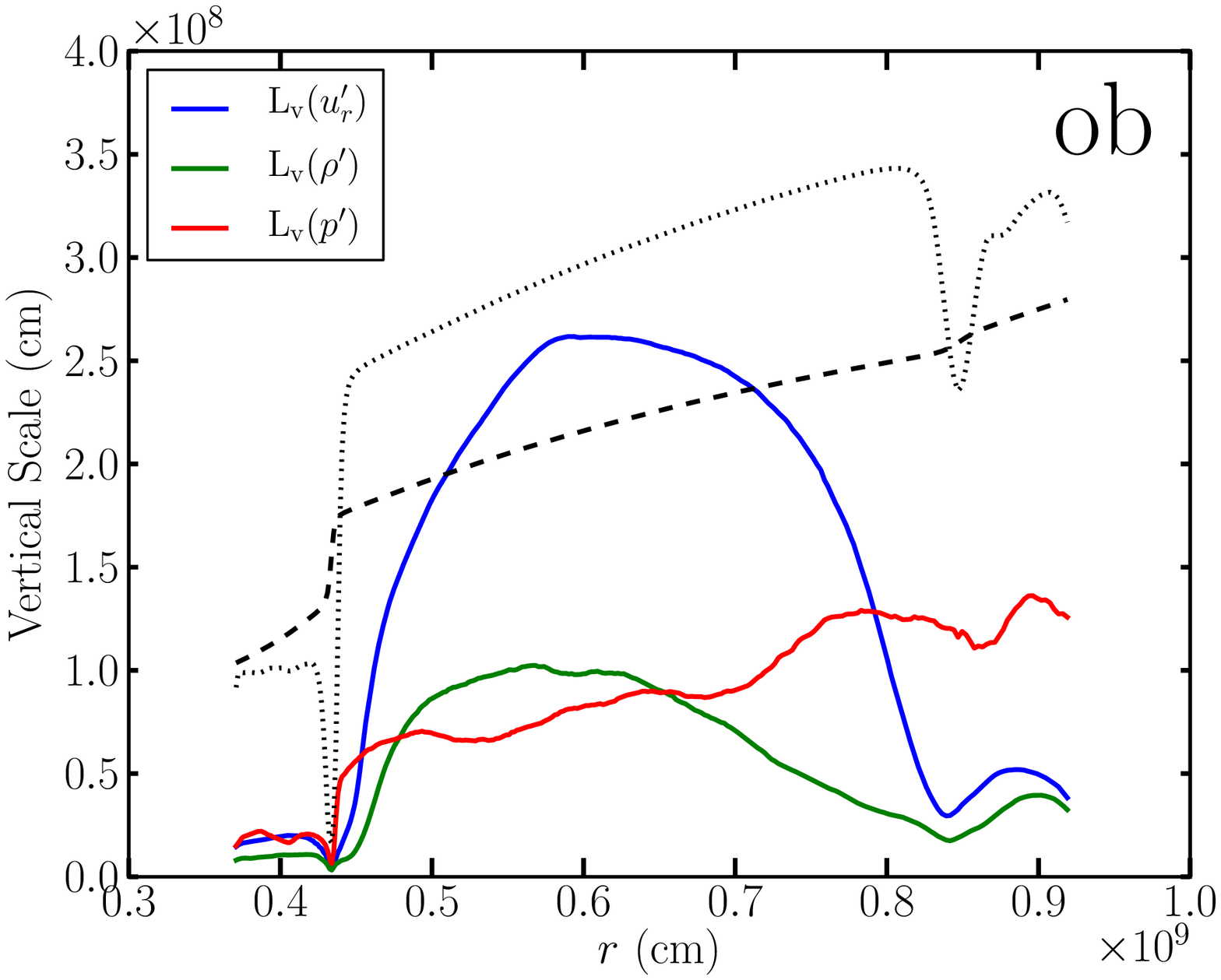}}
\caption{Vertical correlation length-scales for the radial velocity, density and pressure fluctuations in model {\sf rg.3D.mr} (left panel) and model {\sf ob.3D.mr} (right panel). The dashed and dotted lines show the pressure and density scale-heights (respectively).}
\label{fig:Lv}
\end{figure*}

The bottom-right panels in Fig. \ref{fig:rans_rg} and Fig. \ref{fig:rans_ob} show the budget integrals over the convective zone for each model. In both case, a net heating term is present, in the red giant due to radiation, in the oxygen-burning shell due to nuclear burning. In the latter, this is counterbalanced by the evolution of the background state, through the change in time of the total energy  and the $-\av{p}\nabla_r \fav{u_r}$ term which is the rate of work by the mean pressure due to a global expansion/contraction of the star. In the red giant model, the radiative heating is mostly balanced by the cooling at the surface. The terms describing a global evolution of the background roughly balance each others, and describe how the background state is slowly evolving toward thermal equilibrium (see discussion in Sect. \ref{penetration})

\subsection{The turbulent velocity field}
\label{turbulent_velocity_field}

In this section, we analyze the properties of the velocity field, and consider relevant approximations for this field. We analyze in some detail the anisotropy of the Reynolds stresses. Our results motivate a two component decomposition to the flow, with plumes at large scales and isotropic turbulence at small scales. We show how the isotropic component is related to the observed kinetic energy dissipation, consistent with  the Kolmogorov cascade.

\subsubsection{Approximations for deep and shallow convection}
\label{boussinesq_anelastic}

Figures \ref{fig:rg_thermodynamical_perturbations} and \ref{fig:ob_thermodynamical_perturbations} show that the thermodynamical fluctuations in both stellar models are small relative to the background values. Therefore, for the purpose of analysis, we will often use simplified equations resulting from a linearization around the background state. In such a case, we shall use $\rho_0$, $P_0$, etc, instead of the notation $\av{\rho}$, $\av{P}$, etc, to emphasize that we are considering the linearized equations.

The linearized continuity equation reads

\begin{align}
\partial_t \rho' + \vec \nabla \cdot ( \rho_0 \vec u') = 0.
\end{align}

Both the red giant and oxygen-burning shell models are characterized by low Mach flows. Low Mach turbulence is an inefficient producer of sound \citep{lighthill_sound_1952}. Therefore, we can neglect $\partial_t \rho'$ in the above equation and focus on fluctuations having a time scale longer than the acoustic time scale \citep[][]{gough_anelastic_1969,dutton_approximate_1969}. Furthermore,

\begin{align}
\vec \nabla \cdot ( \rho_0 \vec u') = \rho_0 \vec \nabla_h \cdot \vec u_h' +  \rho_0 \nabla_r u_r' + u_r' \partial_r \rho_0,
\end{align}

\noindent where we have separated the horizontal flow (subscript $h$) from the vertical flow. The order of magnitude of the ratio of the third term to the second  is:

\begin{equation}
\frac{| u_r' \partial_r \rho_0 |}{|\rho_0 \nabla_r u_r'|} \sim \frac{L_u}{H_\rho},
\end{equation}

\noindent where $L_u$ is the characteristic vertical length-scale of radial velocity perturbations, and $H_\rho = - dr/d\ln \rho_0$ is the density scale-height. For ``shallow" convection,  this ratio is small and the following approximation is justified:

\begin{align}
\label{eq:boussinesq}
\vec \nabla \cdot  \vec u' &= 0,
\end{align}

\noindent i.e., the turbulent velocity field is solenoidal. Otherwise, for ``deep" convection, where $L_u \gtrsim H_\rho$, the appropriate approximation is:

\begin{align}
\label{eq:anelastic_momentum}
\vec \nabla \cdot \big ( \rho_0 \vec u' \big ) &= 0.
\end{align}

\noindent From this relation, we can deduce that

\begin{equation}
\label{eq:anelastic_dilatation}
\vec \nabla \cdot  \vec u'  = \frac{u_r'}{H_\rho}.
\end{equation}

\noindent The dilatation of the velocity field is due to the vertical motion of parcels in the background stratification.

\cite{meakin_turbulent_2007} compute the two-point correlation function of the radial velocity fluctuations:

\begin{equation}
C^V(r, \delta r) = \frac{\av{u_r'(t; r, \theta, \phi) u_r'(t; r + \delta r , \theta, \phi)}}{u_\mathrm{rms}(r) u_\mathrm{rms}(r+\delta r)},
\end{equation}

\begin{figure*}[t]
\parbox{0.49\linewidth}{\center \includegraphics[width=0.8\linewidth]{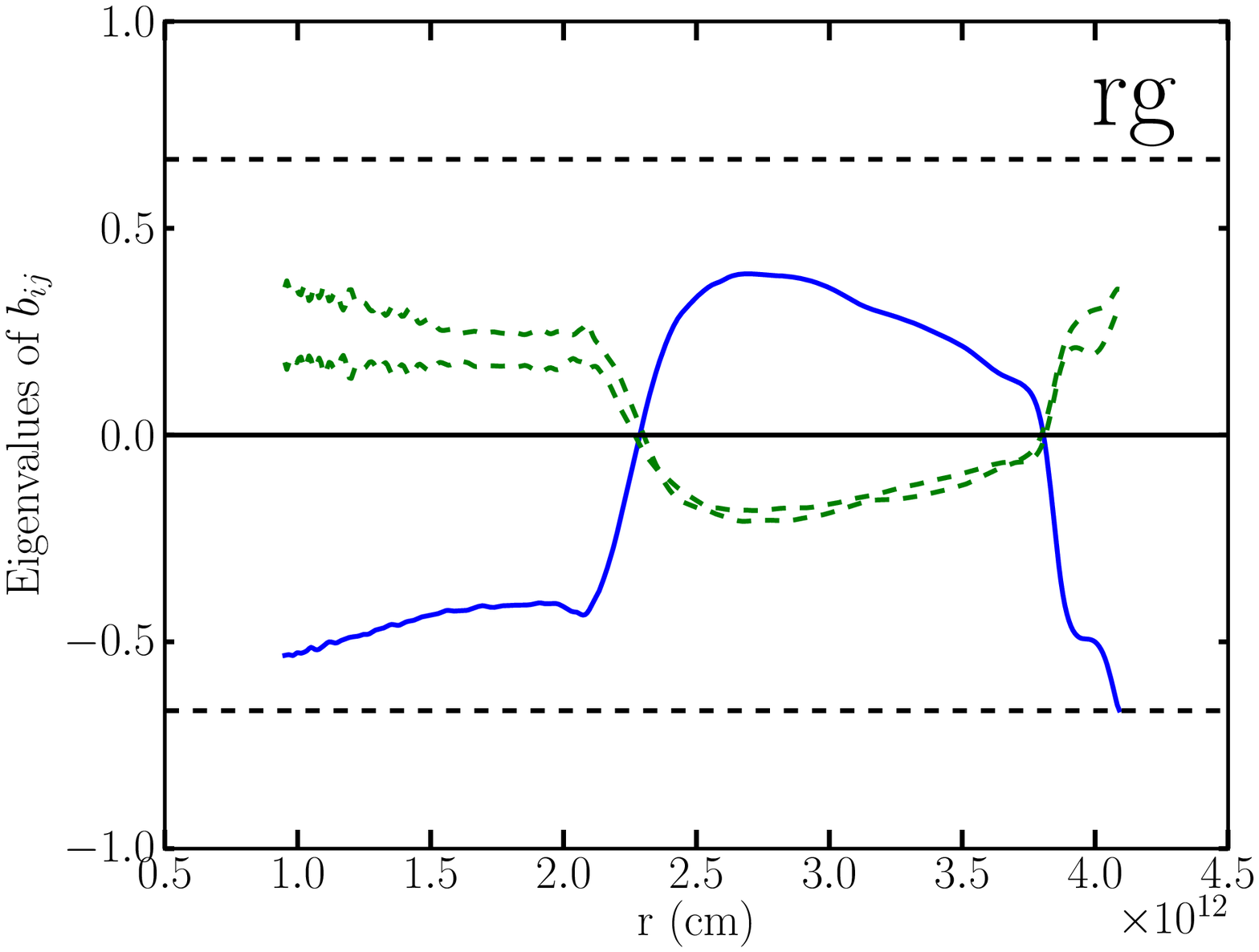}}
\parbox{0.49\linewidth}{\center \includegraphics[width=0.8\linewidth]{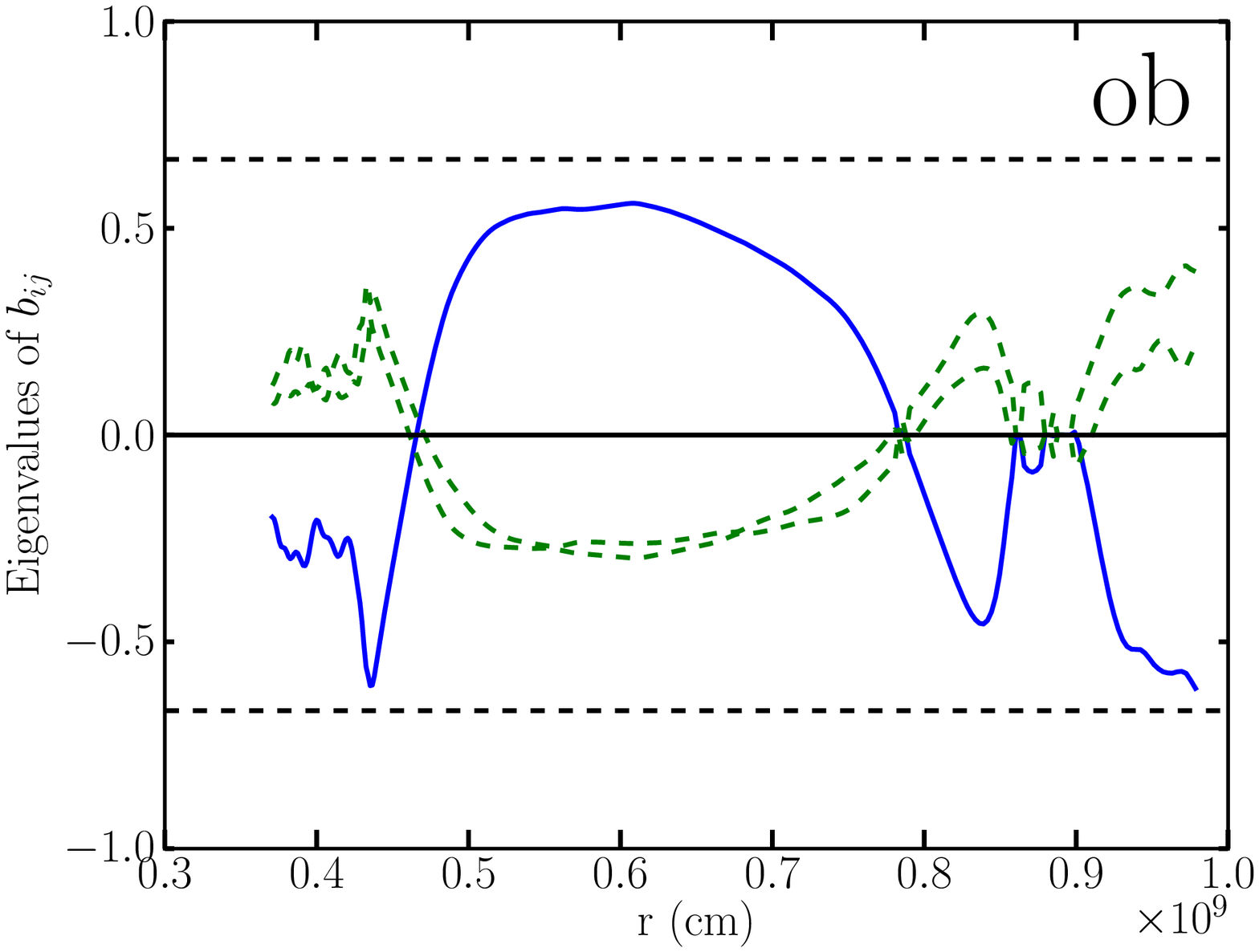}}
\caption{Eigenvalue analysis of the anisotropy tensor $b_{ij}$for model {\sf rg.3D.mr} (left panel) and model {\sf ob.3d.mr} (right panel). The figures show the radial profiles of the eigenvalues. The continuous line corresponds to the radial eigenvalue, the dashed lines correspond to the horizontal eigenvalues (see text). The two horizontal dashed lines indicate the range $[-2/3, 2/3]$.}
\label{fig:anisotropy_eigenvalues}
\end{figure*}

\noindent where $u_\mathrm{rms}^2 = \av{u_r'^2}$. The vertical correlation length-scale is defined as the width at half maximum of the two-point correlation function $C^V$. Figure \ref{fig:Lv} shows the vertical correlation length-scale of the radial velocity fluctuations (among others) computed at each radii in models {\sf rg.3D.mr} and {\sf ob.3D.mr}. The figure emphasizes an important difference between the models: in the oxygen-burning shell model, the vertical correlation length-scale is everywhere smaller than the density scale-height, whereas in the red giant model it is larger in most of the convective zone. This suggests that the flow in the oxygen-burning shell model is much less affected by the stratification than the red giant model. Therefore, we will adopt the shallow convection approximation Eq. (\ref{eq:boussinesq}) to describe the turbulent velocity field in the oxygen-burning shell models. For the red giant models, the appropriate approximation is the one given by Eq. (\ref{eq:anelastic_momentum}).

Equations (\ref{eq:boussinesq}) and (\ref{eq:anelastic_momentum}) are the basis for the Boussinesq and anelastic approximations. We stress that for both the oxygen-burning shell and the red giant models, these approximate models of the hydrodynamical equations \emph{should not} be used to model the flow. For instance, the computational domain in the oxygen-burning shell models is not ``shallow", i.e. $(r_\mathrm{out} -r_\mathrm{in})/H_p > 1$, and in the red giant models the Mach numbers are too large near the surface. However, these approximations provide a useful theoretical framework for the analysis of the results.

\subsubsection{Anisotropy of the Reynolds stresses}
\label{anisotropy}

We decompose the Reynolds stresses as

\begin{equation}
\fav{u_i''u_j''} = \frac{2}{3} \fav{k} \delta_{ij} + \fav{k} b_{ij},
\end{equation}

\noindent where $\fav{k} = \frac{1}{2}\fav{u_i''u_i''} $ is the specific turbulent kinetic energy and $b_{ij}$ is a trace-less and symmetric tensor that characterizes the anisotropy of the Reynolds stresses. We extract the three eigenvalues  $(\lambda_r, \lambda_\theta, \lambda_\phi)$ and eigenvectors of $b_{ij}$ at each radii. It can be shown that, for any two eigenvalues $\alpha$, $\beta$, one has $\alpha \ge -2/3$, $\beta \ge -2/3$, and $\alpha+\beta \leq 2/3$.

At each radius, one of the eigenvalue is unambiguously associated with a purely radial eigenvector, and the two other eigenvalues are roughly equal and associated with two purely horizontal eigenvectors. Therefore, the Reynolds stresses are axisymmetric, which is expected as the angular directions are homogeneous in the absence of rotation or magnetic field. The orientation of the horizontal eigenvectors is therefore not physically relevant. The radial profiles of the radial eigenvalue $\lambda_r$ and of the horizontal eigenvalues $\lambda_\theta$ and $\lambda_\phi$ are shown in Fig. \ref{fig:anisotropy_eigenvalues} for models {\sf rg.3D.mr} and {\sf ob.3D.mr}. These quantities describe the shape of the Reynolds stress tensor. In the bulk of the convective zone, where $\lambda_r > 0$ and $\lambda_\theta = \lambda_\phi <0$, the shape is ``rod-like" as the stress is maximum in the radial direction. Otherwise, when $\lambda_r < 0$  and $\lambda_\theta = \lambda_\phi > 0$, the shape is ``disc-like". The radii at which the transition between the two shapes occurs (and where all eigenvalues cancel) are $r_\mathrm{in} \sim 2.2 \times 10^{12}$ cm and $r_\mathrm{out} \sim 3.75 \times 10 ^{12}$ cm in the red giant model, and  $r_\mathrm{in} = 0.45\times 10^9$ cm and $r_\mathrm{out} \sim 0.85 \times 10 ^{9}$ cm in the oxygen-burning shell model. At these radii, the vertical motions are deflected horizontally as they approach the convective boundaries. A second transition radius is seen at $r=2\times 10^{12}$ cm in the red giant model and at $r=0.41\times 10^9$ cm in the oxygen-burning shell model, below this radius the flow is dominated by waves rather than turbulence.

The anisotropy of the Reynolds stresses is another manifestation of the two components character of the flow, with convective plumes at large scales that dominate the vertical motion and an isotropic component at small scales. It is essential for the design of turbulence models to better understand the role of the different scales of the flow, for instance in terms of the transport of energy \citep{cattaneo_turbulent_1991, 2011ApJ...728..115B}. We leave a detailed analysis for a future publication. 

Here, we outline an approach for decomposing the velocity field. We write:

\begin{equation}
u_i = \fav{u_i} + v_i'' + w_i'',
\end{equation}

\noindent where we have split the velocity fluctuation ($u_i''$ in our usual notation) into a large scale, coherent structures component $v_i''$, and an isotropic component $w_i''$ characterizing the small scales. We then assume that both components have zero averages and are not correlated:

\begin{align}
\fav{v_i''} &= 0, \forall i,\\
\fav{w_j''} &= 0, \forall j,\\
\fav{v_i'' w_j''} &= 0, \forall (i,j).
\end{align}

With these assumptions, the velocity components correlation tensor can be split into two contributions:

\begin{equation}
\fav{u_i'' u_j''} = \fav{v_i'' v_j''} + \fav{w_i'' w_j''}.
\end{equation}
 
Furthermore, we have

\begin{align}
\fav{v_i'' v_i''}   &= 2\tilde{k}_\mathrm{plumes},\\
\fav{w_i'' w_j''} &= \frac{2}{3} \tilde{k}_\mathrm{iso}  \delta_{ij},
\end{align}

\noindent with $\fav{k} = \tilde{k}_\mathrm{plumes} + \tilde{k}_\mathrm{iso}$, where $\tilde{k}_\mathrm{plumes}$ and $\tilde{k}_\mathrm{iso}$ are the specific kinetic energy of the plumes and of the isotropic turbulence (respectively).  The above hypothesis are not sufficient to determine uniquely the decomposition. Here, it will be enough for our purpose to adopt the approach of \cite{meakin_turbulent_2007}: far from the boundaries, we identify the horizontal flow with the isotropic component. Therefore, as a proxy, we can define $\tilde{k}_\mathrm{iso}$ by

\begin{align}
2 \tilde{k}_\mathrm{iso} &= \frac{3}{2} \Big( \fav{{u_\theta''}^2} + \fav{{u_\phi''}^2} \Big).
\end{align}

Furthermore, since $\fav{{u_\theta''}^2} \approx \fav{{u_\phi''}^2}$, one has

\begin{equation}
\fav{v_i'' v_j''}   \approx 2 \tilde{k}_\mathrm{plumes} \delta_{i1} \delta_{j1},\\
\end{equation} 

\noindent with 

\begin{align}
2 \tilde{k}_\mathrm{plumes} = \fav{{u_r''}^2}  - \frac{\fav{{u_\theta''}^2} + \fav{{u_\phi''}^2}}{2}.
\end{align}

\subsubsection{Kinetic energy damping}
\label{global:ke_damping}
\label{turbulent_heating}

The total kinetic energy dissipated per unit time is $L_d = \int \av{\rho \epsilon_{d}} dV$, see Tables \ref{table:rg_runs} and  \ref{table:ob_runs}. The rate of dissipation is not constant in time, but evolves with the flow. In the red giant model, we see a decrease in time of $L_d$ toward the value quoted in the table as the model relaxes toward a quasi-steady state (see left panel of Fig. \ref{fig:rg_ekin}). In the oxygen burning-shell model, $L_d$ increases slowly with time as a result from the global heating due to the imbalance between nuclear heating and neutrino cooling.

As in \cite{arnett_turbulent_2009}, we compute a dissipation length-scale $l_d$ and timescale $\tau_d$ as:

\begin{align}
\int \av{\rho \epsilon_\mathrm{d}} dV  = M_\mathrm{CZ} \frac{v_\mathrm{rms}^3}{l_d}, \label{eq:global_lD}\\
\tau_d = \frac{E_\mathrm{K,CZ}}{\int \av{\rho \epsilon_{d}} dV} = \frac{1}{2} \frac{l_d}{v_\mathrm{rms}},
\end{align}

\noindent where $v_\mathrm{rms}$ is computed from the isotropic kinetic energy defined in the previous section, i.e. $v_\mathrm{rms}= (2 \tilde{k}_\mathrm{iso})^{1/2}$. This yields $l_d = 7.7 \times 10^{11}$ cm in the red giant model, and  $l_d \sim 0.4 l_\mathrm{CZ}$ ; and $l_d= 5.5 \times 10^{8}$ cm in the oxygen burning-shell model, so $l_d \sim l_\mathrm{CZ}$. Furthermore, $\tau_d \sim 19$~d in the red giant model and $\tau_d \sim 29$~s in the oxygen-burning shell model. In both cases this is much shorter than the convective turnover timescale. As discussed in \cite{arnett_turbulent_2009}, this illustrates the strong dissipative character of turbulent convection. 

\begin{figure}[t]
\center
\includegraphics[width=0.8\linewidth]{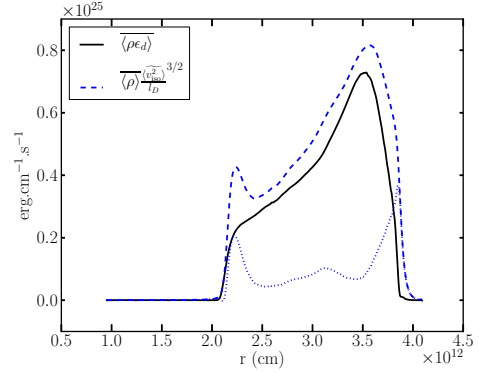}
\caption{Comparison of the dissipation described by a $\epsilon_k = u'^3/\Lambda$ law (dashed line) with the dissipation inferred from the data in model {\sf rg.3D.mr} (continuous line). The dotted line shows the residual.}
\label{fig:dissipation_modeling}
\end{figure}

We now relate the kinetic energy dissipation inferred from the numerical simulations with global properties of the flow. The simplest approach is to use the formula for the rate of dissipation in isotropic and homogeneous incompressible turbulence \cite[][]{pope_turbulent_2000}:

\begin{equation}
\label{eq:isotropic_dissipation}
\epsilon_d = \frac{u'^3}{\Lambda},
\end{equation}

\noindent where $u'$ and $\Lambda$ are the rms velocity and length-scale of the most energetic eddies. 
\cite{arnett_turbulent_2009} fit the dissipation in their oxygen-burning shell data with $u' = (2 \tilde{k}_\mathrm{iso})^{1/2}$ and by setting $\Lambda$ to the dissipation length-scale $l_d$ derived above, see the right panel in their Fig. 2. In Fig. \ref{fig:dissipation_modeling}, we show that the same approach gives fairly good results for the red giant model as well. The amplitude of the dissipation is slightly overestimated, but the overall shape of the profiles agree very well. This suggests that a somewhat larger value of $l_d$ would give a better fit. The largest discrepancies are found close to the convective boundaries, where $\tilde{k}_\mathrm{iso}$ is affected by the horizontal flow due to the interaction with the boundaries.

It should be emphasized that this approach works well in our case because we are in a statistically steady state. In situations where the kinetic energy changes rapidly, the dissipation rate lags behind the cascade rate given by $u'^3/\Lambda$ due to the time needed to redistribute the kinetic energy over the inertial range \citep{livescu_high-reynolds_2009}.

\subsection{Magnitude of pressure fluctuations}
\label{pressure_fluctuations}

A comparison of Fig. \ref{fig:rg_thermodynamical_perturbations} and Fig. \ref{fig:ob_thermodynamical_perturbations} shows that $p'/p_0 \sim \rho'/\rho_0$ in the red giant model, whereas the oxygen-burning shell model is characterized by $p'/p_0 < \rho'/\rho_0$. This is related to the background stratification.

Within the framework of the approximations presented in Sect. \ref{boussinesq_anelastic}, it is possible to obtain an elliptic equation for the pressure fluctuations (see Appendix \ref{elliptic_pressure_equation}): 

\begin{equation}
\label{parabolic_pressure}
\Delta  p' = - \vec \nabla : \big ( \rho_0 \vec u' \otimes \vec u'  \big )  - g \frac{\partial \rho'}{\partial r}.
\end{equation}

The first term on the r.h.s. describes the generation of pressure fluctuations by the Reynolds stresses, also called the ``pseudo-sound". The second term describes the generation of pressure fluctuations by buoyancy effects. The relative orders of magnitude of the different terms can be written as:

\begin{figure*}[t]
\centering
 \parbox{0.49\linewidth}{\center \includegraphics[width=0.8\linewidth]{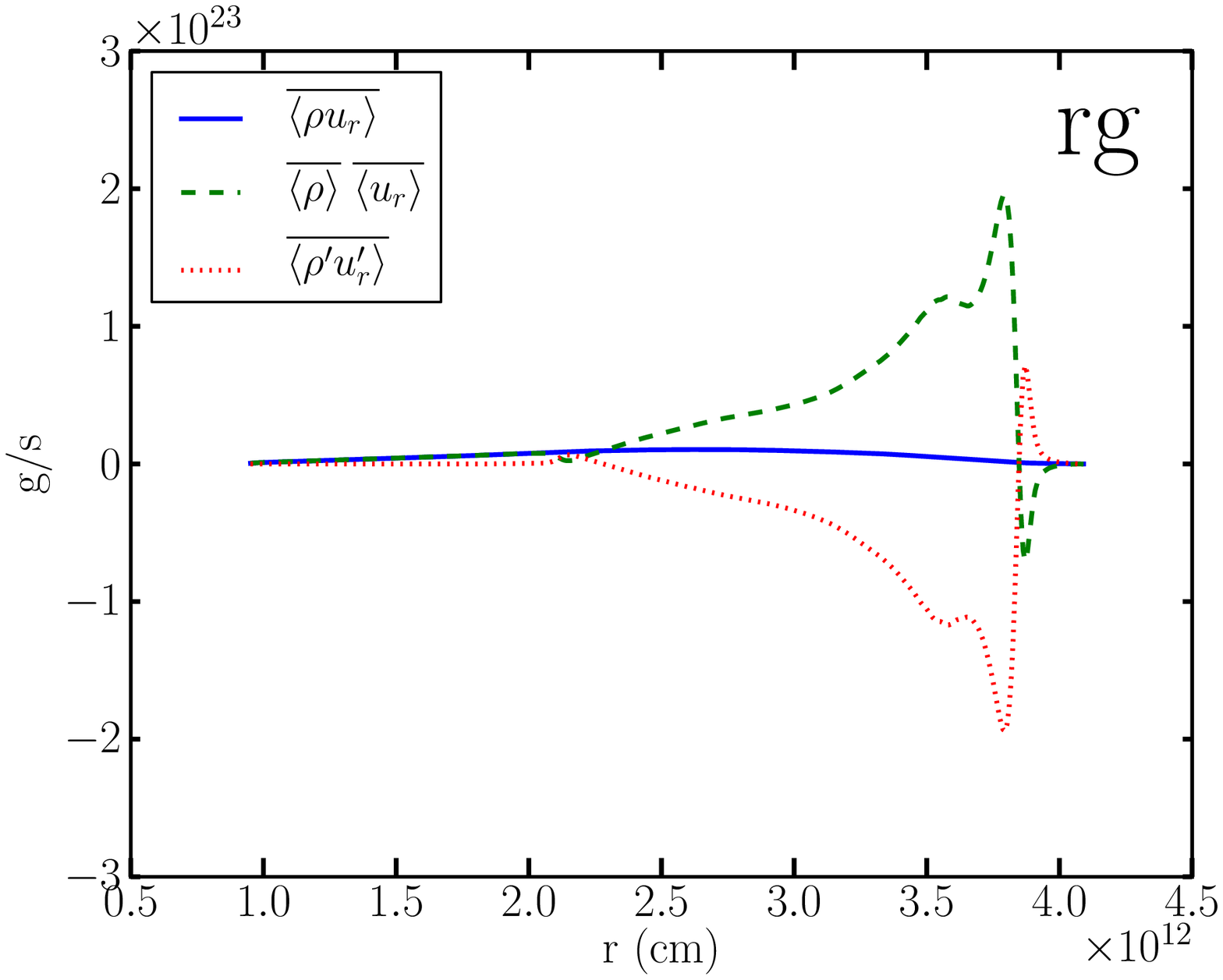}}
 \parbox{0.49\linewidth}{\center \includegraphics[width=0.8\linewidth]{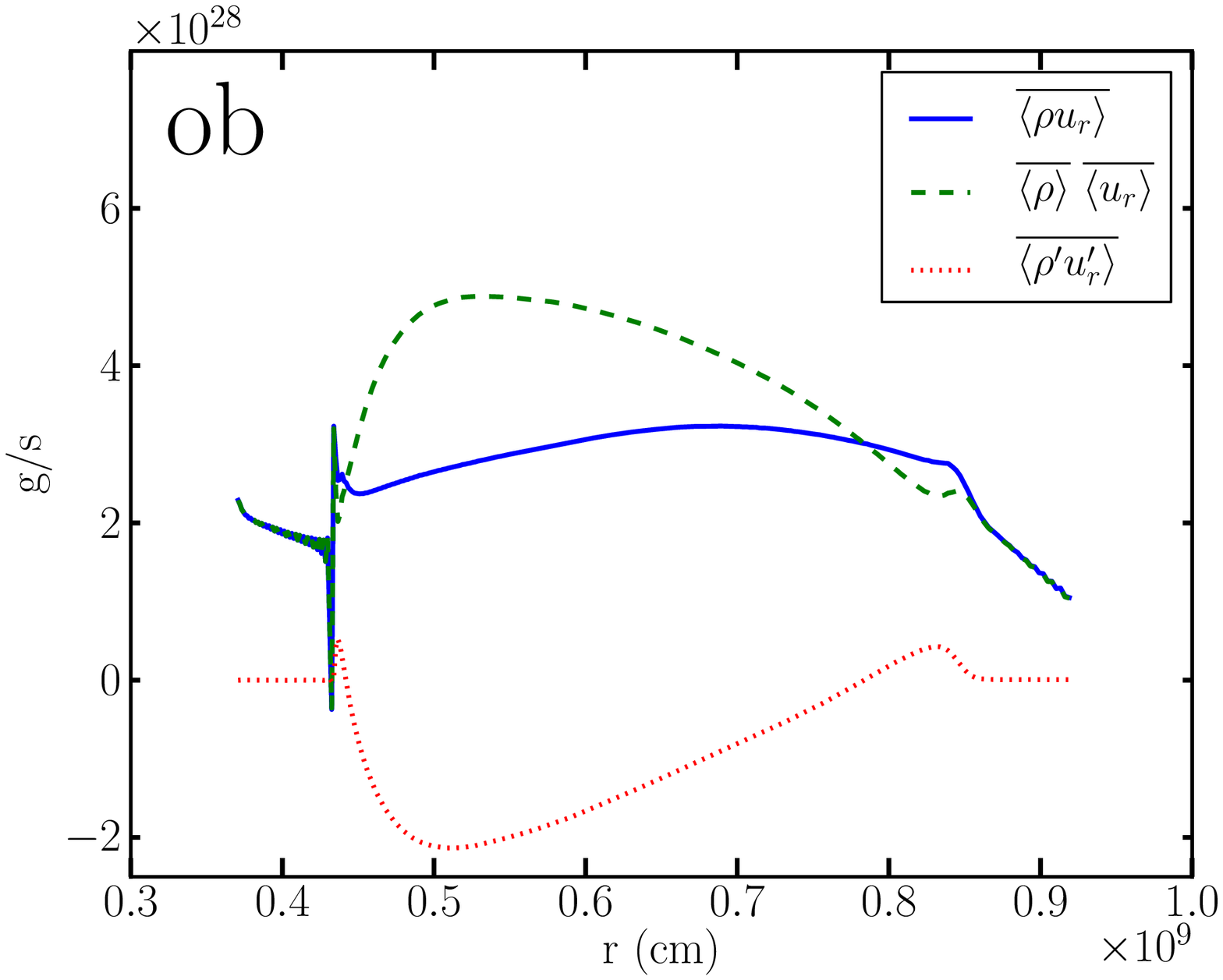}}
 \parbox{0.49\linewidth}{\center \includegraphics[width=0.8\linewidth]{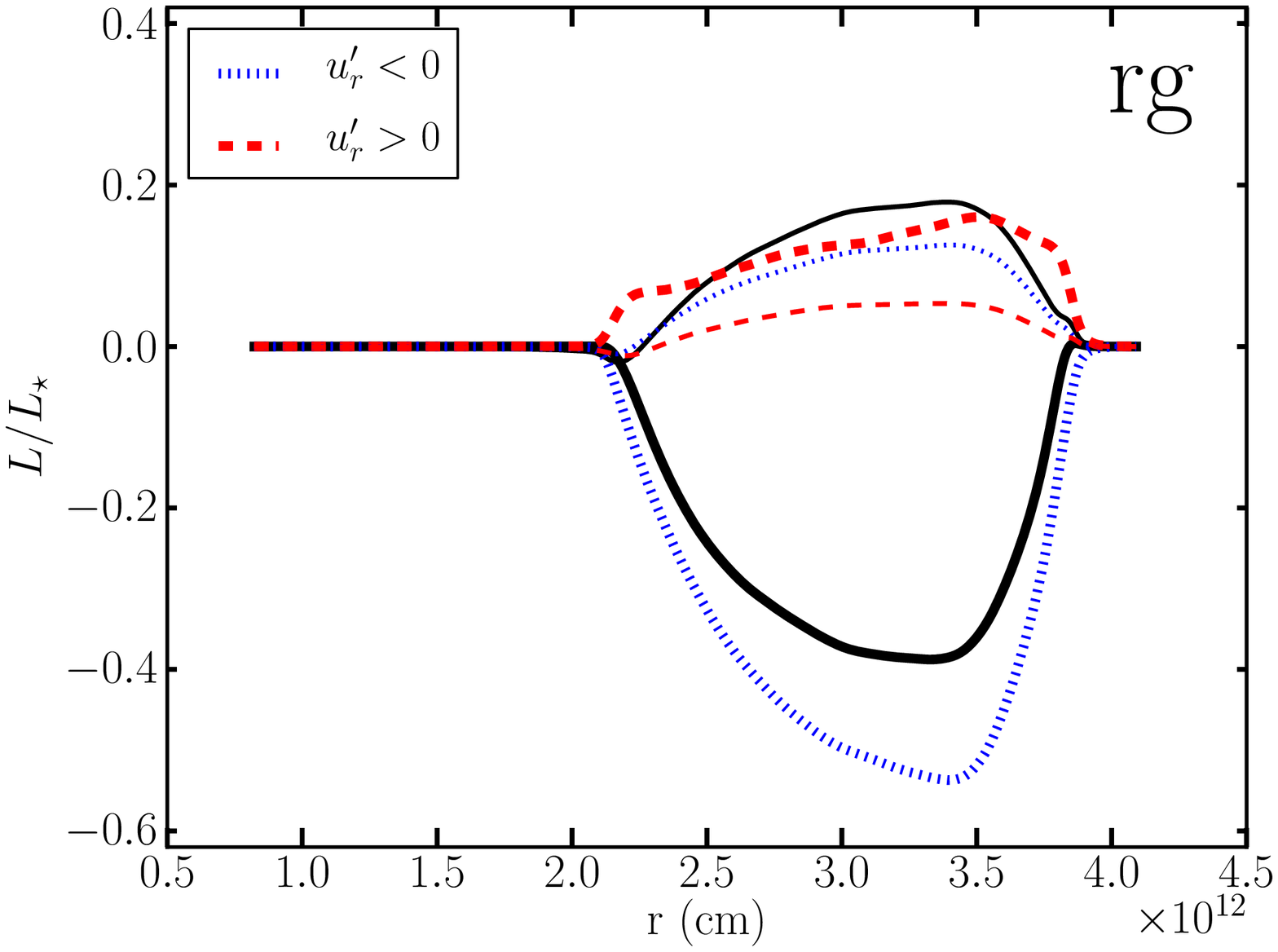} }
 \parbox{0.49\linewidth}{\center \includegraphics[width=0.8\linewidth]{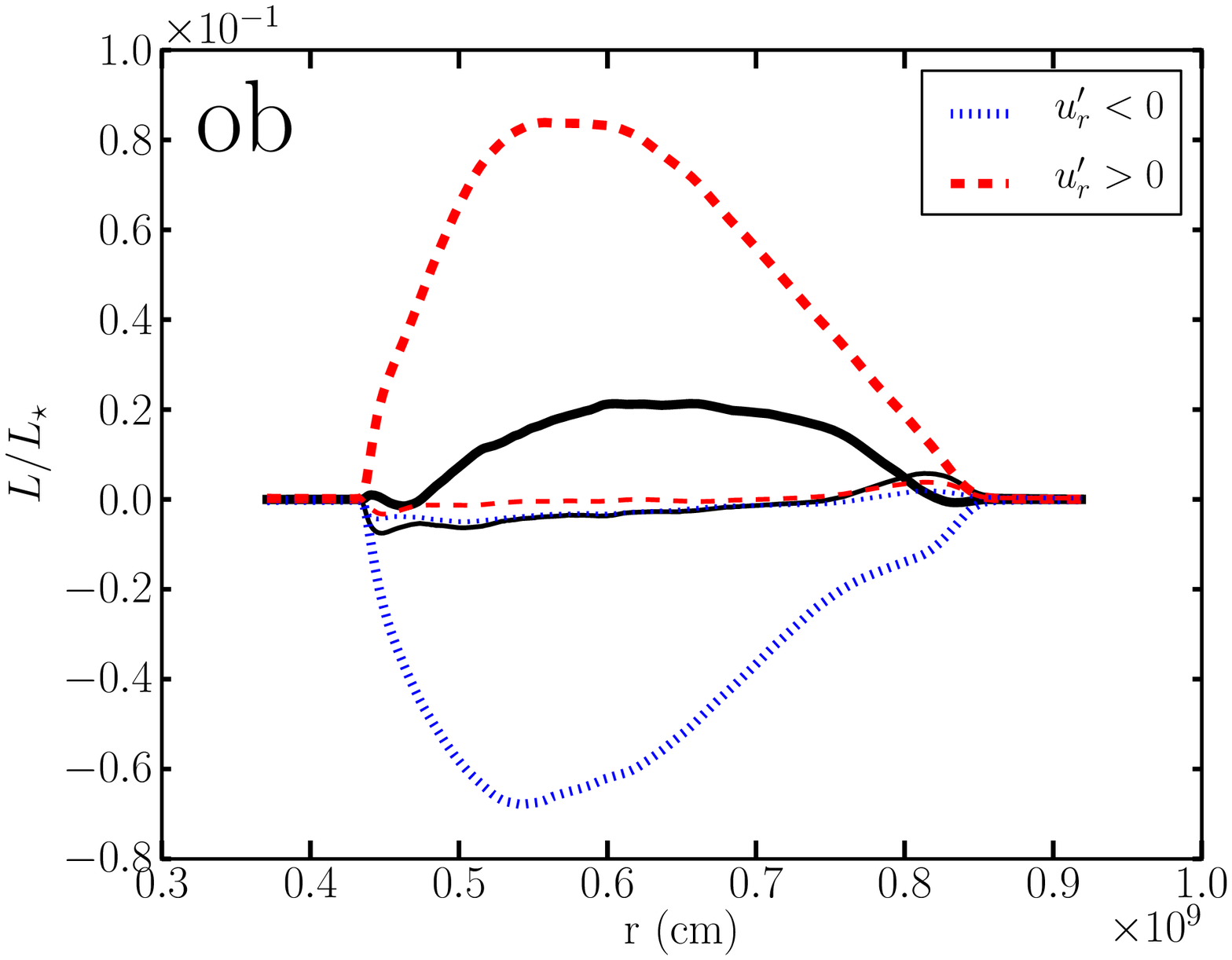} }
\caption{Top panels: decomposition of the average mass flux (see text). Bottom panels: Splitting of the kinetic energy (thick lines) and acoustic (thin lines) fluxes into up-flow and down-flow components (the fluxes have been multiplied by $4 \pi r^2$ and normalized by the luminosity). The continuous lines are the total fluxes. Left panels: model {\sf rg.3D.mr}, right panels: model {\sf ob.3D.mr}.}
   \label{fig:rg_fkp_updown}
\label{fig:turbulent_fluxes}
\end{figure*}

\begin{align}
\frac{|p'|}{p_0} :  \Big ( \frac{L_p}{L_u} \Big )^2 \frac{\rho_0 |u'|^2}{p_0}  :  \frac{L_p^2}{H_p L_\rho} \frac{|\rho'|}{\rho_0},
\end{align}

\noindent where $L_p$, $L_u$, and $L_\rho$ are typical length-scales for variation of the pressure, velocity, and density fluctuations. We use the vertical correlation length-scales for $u_r'$, $\rho'$ and $p'$ shown in Fig. \ref{fig:Lv} to estimate the relative magnitude of these effects. We estimate that $L_p^2/L_u^2 \sim 0.2$, $L_p^2/(H_p L_\rho) \sim 2$ in the red giant model, and $L_p^2/L_u^2 \sim 0.1$, $L_p^2/(H_p L_\rho) \sim 0.4$ in the oxygen-burning shell model. Although these estimates are only qualitative, they suggest that the pseudo-sound term is not the dominant source of pressure fluctuations in the convective zone (the dashed line in Figs. \ref{fig:rg_thermodynamical_perturbations} and \ref{fig:ob_thermodynamical_perturbations} falls below the pressure curve after multiplication by $L_p^2/L_u^2$). Therefore, pressure fluctuations are mainly due to buoyancy, and the order of magnitude analysis above shows that magnitude is related to the background stratification. If we loosely assume $L_\rho \sim L_p$ (which is not quite correct in the red giant models), we have

\begin{equation}
\label{eq:relation_pressure_density_fluctuations}
\frac{|p'|}{P_0} \sim \frac{L_p}{H_p}\frac{|\rho'|}{\rho_0},
\end{equation}

\noindent which is in qualitative agreement with the numerical models.

Although the order of magnitude estimates we carried out in this section have no predictive power, they suggest that Eq. (\ref{parabolic_pressure}) provides a valuable framework to analyse pressure fluctuations \citep[][]{chassaing2002variable}. We plan to push the analysis based on this equation further, with the aim of obtaining more quantitive predictions.

\subsection{The turbulent fluxes}
\label{turbulent_fluxes}

\subsubsection{The turbulent mass flux}

The turbulent mass flux is defined as

\begin{equation}
f_m = \av{\rho' u_r'}.
\end{equation}

By definition, we have

\begin{equation}
\label{eq:fm_decomposition}
\av{\rho} \fav{u_r} = \av{\rho u_r} = \av{\rho}\ \av{u_r} + \av{\rho' u_r'}.
\end{equation}

The top panels of Figure \ref{fig:turbulent_fluxes} illustrate this relation in models {\sf rg.3D.mr} and {\sf ob.3D.mr}. In the red giant model, one has $ \fav{u_r} \approx 0$ as the model is close to equilibrium. In this case, Eq. (\ref{eq:fm_decomposition}) implies that the mean flow $\av{u_r}$ counter-balances the mass displaced by turbulence. The oxygen burning-shell model shows a quite significant expansion of the background, i.e.  $ \fav{u_r} > 0$, driven by the imbalance between nuclear burning and neutrino cooling. There is still a mean flow which tends to counter-act the effect of the turbulent mass flux.

In the gravity field, the mass displaced by the turbulence induces a work which is one of the kinetic energy driving term introduced in Sect. \ref{rans:analysis_ek}:

\begin{equation}
\label{eq:WbFm}
W_b = \av{\rho' \vec u' \cdot \vec g} = - g f_m,
\end{equation}

\noindent where we took $g$ outside of the averaging operator (the Cowling approximation). Note that energetically, Eq. (\ref{eq:fm_decomposition}) implies that when $ \fav{u_r} \approx 0$, the gravitational work done by the turbulence is canceled by the gravitational work done by the mean flow, so that the total work done by gravity is zero. This is a direct consequence of mass conservation.

\begin{figure*}[t] 
   \centering
   \parbox{0.49\linewidth}{\center \includegraphics[width=0.8\linewidth]{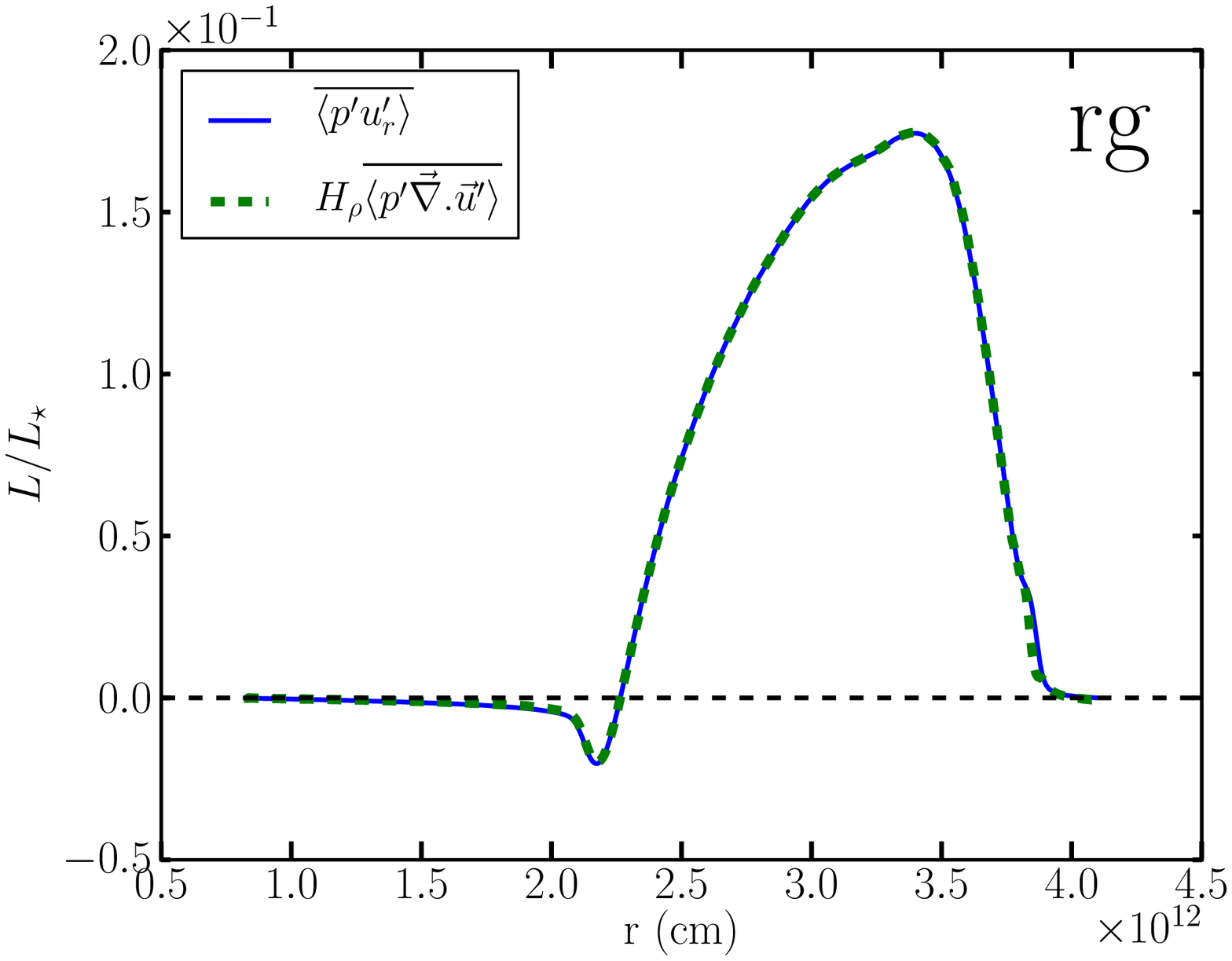}}
    \parbox{0.49\linewidth}{\center \includegraphics[width=0.8\linewidth]{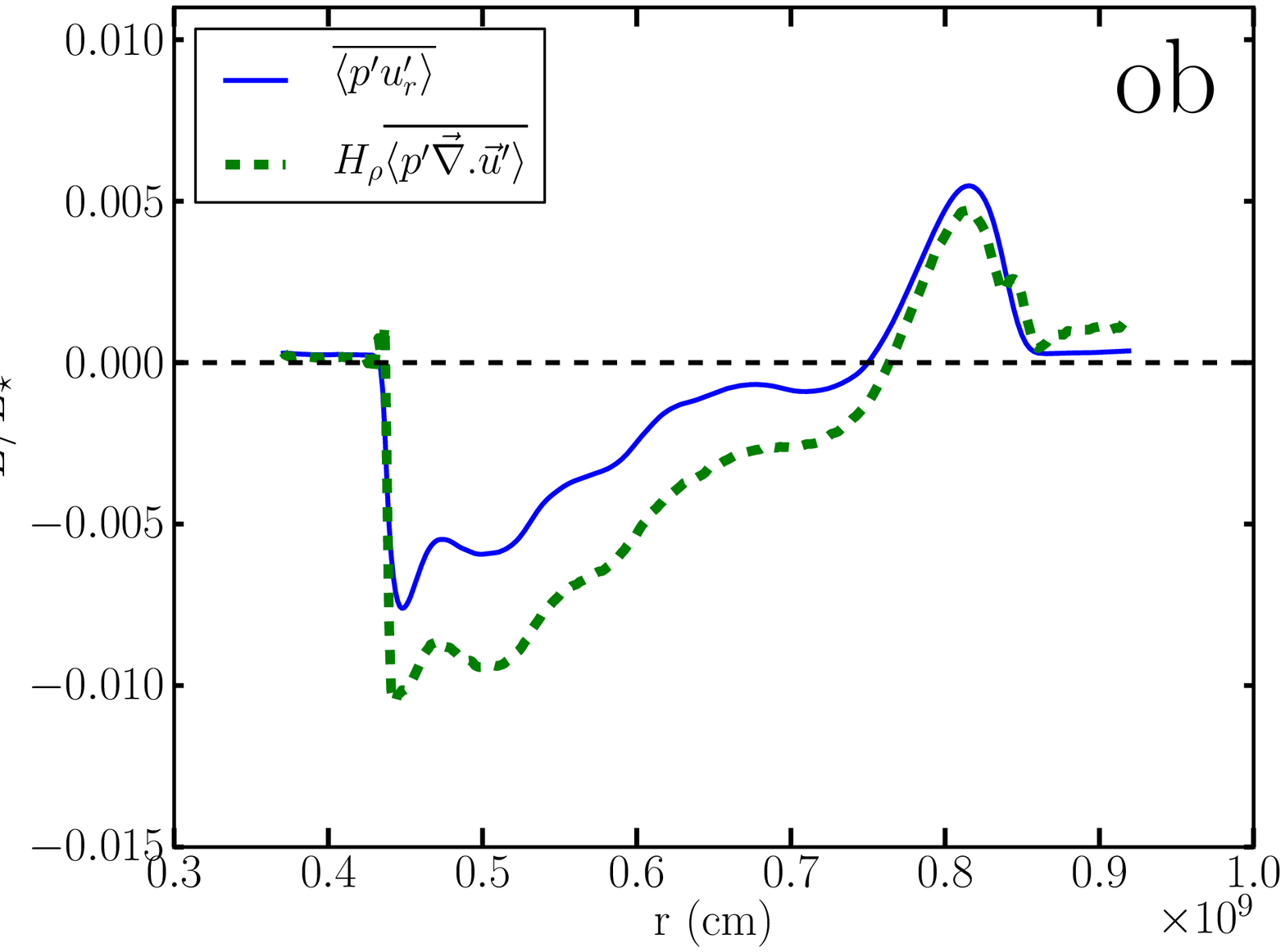}}
   \caption{Comparison between the acoustic flux $f_p = \av{p' u_r'}$ and pressure-dilatation $W_p = \av{p' \vec \nabla \cdot \vec u'}$ multiplied by the density scale-height $H_\rho$. Both terms were multiplied by $4\pi r^2$ and normalized by the stellar luminosity. Left panel: model {\sf rg.3D.mr}, right panel: model {\sf ob.3D.mr}.}
\label{fig:fp_decomposition}   
\end{figure*}

\subsubsection{Acoustic and kinetic energy fluxes}

Assuming that $\av{\rho} \fav{D_t} \fav{\epsilon_k}=0$, which is justified by the analysis in Sect. \ref{rans:analysis_ek}, the integral version of the kinetic energy balance, Eq. (\ref{eq:rans_ekin}), reads

\begin{equation}
\label{eq:ke_transport}
4 \pi r^2 (f_k + f_p) = \int_0^r \big ( W_b + W_p - \av{\rho \epsilon_d} \big) \dV,
\end{equation}

\noindent which simply says that transport by combined kinetic energy and acoustic fluxes is the residual between driving and dissipation. It is not an explicit formula for the fluxes since the velocity field also enters in the r.h.s, both in the driving and in the dissipation. When $f_p$ and $W_p$ are negligible, the equation simplifies to Eq. (4) of \cite{meakin_properties_2010}. 

The bottom-left panel in Fig. \ref{fig:rg_fkp_updown} shows the radial profiles of $f_k$ and $f_p$ in model {\sf rg.3D.mr}. As discussed in Sect. \ref{rans:analysis_et}, the kinetic energy flux $f_k$ is large and downward directed. The figure shows that $f_p$ is smaller, but not negligible, and upward directed. Furthermore, we show in the figure the splitting of $f_k$ and $f_p$ into the downflow ($u_r' <0$) and upflow components ($u_r' >0$). In both cases, the contribution from the downdrafts dominates significantly. This emphasizes the strong asymmetry of the flow which results from the large stratification. The figure shows that both components of the acoustic flux are positive, i.e. down-flows have $p' <0$ and up-flows have $p'>0$. Since $f_p$ is not negligible and is upward directed (opposite to the kinetic energy flux), a better understanding of $f_p$ is necessary to understand what is setting the amplitude of the kinetic energy flux. The right panel in Fig. \ref{fig:rg_fkp_updown} shows the kinetic energy and acoustic flux in model {\sf ob.3D.mr}, which are both small compared to the enthalpy flux. The acoustic flux shows a more complex behavior than in the red giant models. The kinetic energy flux is upward directed, and shows a large cancelation due to the approximate symmetry between upflows and downflows. \cite{meakin_properties_2010} experimented with the oxygen burning shell models by changing from a heating from below to a cooling from above, and found that the kinetic energy flux reversed direction (see their {\sf c1} model). Qualitatively, this change of behavior is explained by the different direction of propagation of plumes, which propagate downward when triggered by cooling at the top, as in the red giant model.

Finally, multiplying Eq. (\ref{eq:anelastic_dilatation}) by $p'$ and taking the average one obtains:

\begin{equation}
\label{eq:fp_decomposition}
f_p = \av{p' u_r'} = H_\rho \av{p' \vec \nabla \cdot \vec u'} = H_\rho W_p.
\end{equation}

The left panel of Fig. \ref{fig:fp_decomposition} shows that this relation holds to a very good degree in the red giant model, validating a posteriori the use of the anelastic approximation (Sect. \ref{boussinesq_anelastic}). Equation (\ref{eq:fp_decomposition}) is very similar to Eq. (\ref{eq:WbFm}): they both connect a flux, $f_m$ and $f_p$, to a kinetic energy source term, $W_b$ and $W_p$. For $f_p$ and $W_p$ this is true only within the anelastic approximation. We will come back to this in Sect. \ref{KEdriving}. For the oxygen-burning shell model, consistently with the approximation $\vec \nabla \cdot \vec u'=0$ introduced in Sect. \ref{boussinesq_anelastic}, we consider that $W_p \approx 0$ which is a good approximation. $f_p$ is more complex than in the red giant model, where it is dominated by stratification effects. The right panel of Fig. \ref{fig:fp_decomposition} shows that the effects of stratification, as described by the anelastic approximation, reproduce the gross features of $f_p$, but that other effects contribute. This could be due to pressure perturbations related to boundary effects, or to the background expansion, and requires further investigation.

Note that we have the following relation:

\begin{equation}
\label{eq:anelastic_pressure_dilatation}
\av{\frac{p'}{\rho_0} \vec \nabla \cdot (\rho_0 \vec u')} = W_p - \frac{f_p}{H_\rho},
\end{equation}

\noindent which characterizes the deviation from the anelastic approximation. 

\begin{figure*}[t]
\centering
\parbox{0.49\linewidth}{\center \includegraphics[width=0.8\linewidth]{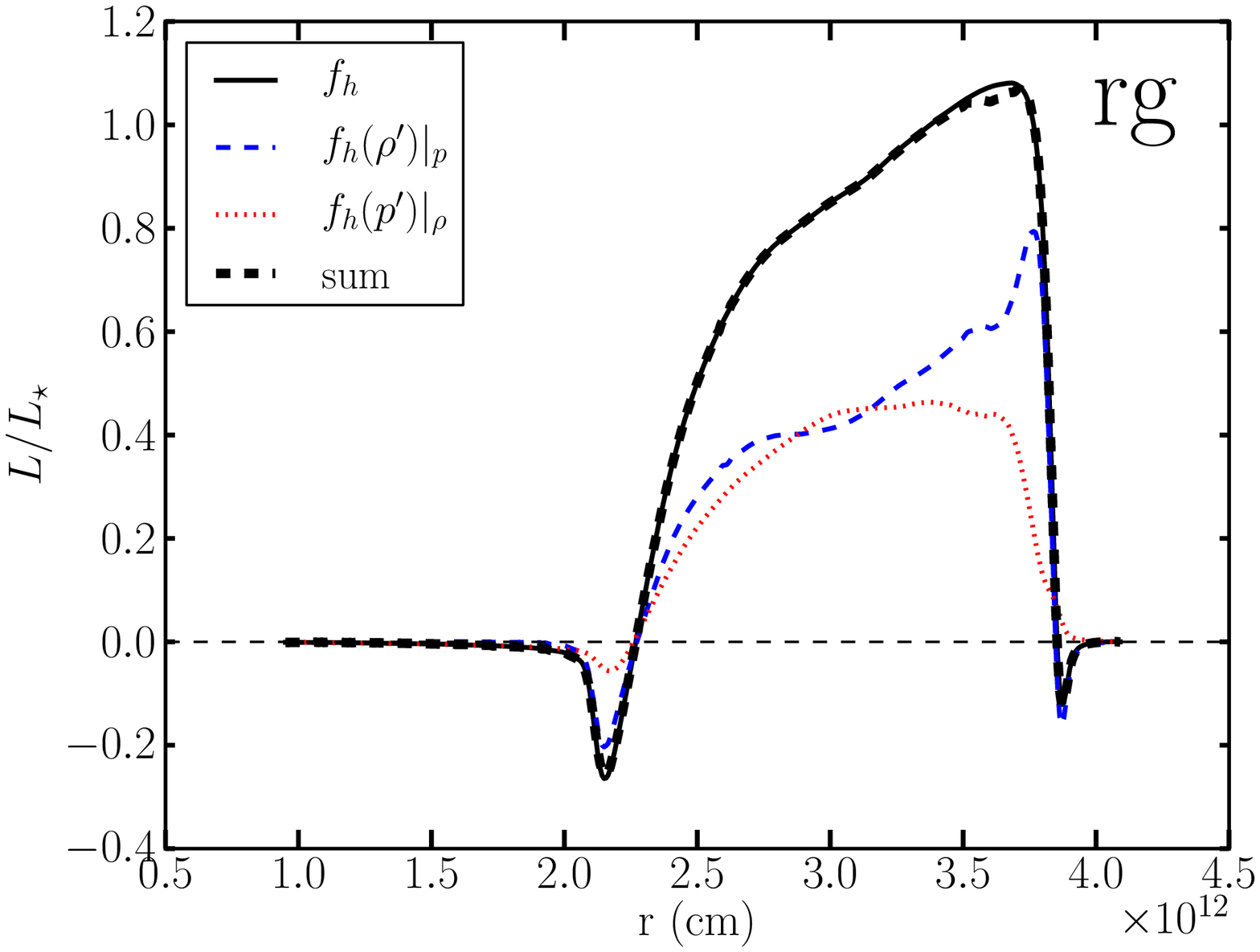}}
\parbox{0.49\linewidth}{\center \includegraphics[width=0.8\linewidth]{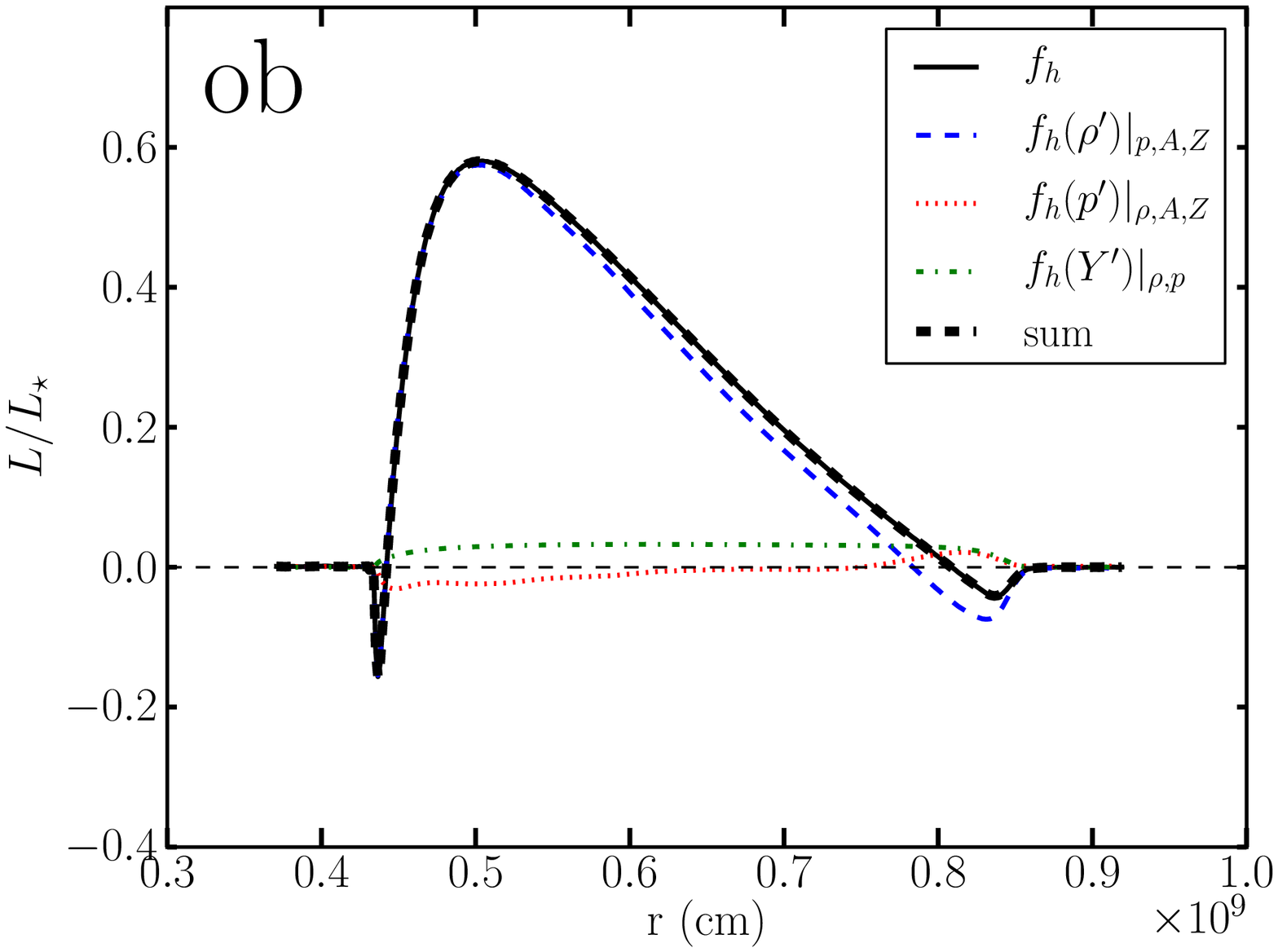}}
\parbox{0.49\linewidth}{\center \includegraphics[width=0.8\linewidth]{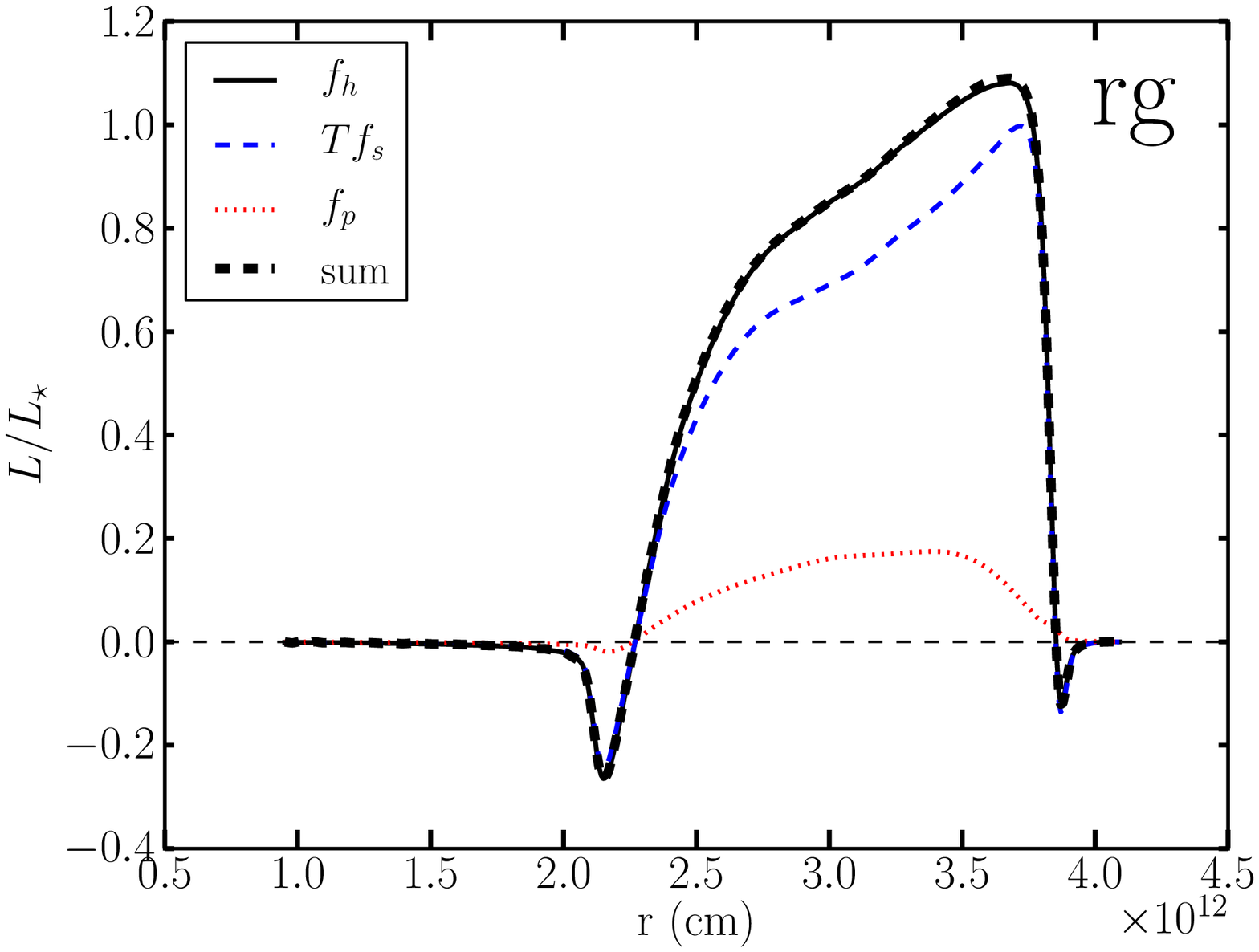}}
\parbox{0.49\linewidth}{\center \includegraphics[width=0.8\linewidth]{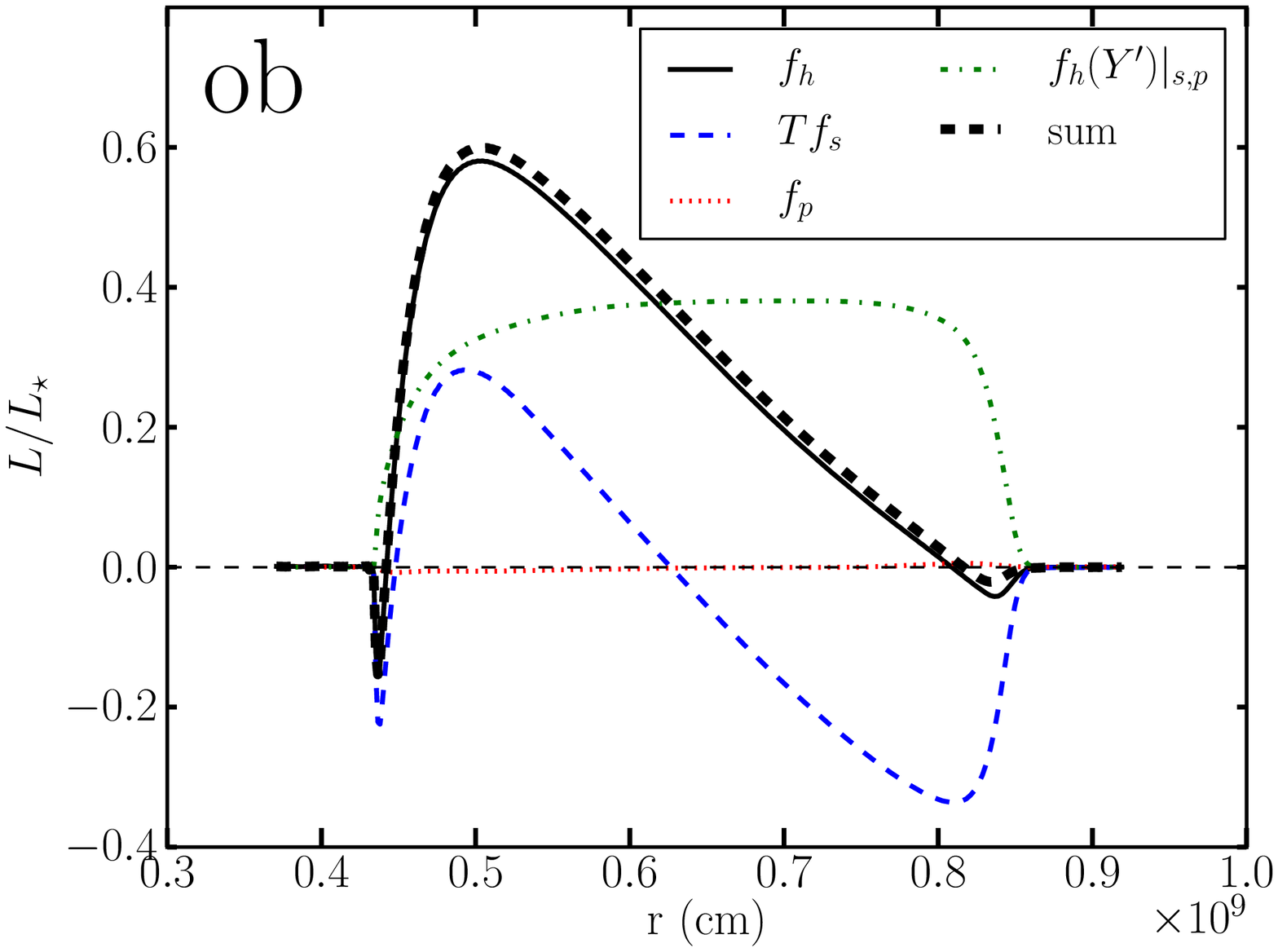}}
\parbox{0.49\linewidth}{\center \includegraphics[width=0.8\linewidth]{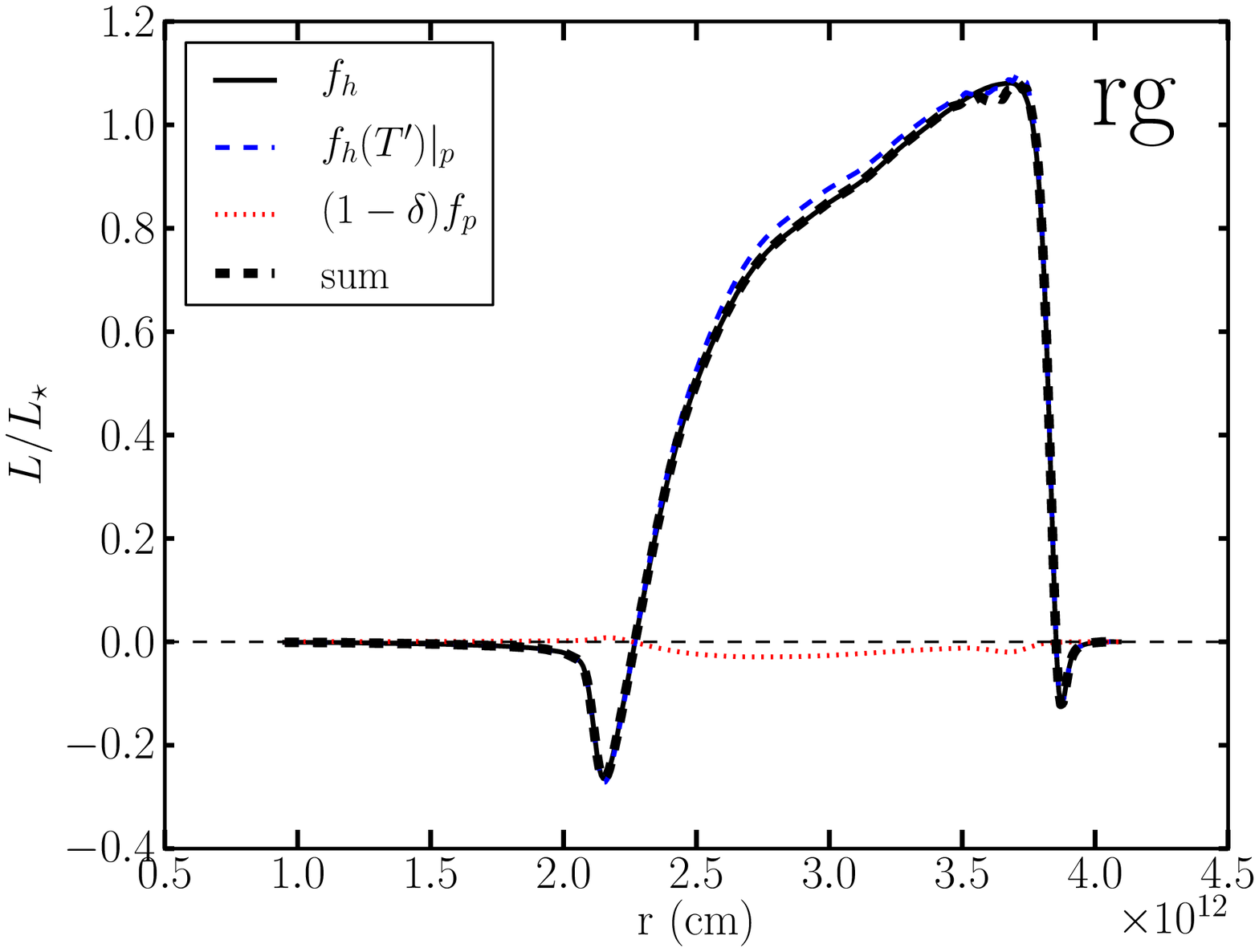}}
\parbox{0.49\linewidth}{\center \includegraphics[width=0.8\linewidth]{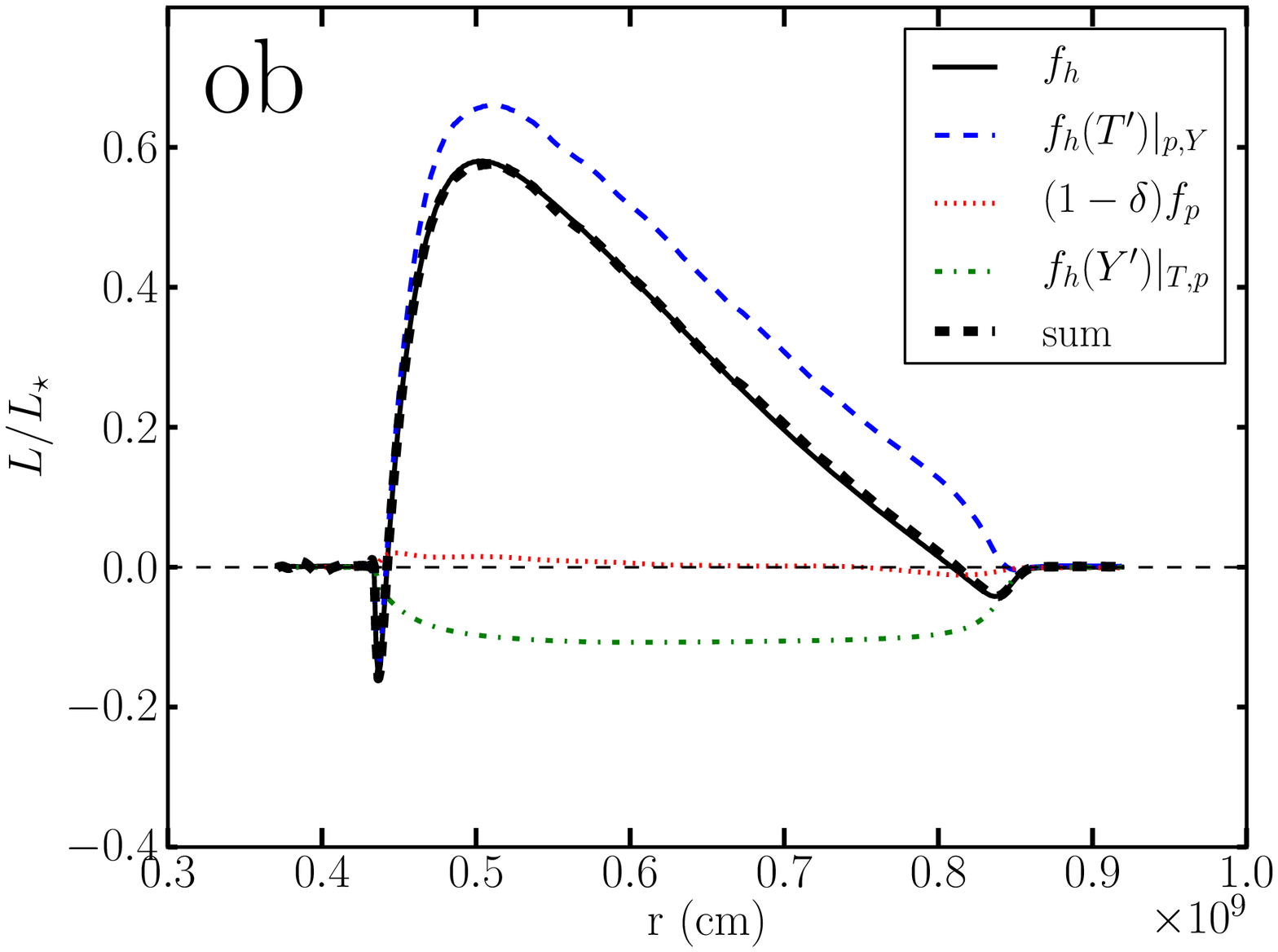}}
\caption{Splitting of the enthalpy flux in models {\sf rg.3D.mr} (left) and {\sf ob.3D.mr} (right). Top panels: decomposition with Eq. (\ref{eq:fe_decomposition}). Middle panels: decomposition with Eq. (\ref{eq:fe_decomposition2}). Bottom panels: decomposition with Eq. (\ref{eq:fe_decomposition3}). $f_h(Y')$ represents the sum of the composition terms due to $A'$ and $Z'$. The different terms are multiplied by $4\pi r^2$, and normalized to the model luminosity.}
\label{fig:fh_decomposition}
\end{figure*}

\subsubsection{Splitting of the enthalpy flux}
\label{enthalpy_flux_splitting}

The enthalpy flux describes the transport of heat by convection and is therefore important for stellar interior modeling. In this section, we use thermodynamical relationships to study its relation to other turbulent fluxes. As two state variables are sufficient to determine the thermodynamic state, different expressions for the enthalpy flux can be derived. We choose the pressure as one of the variables, as pressure fluctuations play a different role in the two stellar models presented here. For the other variable, we will consider density, entropy, and temperature, respectively. For the oxygen-burning shell model, we describe composition effects in terms of the average number of nucleons $A$ and free electrons $Z$ per nucleus\footnote{If the composition variables $Y=1/A$ and $Y_e=Z/A$ are used, most of the effect is concentrated in the single variable $Y$. $Y_e$ is almost constant in the oxygen-burning shell.}. For the red giant model, these terms are zero  because the composition was uniform.

Formally, the turbulent enthalpy flux is defined as $f_h = \av{\rho} \fav{h'' u_r''}$, but we have checked that it is identical to $f_h = \av{\rho}\ \av{h' u_r'}$ (the same holds for the other turbulent fluxes). Based on this second form, we first split the enthalpy flux in terms of density, pressure, and composition fluctuations. Assuming that fluctuations of these variables are small, 

\begin{align}
h' = & \frac{\partial h}{\partial \rho}\Big|_{p, A, Z} \rho' +  \frac{\partial h}{\partial p}\Big|_{\rho, A, Z} p' \notag \\
+ &\frac{\partial h}{\partial A}\Big |_{\rho, p, Z} A' + \frac{\partial h}{\partial Z}\Big |_{\rho, p, A} Z',
\end{align}

\noindent which leads to the following expression: 

\begin{align}
\label{eq:fe_decomposition}
f_h = & - \frac{P}{\rho} \frac{\Gamma_1}{\Gamma_3 -1} \av{\rho' u_r'} +  \frac{\Gamma_3}{\Gamma_3 -1} \av{p' u_r'} \notag \\
         & + \rho \frac{\partial e}{\partial A}\Big |_{\rho, p, Z} \av{ A' u_r'} + \rho \frac{\partial e}{\partial Z}\Big |_{\rho, p, A} \av{ Z' u_r'},
\end{align}

\noindent where the coefficients are evaluated using the background state. This relation illustrates how the enthalpy flux can be decomposed into separate contributions from the turbulent mass flux, the acoustic flux, and composition fluxes. The top panels in Figure \ref{fig:fh_decomposition} show how this decomposition compares with the numerical data. We find that decomposition (\ref{eq:fe_decomposition}) holds to an excellent degree of approximation. In the red giant model, both the terms related to density and pressure fluctuations contributes, whereas in the oxygen-burning shell model, the term related to density fluctuations provides the main contribution. 

A second way to split the enthalpy flux is by introducing the entropy flux in place of the turbulent mass flux. In this approach, we start from

\begin{align}
h' = & T s' +  \frac{1}{\rho} p' + \frac{\partial h}{\partial A}\Big |_{s, p, Z} A' + \frac{\partial h}{\partial Z}\Big |_{s, p, A} Z',
\end{align}

\noindent which, by the same arguments as above, leads to 

\begin{align}
\label{eq:fe_decomposition2}
f_h =& T f_s + f_p \notag \\
 & + \rho \frac{\partial h}{\partial A}\Big |_{s, p, Z} \av{A' u_r'} + \rho \frac{\partial h}{\partial Z}\Big |_{s, p, A} \av{Z' u_r'}.
\end{align}

The middle panels in Fig. \ref{fig:fh_decomposition} shows how this relation compares with the numerical data. In both models, the agreement is very good. In the red giant model, the enthalpy flux mainly results from the entropy flux, with a non-negligible contribution from the acoustic flux. In the oxygen-burning shell models, the contributions from both the entropy and composition fluxes are important. Comparing this with the first splitting, it illustrates that density fluctuations are here both due to thermal effects (i.e. entropy fluctuations) and composition effects.

Finally, it is useful to write the decomposition of the enthalpy flux by introducing temperature fluctuations, one obtains:

\begin{align}
\label{eq:fe_decomposition3}
f_h &= \rho c_p \av{T' u_r'}  + (1-\delta) f_p \notag \\
& + \rho \frac{\partial h}{\partial A}\Big |_{T, p, Z} \av{A' u_r'} + \rho \frac{\partial h}{\partial Z}\Big |_{T, p, A} \av{Z' u_r'}.
\end{align}

\noindent where $\delta = \alpha T$, with $\alpha$ the coefficient of thermal expansion at constant pressure. In the literature, $\rho c_p \av{T' u_r'}$ is often considered as the enthalpy flux. Formally this is correct only when $\delta=1$ (e.g. polytropic gas) or $f_p=0$, and when composition effects are neglected. The left-bottom panel in Fig. \ref{fig:fh_decomposition} shows that $\rho c_p \av{T' u_r'}$ yields a very good approximation of $f_h$ in the red giant model. The right-bottom panel illustrates that in the oxygen-burning shell model, composition effects are significant in this decomposition.

\subsection{Kinetic energy driving in compressible fluids and turbulent dissipation in the convective zone}
\label{KEdriving}

In Section \ref{rans:analysis_ek}, we expressed the driving of kinetic energy in terms of $W_b$, related to density fluctuations, and $W_p$, related to pressure fluctuations. However, both density and pressure fluctuations arise from different physical processes: thermal effects (i.e. entropy fluctuations), dynamical effects (e.g. compressibility), and composition effects. This leads to a different expression of kinetic energy driving, as shown below.

In the linearized velocity equation, the acceleration on the r.h.s. is

\begin{equation}
\vec a = - \frac{1}{\rho_0} \vec \nabla p' + \frac{\rho'}{\rho_0} \vec g,
\end{equation}

\noindent see Appendix \ref{elliptic_pressure_equation}. It can be also written as 

\begin{equation}
\label{eq:linear_force}
\vec a =  - \vec \nabla \frac{p'}{\rho_0} + \Big ( \frac{\rho'}{\rho_0} - \frac{H_p}{H_\rho} \frac{p'}{P_0} \Big ) \vec g,
\end{equation}

\noindent \citep{braginsky_equations_1995}. The physical meaning of the first term can be elucidated by computing its work:

\begin{align}
- \rho_0 \vec u' \cdot \vec \nabla \frac{p'}{\rho_0} &=- \vec \nabla \cdot ( p'\vec u') +\frac{p'}{\rho_0} \vec \nabla \cdot (\rho_0 \vec u')
\end{align}

This term gives rise to the transport by the acoustic flux and to a source term which is related to the deviation from the anelastic approximation (see Eq. \ref{eq:anelastic_pressure_dilatation}). This last term will contribute only when compressible effects become important, i.e. for $M_s \gtrsim 1$.

We identify the second term in Eq. (\ref{eq:linear_force}) with the buoyancy force, written in terms of density and pressure fluctuations. 
%
%
%
\begin{figure*}[t]
\centering
\parbox{0.49\linewidth}{\center \includegraphics[width=0.8\linewidth]{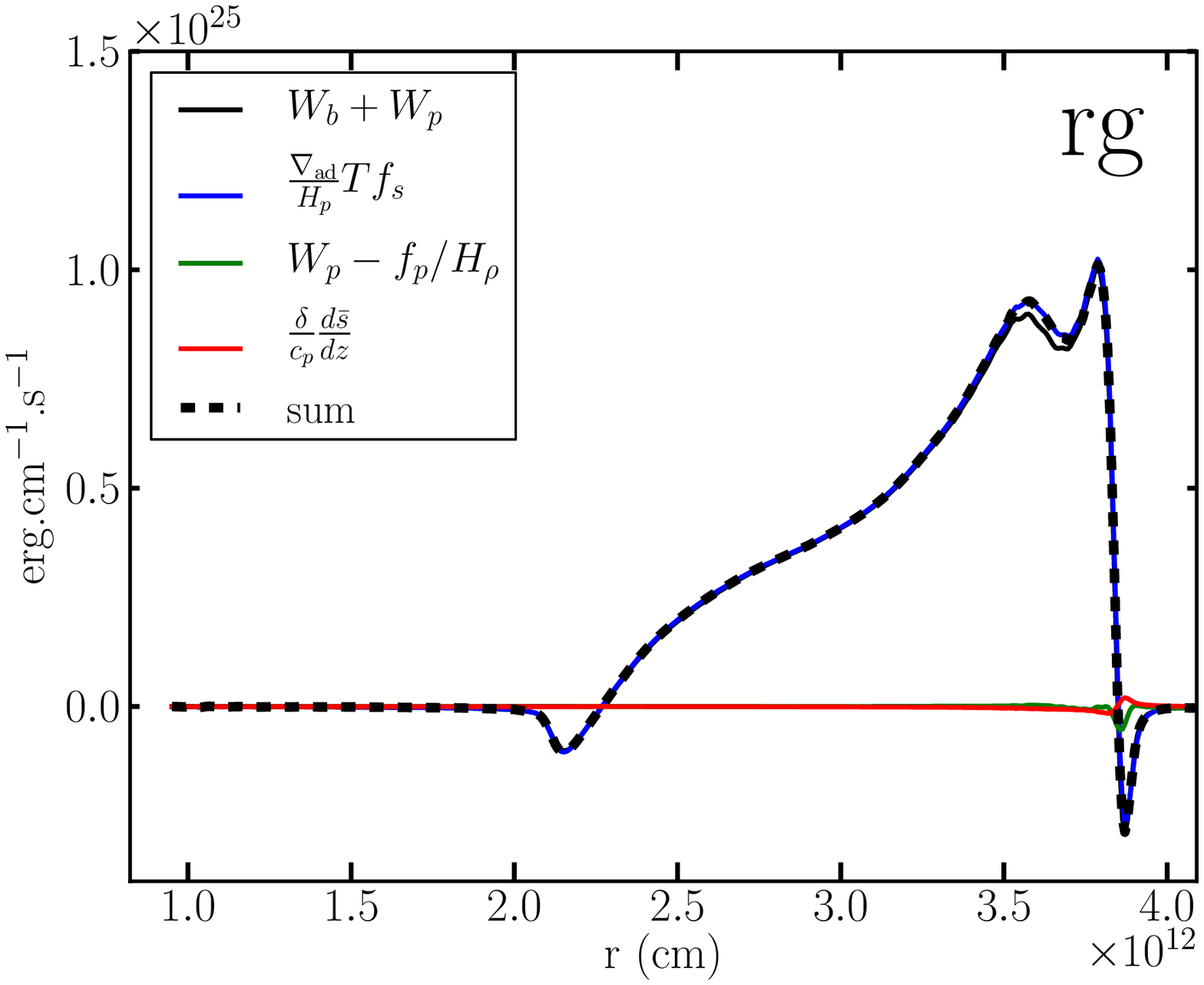}}
\parbox{0.49\linewidth}{\center \includegraphics[width=0.8\linewidth]{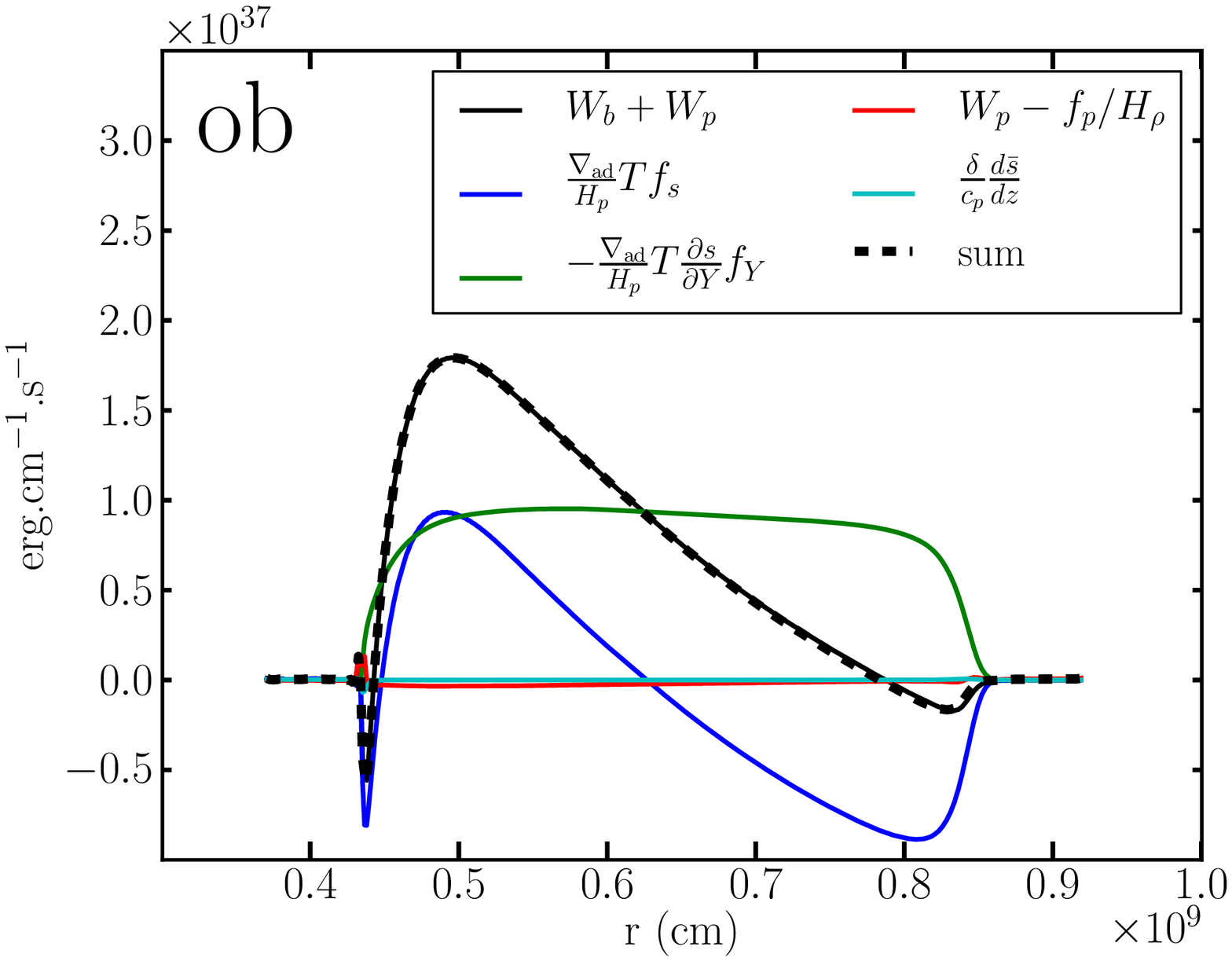}}
\caption{Splitting of the kinetic energy driving for model {\sf rg.3D.mr} and {\sf ob.3d.mr}.}
\label{fig:kedriving_splitting}
\end{figure*}
It can be written as:

\begin{align}
\label{eq:buoyancy_force}
\frac{\rho'}{\rho_0} - \frac{H_p}{H_\rho}\frac{p'}{P_0} & = \Big ( \frac{\rho'}{\rho_0} - \frac{1}{\Gamma_1}\frac{p'}{P_0} \Big ) - \frac{\delta}{\rho_0 c_p g} \frac{d s_0}{dz} p',
\end{align}

\noindent where we used 

\begin{equation}
\frac{1}{H_\rho} - \frac{1}{\Gamma_1 H_p} = \frac{\delta}{c_p} \frac{d s_0}{dz}.
\end{equation}

This term characterizes the deviation from adiabaticity of the background. We can make the connection with entropy and composition fluctuations, since we have the thermodynamical relationship

\begin{align}
\label{eq:buoyancy_rhoT_sY}
\frac{\rho'}{\rho_0} - \frac{1}{\Gamma_1} \frac{p'}{p_0} = - \frac{\delta}{c_p} s' + \frac{\delta}{c_p} \frac{\partial s}{\partial y_i}\Big |_{\rho, p} y_i',
\end{align}

\noindent with $y_i' \equiv A, Z$ (with an implicit summation, $y_1=A$, $y_2=Z$).

Using these relations, the total kinetic energy driving $W_b+W_p$ can be written as

\begin{align}
\label{eq:KEdriving_splitting}
W_p + W_b=  & \frac{\nabla_\mathrm{ad}}{H_p} T f_s & \mathrm{Thermal\ effects} \notag \\
 & - \frac{\nabla_\mathrm{ad}}{H_p} T \frac{\partial s}{\partial y_i} \Big |_{P,\rho} f_{y_i} &\mathrm{Composition\ effects} \notag \\
 &+  \frac{\delta}{c_p} \frac{d \av{s}}{dz} f_p &\mathrm{Background\ effects} \notag \\
 & + \Big( W_p - \frac{f_p}{H_\rho} \Big ), &\mathrm{Compressibility\ effects}
\end{align}

\noindent where we used $\frac{\delta g}{c_p T} = \frac{\nabla_\mathrm{ad}}{H_p}$. The first two contributions in Eq. (\ref{eq:KEdriving_splitting}) can be also written in terms of the turbulent mass flux and acoustic flux thanks to Eq. (\ref{eq:buoyancy_rhoT_sY}).  The first three terms are related to buoyancy, which contrasts which the usual designation of $W_b$ as the ``buoyancy" driving. Figure \ref{fig:kedriving_splitting} illustrates this splitting of the driving for the numerical models. In our cases, both the deviation from adiabaticity and the compressibility effects are negligible. In the red giant model, there is no compositional effects and the above formulation is the most convenient as it expresses the driving only in terms of the entropy flux. In the oxygen-burning shell model, composition effects are important. However, since $p'/P_0 < \rho' / \rho_0 $, the driving can be expressed in terms of density fluctuations mainly. This is similar to the comment made with the right panels in Fig. \ref{fig:fh_decomposition}.

Finally, we can relate the enthalpy flux to the kinetic energy driving. In the oxygen-burning shell case, we have from Eq. (\ref{eq:fe_decomposition}):

\begin{equation}
\label{eq:ob_fh_driving}
f_h \approx - \frac{P}{\rho} \frac{\Gamma_1}{\Gamma_3 -1} \av{\rho' u_r'} = \frac{H_p}{\nabla_\mathrm{ad}} W_b,
\end{equation}

\noindent which connects kinetic energy driving (since $W_p$ is negligible) with the enthalpy flux. For the red giant case, we have from Eq. (\ref{eq:KEdriving_splitting}):

\begin{equation}
W_b + W_p = \frac{\nabla_\mathrm{ad}}{H_p} T f_s,
\end{equation}

\noindent which we use Eq. (\ref{eq:fe_decomposition2}) to obtain:

\begin{align}
\label{eq:rg_fh_driving}
f_h = \frac{H_p}{\nabla_\mathrm{ad}} \Big (  W_b + \Gamma_3 W_p \Big ).
\end{align}

\noindent When $W_p$ is negligible, this gives Eq. (\ref{eq:ob_fh_driving}). 

In a quasi-steady state, we have $L_d = \int (W_b+W_p) \dV$, see Eq. (\ref{eq:ek_balance}). Therefore, we can estimate $L_d$ from

\begin{align}
L_d &\approx \int \frac{\nabla_\mathrm{ad}}{H_p} f_h dV \notag \\
        &\approx \bar{\nabla}_\mathrm{ad} \bar{L}_c \int_\mathrm{CZ} \frac{dr}{H_p} \notag \\
        &\approx \bar{\nabla}_\mathrm{ad} \bar{L}_c n_{H_p},
\end{align}

\noindent where $\bar{\nabla}_\mathrm{ad}$ is the average value of the adiabatic gradient, $\bar{L}_c$ is the average value of the enthalpy luminosity ($4 \pi r^2 f_h$) over the convective zone, and $n_{H_p}$ is the number of pressure scale-heights in the convective zone. This overestimates $L_d$ when $W_p$ is not negligible, because of the factor $\Gamma_3>1$ in Eq. (\ref{eq:rg_fh_driving}). Nevertheless, it shows that the turbulent dissipation is of the same order as the convective luminosity. This result was suggested by \cite{hewitt_dissipative_1975}. For instance, in the red giant models, we have: $\bar{L}_c \sim 4\times10^{36}$ erg/s, $\nabla_\mathrm{ad} \sim 0.35$, $n_{H_p} \sim 7.8$, which gives $L_d \sim 11\times10^{36}$ erg/s. In the oxygen-burning shell model, we have: $\bar{L}_c \sim 5\times10^{45}$ erg/s, $\nabla_\mathrm{ad} \sim 0.245$, $n_{H_p} \sim 2$, which gives $L_d \sim 2.45\times10^{45}$ erg/s. Both values agree well with the inferred values (see Tables \ref{table:rg_runs} and \ref{table:ob_runs}). As expected, the value obtained for the red giant is overestimated. 

\begin{figure*}[t]
\centering
\parbox{0.49\linewidth}{\center \includegraphics[width=0.8\linewidth]{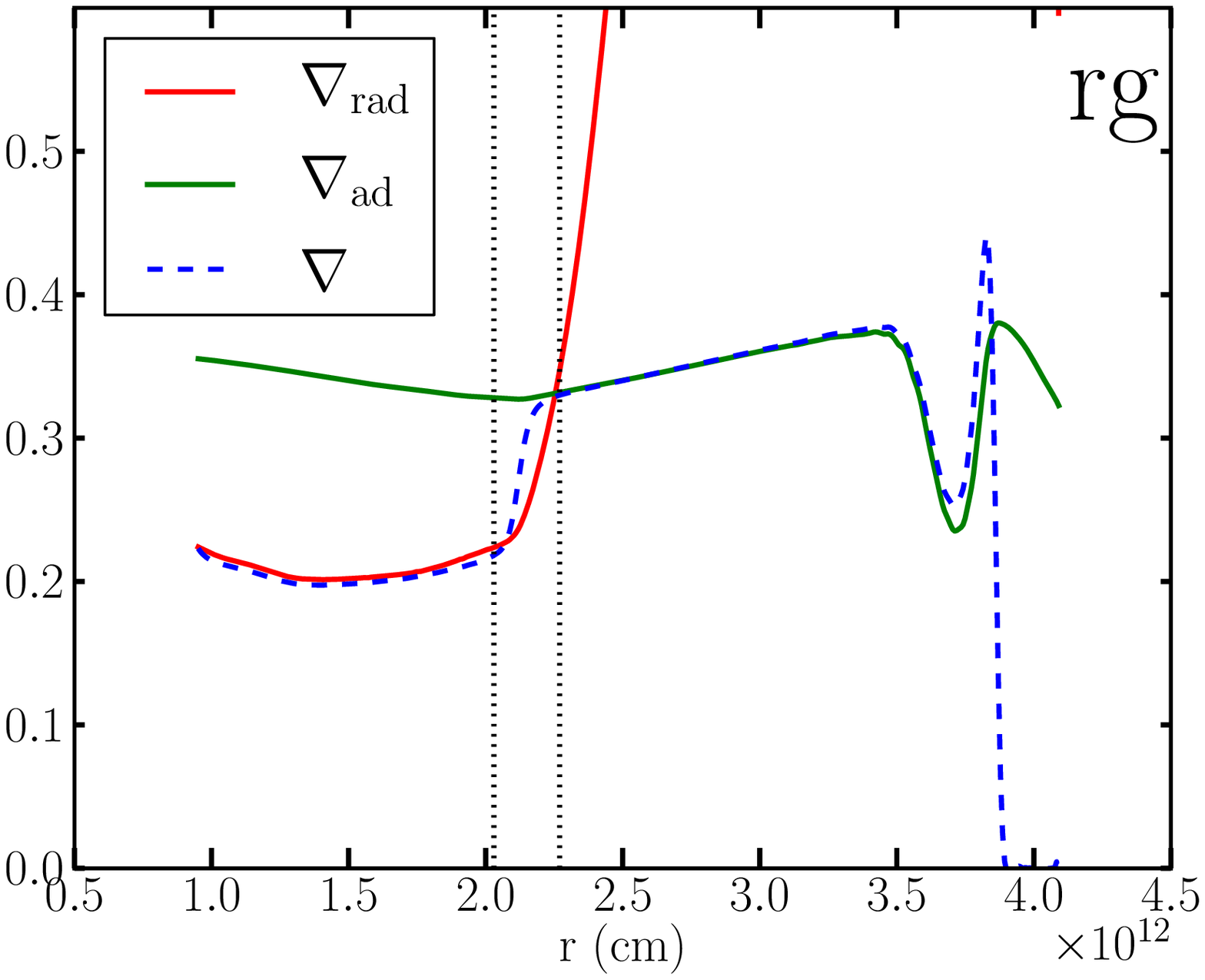}}
\parbox{0.49\linewidth}{\center \includegraphics[width=0.8\linewidth]{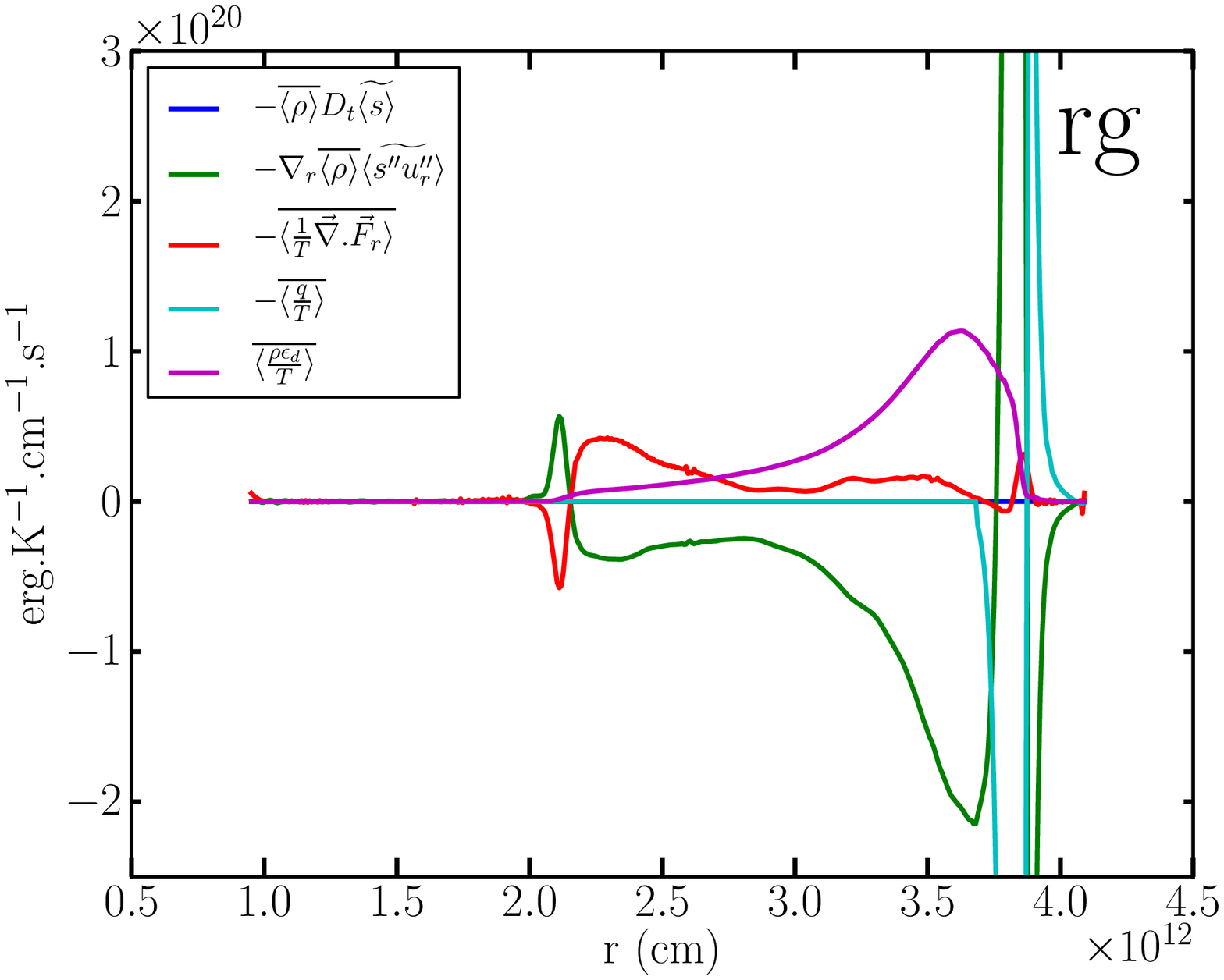}}
\caption{Left panel: Dimensionless temperature gradients in model {\sf rg.3D.mr}, the vertical dotted lines show the bottom convective boundary layer (see text). Right panel: entropy balance in model {\sf rg.3D.mr}.}
\label{fig:rg_penetration}
\end{figure*}

\subsection{Thermal effects and overshooting in the red giant model}
\label{penetration}

The Kelvin-Helmholtz timescale is defined as the ratio between the thermal energy and the luminosity of the star:

\begin{equation}
\tau_\mathrm{KH} = \frac{E_\mathrm{int}}{L_\star},
\end{equation}

\noindent where $E_\mathrm{int}= \int \rho \epsilon_i dV$ is the total thermal energy. For the red giant the Kelvin-Helmholtz timescale is $2.1\times 10^3$~yr. This is much longer than our simulations, which span roughly 8 years of model time. In fact we are not able to simulate over several thermal timescales, as it would be necessary to ensure that the models have reached thermal equilibrium. This is a common limitation to all numerical simulations of deep convective envelopes which include radiative cooling. One possible way to overcome this problem is to boost the luminosity, thereby bringing the dynamical and thermal timescales closer to each other; e.g., \cite{dobler_magnetic_2006}. A consequence is that the characteristic Mach number of the flow increases. The physical character of turbulent convection becomes very different, with properties closer to photospheric convection, as compressibility and superadiabatic effects become important. Furthermore, increasing the luminosity implies an increase in the thermal diffusivity\footnote{This assumes a given temperature stratification characterizing a stellar structure.}, so that for a given Reynolds number, the P\'eclet number decreases. As discussed below, this will change the behavior at the convective boundaries. For these reasons, we prefer to use a realistic value of the luminosity. As shown in Sect. \ref{rans:analysis}, out of thermal equilibrium behavior can be taken into account in the framework of the mean-field equations.

We characterize our red giant models with a global P\'eclet number, defined as

\begin{equation}
\label{eq:global_peclet}
\mathrm{Pe} = \frac{v_\mathrm{rms} l_\mathrm{CZ}}{\chi},
\end{equation}

\noindent where $\chi$ is the average value of the radiative diffusivity over the convective zone, typically $\chi = 9\times 10^{13}$ cm$^2$/s. We find Pe $\sim 5200$ in model {\sf rg.3D.mr}. This large value characterizes a very efficient transport of heat, and thus a very efficient convection.

\begin{figure*}[t]
\centering
\parbox{0.49\linewidth}{\center \includegraphics[width=0.8\linewidth]{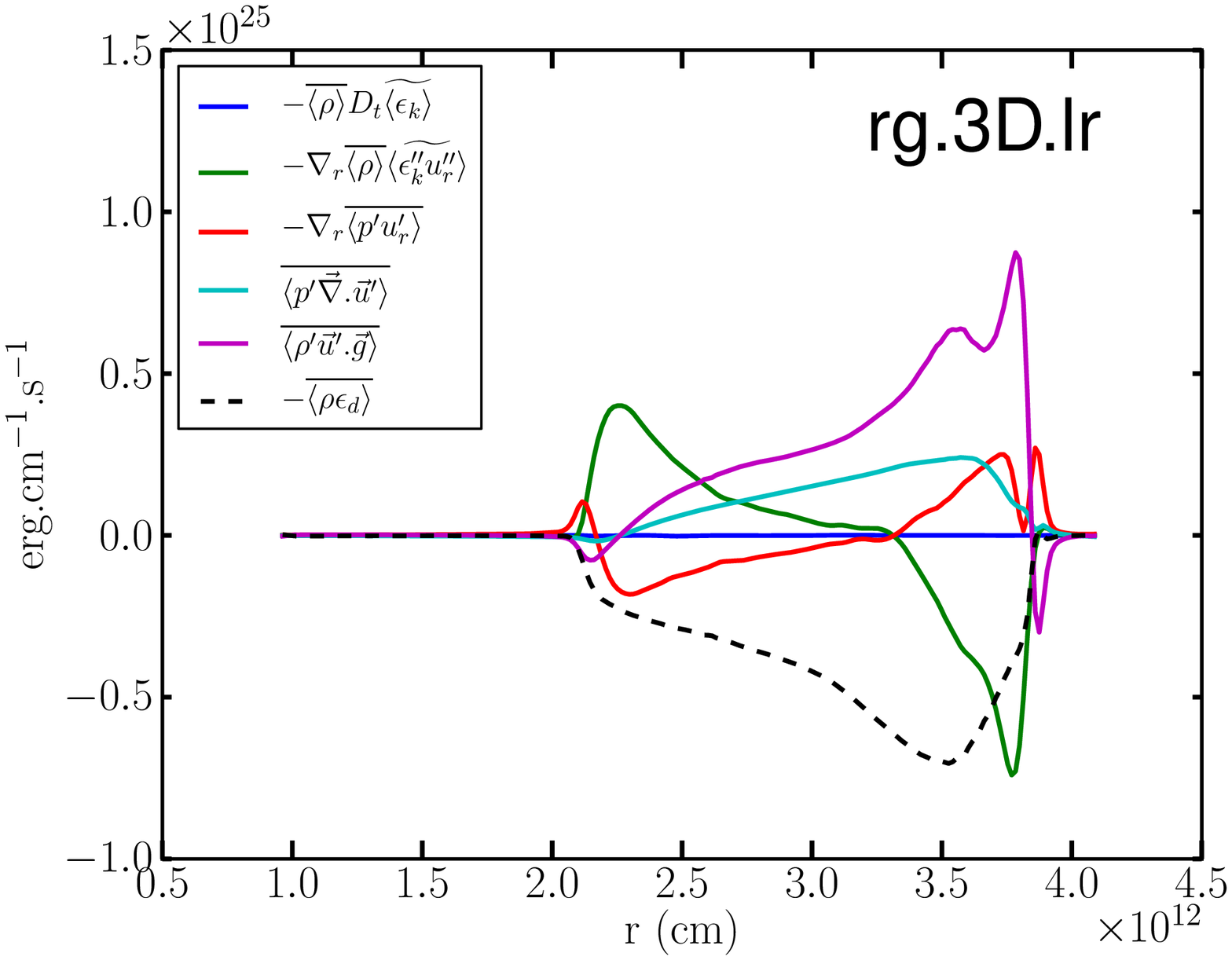}}
\parbox{0.49\linewidth}{\center \includegraphics[width=0.8\linewidth]{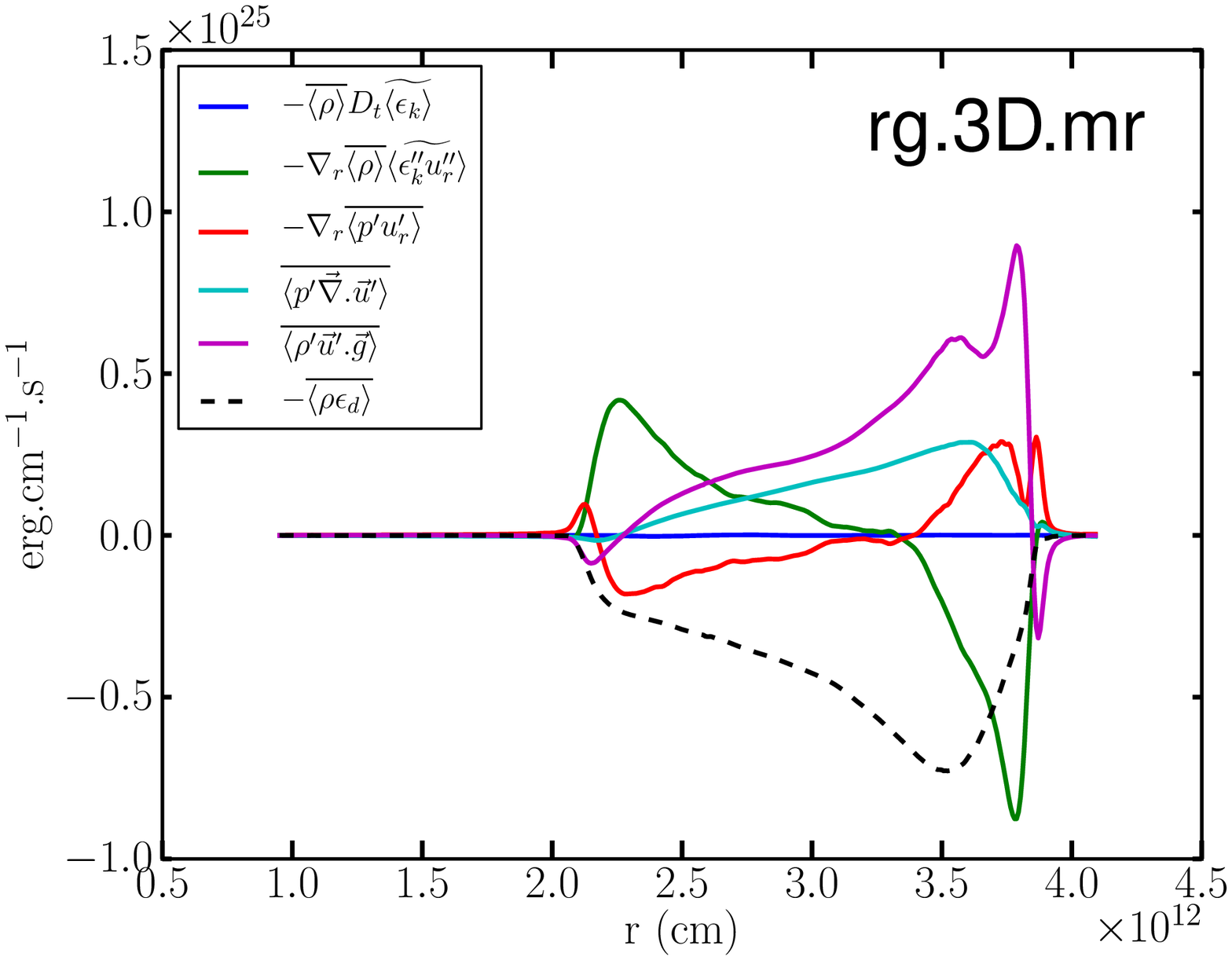}}
\parbox{0.49\linewidth}{\center \includegraphics[width=0.8\linewidth]{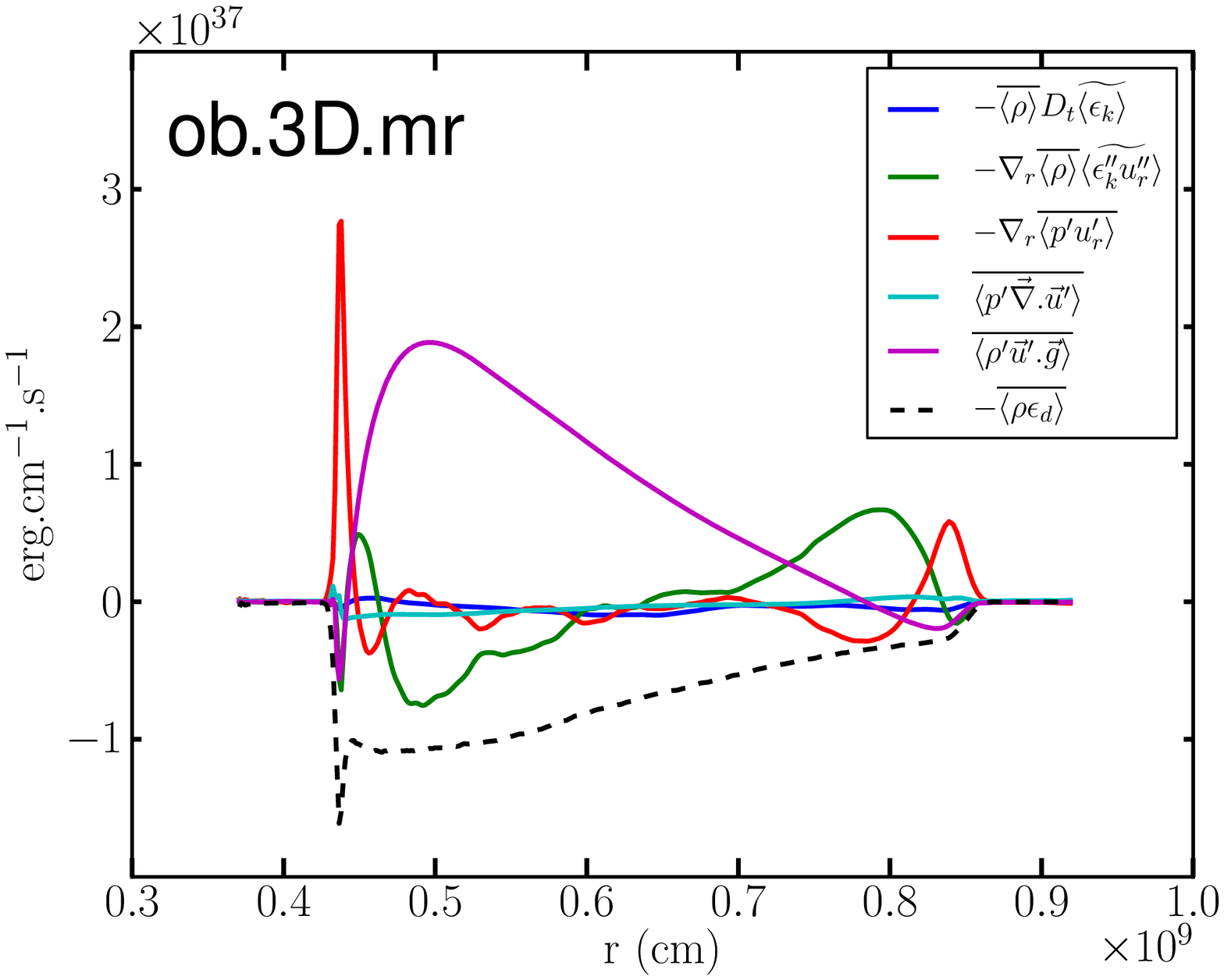}}
\parbox{0.49\linewidth}{\center \includegraphics[width=0.8\linewidth]{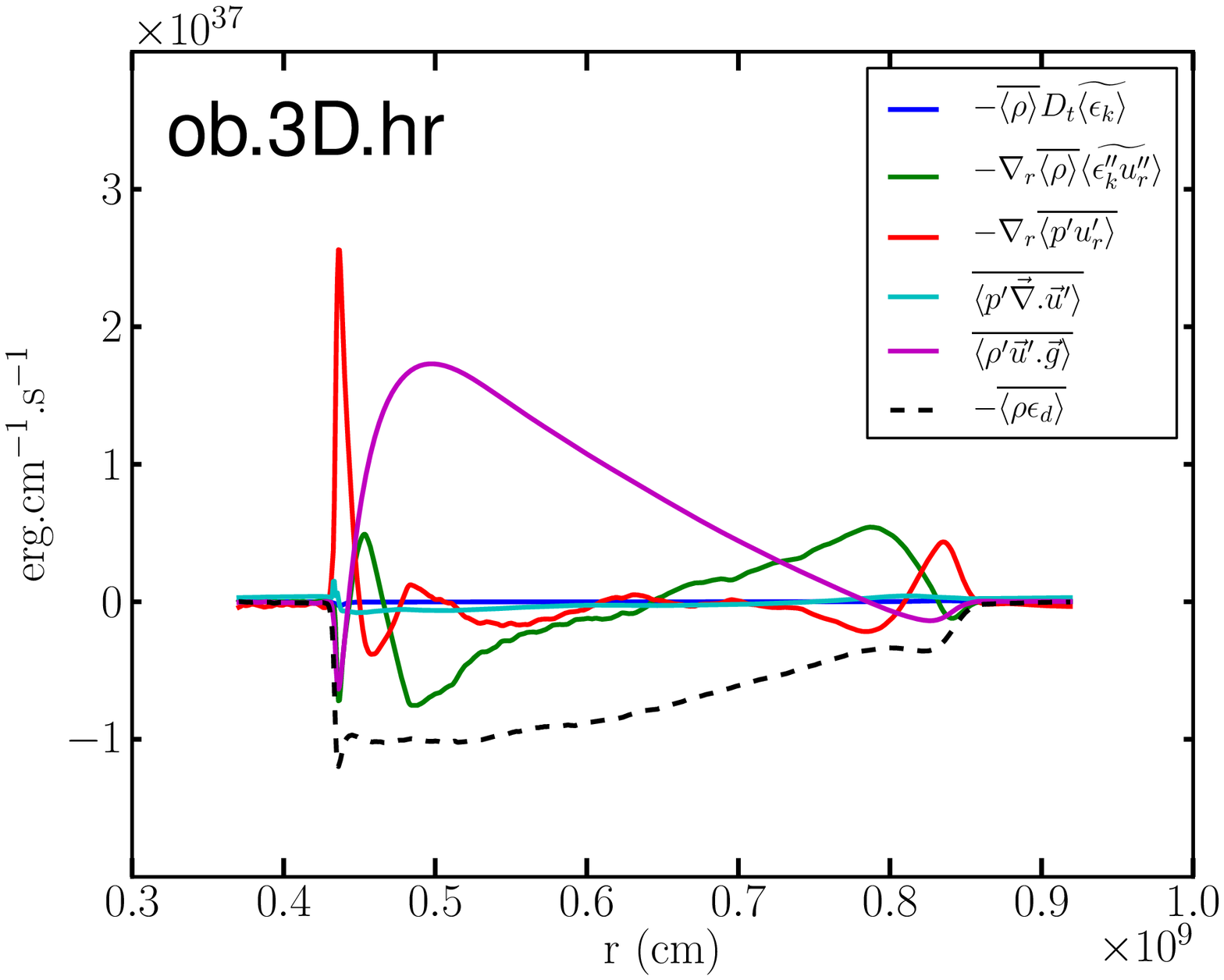}}
\caption{Comparison of the kinetic energy balance for different resolutions. Top panels: models {\sf rg.3D.lr} and {\sf rg.3D.mr}.  Bottom panels: models {\sf ob.3D.mr} and {\sf ob.3D.hr}.}
\label{fig:resolution_ekin_balance}
\end{figure*}

In stellar evolution calculations, the locations of the convective zone boundaries are based on linear criteria for dynamical stability, namely the Schwarzschild or the Ledoux criteria (the latter takes composition gradients into account). However, due to inertia, fluid parcels can cross this limit. \cite{zahn_convective_1991} presents an analytical investigation of the problem \citep[see also][]{schmitt_overshoot_1984,rempel_overshoot_2004}.  Zahn describes as ``penetrative convection" the process in which the superadiabatic region, grows in size due to an efficient thermodynamical mixing at the convective boundary. This is the case on the earth, where the planetary boundary layer grows in size during the day. Regarding the connection of the unstable region to the stable interior, \cite{zahn_convective_1991} distinguishes between ``overshooting", in which the transition is made directly in a shallow thermal boundary layer, and ``subadiabatic penetration", in which the penetrative flow first establishes a nearly adiabatic, yet stable, region below the convective zone. \cite{zahn_convective_1991} suggests that in stellar interiors, owing to large values of the P\'eclet number, the conditions for subadiabatic penetration are fulfilled. 

Pioneering numerical studies of the problem are presented in \cite{hurlburt_nonlinear_1986, hurlburt_penetration_1994-1}. \cite{hurlburt_penetration_1994-1} study how subadiabatic penetration/overshooting changes depending on the ``stiffness" of the interface, a free parameter of their models. Their models show subadiabatic penetration for low values of the stiffness, whereas for large stiffness they only have an overshooting layer. More recent 2D work by \cite{rogers_penetrative_2005} arrive to a similar conclusion. \cite{brummell_penetration_2002-4} revisit the problem in 3D, using a similar approach as \cite{hurlburt_penetration_1994-1}. None of their 3D models show evidence for a subadiabatic region. The authors argue the reason is the lower filling factor of plumes in 3D turbulent convection, resulting in lower local P\'eclet numbers. Surprisingly, their numerical models have a smaller convective region than in the initial model.
 
How do the red giant models compare with these studies ? We discuss here only the bottom boundary, as the analysis of the top boundary is undermined by our artificial treatment of the surface cooling. The left panel of Fig. \ref{fig:rg_penetration} shows the dimensionless temperature gradients in model {\sf rg.3D.mr}. The stratification is very close to adiabatic in the bulk of the convective zone, owing to efficient convection. Note that the structure is taken from the averaged model, and that it does not evolve over the time of the simulation.
We show with two dotted lines the region where the enthalpy flux has a negative bump (see left panel of Fig. \ref{fig:energyflux_balance}). We identify this region with the convective boundary layer. It has a radial extent of 60\% of the local pressure-scale height. The change in sign of the enthalpy flux marks the start of the stably stratified (subadiabatic) region. The radius at which it happens is only slightly smaller than the location of the convective boundary in the initial model: our models do not show evidence for strong convective penetration. Do our models show evidence for subadiabatic penetration? Based on an inspection of the left panel in Fig. \ref{fig:rg_penetration} and of the profile of $N^2$, see right panel in Fig. \ref{fig:rg_structure}, we can see a shallow, nearly adiabatic region, which occupies roughly 30\% of the convective boundary region. The thermal boundary layer, where the temperature gradient connects smoothly to the radiative value, occupies the remaining 70\%. Therefore, we obtain a similar result to \cite{brummell_penetration_2002-4}: this model is characterized by overshooting. The size of our convective boundary layer is on the low side of the range of values resulting from the parameter study of \cite{brummell_penetration_2002-4}. This suggests that our boundary is rather ``stiff". Note that this stiffness is a natural outcome of our models, and not an input parameter as in the above mentioned studies. Furthermore, we have a realistic thermal conductivity profile, depending on density and temperature rather than only on depth. 

Following \cite{zahn_convective_1991}, \cite{brummell_penetration_2002-4} suggest that they would obtain subadiabatic penetration by modeling higher P\'eclet number flows. The oxygen-burning shell models have a formally infinite P\'eclet number\footnote{Being negligible,  radiative diffusion was removed from the code for increased efficiency.}, as thermal diffusion is negligible, and give an insight into the large P\'eclet number limit.  As discussed in detail in \cite{meakin_turbulent_2007}, the oxygen-burning shell models show evidence for ``turbulent entrainment". The physical process is one in which the turbulent kinetic energy present at the boundary is converted to potential energy as it draws material into the convection zone, mainly through shear instabilities and wave breaking. As a result, the stratification is weakened, and this leads a steady increase in the size of the convective zone, see Fig. 4 in \cite{meakin_turbulent_2007}. As such, this is the same effect as the convective penetration discussed above. However, convective penetration seems to be usually thought as being due to the combined effect of large scale plumes, whereas turbulent entrainment describes the continuous erosion by the small scale turbulence at the interface. Both effects have the same signature: a negative buoyancy work ($W_b < 0$). In the first case, it results from buoyancy braking of the plumes; in the second case, it results from the work the flow is doing against gravity in the process of mixing the stable layer material. These effects are not easy to distinguish in numerical simulations, and both may contribute. \cite{meakin_turbulent_2007} characterize the entrainment rate at convective boundaries based on the bulk Richardson number Ri$_B$, which measures the stiffness of the interface by taking turbulence into account. This elucidates the behavior of the convective boundary at very large P\'eclet numbers.

\begin{figure*}[t]
\center
\includegraphics[width=0.8\linewidth]{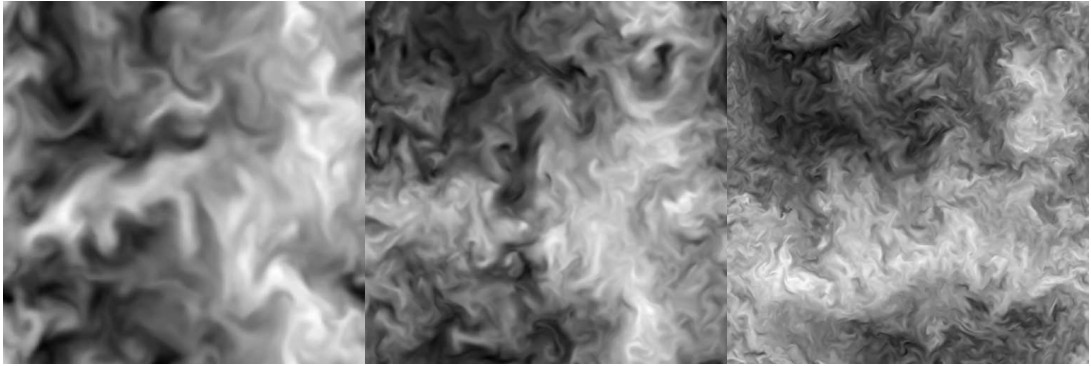}
\caption{Instantaneous snapshots of the radial velocity field in an horizontal plane in the middle of convective zone for models {\sf ob.3D.lr}, {\sf ob.3D.mr},  {\sf ob.3D.hr} (from left to right). Light tones indicate upflows, dark tones indicate downflows. Each subfigures shows the full angular domain $45^\circ \times 45^\circ$.}
\label{fig:resolution_progression}
\end{figure*}

At lower P\'eclet numbers, how do non-adiabatic effects modify this process?
The convective boundary layer in the red giant models is characterized by $W_b<0$ and $W_p<0$; see the middle-left panel of Fig. \ref{fig:rans_rg}. The figure shows that $W_b$ is dominant: kinetic energy is mostly converted into potential energy. Furthermore, the bottom-left panel of Fig. \ref{fig:rans_rg} shows that the divergence of the enthalpy flux leads to heating in the convective boundary layer. This is better analyzed in the framework of the entropy balance, Eq. (\ref{eq:rans_entropy}), which is shown in the right panel of Fig. \ref{fig:rg_penetration}. In the overshooting layer, the divergence of the entropy flux heats (note that in the overshooting layer $f_h \approx T f_s$, see middle-left panel in Fig. \ref{fig:fh_decomposition}). However, it is compensated by cooling from radiation. As a consequence, a quasi-steady state in which non-adiabatic processes counter-balance the effects of turbulent entrainment is possible: the convective region does not increase in size as it does in the oxygen burning case where radiative effects are negligible. The bump in the radiative luminosity is another manifestation of this process. This can be seen also on the left panel in Fig. \ref{fig:rg_penetration}, where the convective boundary is characterized by a temperature gradient that is subadiabatic yet super-radiative:

\begin{equation}
\nabla_\mathrm{rad} < \nabla < \nabla_\mathrm{ad}.
\end{equation}

\cite{zhang_sound-speed_2012} show how taking this effect into account improves the agreement of solar models with helioseismology data \citep[see also][]{christensen-dalsgaard_more_2011}.

Overshooting and subadiabatic penetration correspond to ``thermally inhibited" turbulent entrainment/penetrative convection. Subadiabatic penetration requires large P\'eclet numbers. The estimate given by Eq. (\ref{eq:global_peclet}) maybe misleading, as the scales which are involved in the subadiabatic penetration process can be much smaller, and characterized by lower ``turbulent" P\'eclet numbers. Furthermore, the bulk Richardson number might be relevant for subadiabatic penetration as well. A better understanding of the relative effects of the sporadic plumes \citep{meakin_turbulent_2007,arnett_turbulent_2009} that hit the convective boundary and of the continuous erosion by the turbulence is necessary. We suspect these concepts could shed new light on the results obtained in \cite{brummell_penetration_2002-4}.

\subsection{Comparison of different resolutions}
\label{resolution_effects}

We have performed the various analysis presented in the previous sections for different resolutions and found good agreement, as shown for instance by the various quantities summarized in Tables \ref{table:rg_runs} and \ref{table:ob_runs}. We have not found any significant deviation in the physical results that could stem from resolution issues. The oxygen-burning shell models at the lowest resolution ({\sf ob.3D.lr}) shows spurious oscillations in some averaged quantities. This seems to be related with difficulties Riemann based solvers have with stratification at low resolution. Even in this case, both the profiles and amplitudes of these quantities are actually in good agreement with the results obtained from higher resolutions  which are free of these problems. However, we do not have convergence in the mean fields in the narrow region of steep gradients at the base of the oxygen-burning shell, which are unresolved in the models considered here. There, the dissipation and compositional mixing are affected at the grid scale by the numerical algorithm, which undermines the RANS analysis. The Riemann solver replaces a steep gradient by a contact discontinuity while the RANS analysis requires continuity; this issue requires further study, although the general behavior is relatively sane and the discrepancy localized.

Within the convective region, the mean-field analysis shows robust behavior regarding resolution. The most resolution sensitive diagnostic might be the kinetic energy dissipation, which in our models is purely due to numerics at the grid scale. Figure \ref{fig:resolution_ekin_balance} compares the kinetic energy balance in models {\sf rg.3D.lr} and {\sf rg.3D.mr}, and models {\sf ob.3D.mr} and {\sf ob.3D.hr}. The balance looks nearly the same at different resolutions. The kinetic energy dissipation profiles are very similar, although the resolutions differ by a factor of two. This suggests that the kinetic energy dissipation is set by the large scale properties of the flow, and does not depends on the physics at small scales (here the grid scale). We interpret this as an indication that the dynamics in the turbulent convective zone is governed by the large scale dynamics, characterized by the coherent plumes which are well resolved even at our lowest resolution. This is consistent with the picture of the turbulent cascade \citep{richardson_weather_1922, kolmogorov1941dit}: at large Reynolds numbers the rate of dissipation is set by the energy injection at large scale, and is independent of the value of the viscosity. This is the so-called ``dissipation anomaly". The viscosity sets the scale at which dissipation occurs, here the grid scale.

The turbulent regime obtained in numerical simulations (ours and others) is characterized by coherent plumes which propagate (upward or downward) over a significant fraction of the convective region, if not the whole region. They govern the large scale dynamics and are key in guiding the modeling of the highly non-local and non-isotropic transport properties of the flow \citep[see e.g.][]{rempel_overshoot_2004,lesaffre_two-stream_2005,belkacem_closure_2006,kupka_effects_2007,meakin_turbulent_2007}. However, one should bear in mind that our numerical simulations have non-dimensional numbers (e.g., Re, Pr, Ra) which are orders of magnitude different from the values relevant to stellar hydrodynamics. Figure \ref{fig:resolution_progression} shows snapshots of the flow in the oxygen-burning shell model for three different resolutions. Although the flow is characterized by structures at smaller and smaller scales, we do not see any evidence for a different global behavior of the flow. Whether a transition to a different regime occurs at (much) larger resolution is an outstanding problem. A similar question arises in the study of the simpler, but not less fundamental, Rayleigh-B\'enard convection problem. The quest for an understanding and characterization of the ``ultimate" state of turbulent Rayleigh-B\'enard  convection is the focus of much experimental and theoretical work \citep[see reviews by][]{siggia_high_1994,ahlers_heat_2009}. Although there are significant physical differences with the stellar case, e.g., stemming from the different nature of boundaries, the extremely low values of the Prandtl number, or the effects of compressibility, it can be expected that a better understanding of turbulent Rayleigh-B\'enard convection will provide valuable insight into the turbulent regime at which stellar convection operates \citep[see e.g. discussion in][]{spruit_convection_1997}. Theoretical studies support the existence of these plumes in stellar convective zones \citep{simon_convective_1991,rieutord_turbulent_1995}. Numerical models of the propagation of a plume through the adiabatic stratification show the development of secondary instabilities \citep{rast_compressible_1998,clyne_interactive_2007}, but they deal with an idealized situation as the interaction between plumes clearly dominates in our numerical models. Observationally, these plumes are too deep and too small to be detected by current helioseismology measurements; see \citep{hanasoge_anomalously_2012} and references therein.

\section{Conclusion}
\label{conclusion}

This paper presented 3D models of the turbulent convection in the envelope of a red giant star and in the oxygen-burning shell of a supernova progenitor. The two models differ significantly in their physical properties: they have radically different equations of state, the effects of thermal diffusion is negligible in one but important in the other, one model is multi-fluid and includes nuclear burning, whereas the other is mono-fluid and has cooling at the surface. Their common point, which is the focus of this work, is the presence of a turbulent convective zone which dynamics is controlled by the hydrodynamical equations. Finally, two different numerical methods and codes were used to produce the data. To deal with such a heterogenous set of data, we developed in Sect.~\ref{1DRANS} a set of 1D horizontally-averaged equations that provides a framework for a systematic analysis of hydrodynamical simulations. We showed in Sect. \ref{rans:analysis} that our numerical models show good consistency with the physical equations, although we identified spurious effects localized in the region of the steep, unresolved  gradients present in the oxygen-burning model.

Both our models are characterized by low Mach flows, so that compressible effects are negligible. Similarly, both models have large P\'eclet number (formally infinite in the oxygen-burning shell case) so that the deviation from adiabaticity is small in the convective zone, and has no effect on the dynamics\footnote{In the red giant model, this would not be the case near the surface if it was modeled realistically.}. However, our analysis is general and applies also to flows with larger Mach number and superadiabatic stratifications, two conditions which are found in photospheric convective regions. We plan to apply a similar analysis to photospheric convection. Our mean-field analysis emphasized very similar behavior in both stellar models, without noticeable dependance on the numerical resolution. Both the radial expansion velocity and kinetic energy balances are in a statistically steady state, whereas the total energy balance shows an evolution of the background on a longer timescale. This is due to the clear separation between the dynamical timescale of the system (the turnover timescale) and the nuclear/thermal timescale of the models. The kinetic energy dynamics can be understood as a balance between driving at large scales and dissipation at small scales, connected by the turbulent cascade. Understanding the spatial distribution of the driving and of the dissipation is important for insight into the non-locality of convection.

\begin{figure}[t]
   \center
   \includegraphics[width=0.8\linewidth]{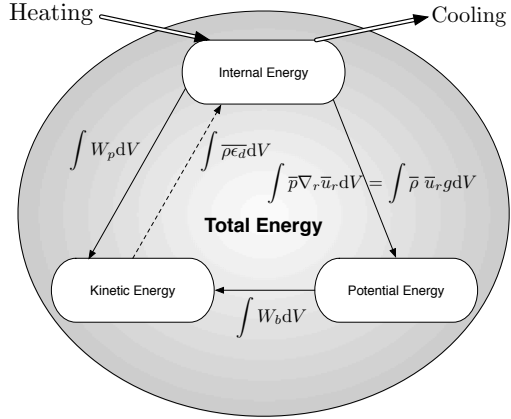} 
   \caption{Global energy balance in turbulent convection. The arrow directions are consistent with the labels, i.e. for positive values the energy flows along the arrows. Continuous lines emphasize the reversible character of the processes, whereas the dashed line denotes the irreversible character of kinetic energy dissipation.}
   \label{fig:total_energy_balance}
\end{figure}

We have shown that the differences between the red giant and oxygen-burning shell models stem mainly from the degree of stratification. This led us to introduce the distinction between ``shallow" convection, in which the velocity correlation length-scales are less than the density scale-height, and ``deep" convection in which they are of the same order, or larger. As shown in Sect.~\ref{pressure_fluctuations}, the effect of stratification on the dynamics can be understood in terms of the magnitude of pressure fluctuations. We discussed how this impacts the mean-field balances: for deep convection pressure-dilatation becomes a non-negligible source of kinetic energy, and the acoustic flux contributes to the transport of kinetic energy and enthalpy. We showed in Sect.~\ref{KEdriving} the connection between the transport of enthalpy, the rate of production of turbulent kinetic energy, and finally the rate of turbulent dissipation at small scales. This should not be surprising: large scale transport of enthalpy needs motion, motion becomes turbulent, and turbulence dissipates kinetic energy at small scales. As a consequence, we find that in a quasi-steady state the rate of dissipation of turbulent kinetic energy is of the order of the convective luminosity. This is consistent with the rate of damping of kinetic energy inferred from our models. How turbulent dissipation affects stellar evolution is an open question. Figure \ref{fig:total_energy_balance} puts the role of turbulent dissipation in perspective regarding the global conservation of energy. The convective instability taps energy of the unstable stratification and converts potential energy into kinetic energy, the work done by the background flow $\av{u_r}$ on the mean background $\av{p}$ converts internal energy into potential energy, and turbulent dissipation closes the loop by converting kinetic energy into internal energy. In a statistically steady state, the amount energy per unit time which is ``flowing" in these channels is $W_b$ (see Eqs. \ref{eq:ek_balance} and \ref{eq:fm_decomposition}).

We have discussed in Sect. \ref{penetration} the overshooting process which is observed in the red giant models. Comparing with the oxygen-burning models, we discussed how turbulent entrainment relates to the more classical concepts of overshooting and subadiabatic penetration used in the stellar context.  Further investigations on the relative effect of the plumes and turbulent entrainment is desirable.

A turbulent model for stellar convection should be able to reproduce the behavior of the mean-field equations without the need to resort to expensive 3D simulations.  In the RANS framework, evolution equations can be derived for the turbulent fluxes. The resulting equations involve higher order terms, for which additional evolution equations can be derived. This leads to a hierarchy of equations which have to be closed with appropriate relations. Future work will apply the same systematic study to higher order equations, aiming at identifying the most important terms to guide the closure strategy  (Moc\'ak et al., in preparation). We intend to release publicly\footnote{\url{http://stellarmodels.org}} our RANS analyzed data and provide analysis and plotting subroutines as open source materials. 

We plan to extend this work by including rotation and magnetic field. As mentioned in Sect. \ref{1DRANS}, the formulation of suitable 1D mean-field equations is not possible when rotation or magnetic field are included. Nevertheless, if non-axisymmetric instabilities are not important, it is possible to average the equations over the azimuthal direction, resulting in a set of 2D mean-field equations. This opens the possibility of performing stellar evolution in 2D, with an appropriate treatment of these effects (in the limit of low rotation rate due to spherical geometry), see e.g. \cite{deupree_stellar_1990,deupree_stellar_2001,li_two-dimensional_2006,li_two-dimensional_2009}. The MUSIC code, which is based on time-implicit methods, provides the ideal framework for that.


\acknowledgments
MV acknowledges support from an International Newton Fellowship from the Royal Society.
CM and WDA acknowledge support from NSF grant 1107445 at the University of Arizona.
This work used the Extreme Science and Engineering Discovery Environment (XSEDE), which is supported by National Science Foundation grant number OCI-1053575.

\clearpage

\appendix

\section{Elliptic equation for the pressure}
\label{elliptic_pressure_equation}

We start from the momentum equation:

\begin{equation}
\rho_0\Big ( \partial_t \vec u + \vec u \cdot \vec \nabla \vec u \Big ) = - \vec \nabla  p' + \rho' \vec g,
\end{equation}

\noindent where we have neglected density fluctuations in front of the Lagrangian derivative, and we have subtracted the hydrostatic background. Taking the divergence of this equation removes the time derivative both in the Boussinesq ($ \vec\nabla  \cdot \vec u$=0) and in the anelastic  ($\vec \nabla  \cdot ( \rho_0 \vec u)$=0) approximations:

\begin{equation}
\vec \nabla \cdot \Big ( \rho_0 \vec u \cdot \vec \nabla  \vec u \Big ) = - \Delta  p'  - g \frac{\partial \rho'}{\partial r},
\end{equation}

\noindent where we considered that $g$ was constant. We write this equation as

\begin{equation}
\Delta  p' = - \vec \nabla : \big ( \rho_0 \vec u' \otimes \vec u'  \big )  - g \frac{\partial \rho'}{\partial r}.
\end{equation}

\bibliography{references}

\end{document}